\documentclass[aps,prd,twocolumn,superscriptaddress,nofootinbib]{revtex4-1}

\usepackage{latexsym}
\usepackage{amsmath}
\usepackage{amssymb}
\usepackage{amsfonts}
\usepackage{bm}

\usepackage{color}

\usepackage{supertabular} 
\usepackage{placeins}
\usepackage{epsfig}
\usepackage{graphicx}

\begin{document}


\title{Exotic resonances of fully-heavy tetraquarks in a lattice-QCD insipired quark model}


\author{Gang Yang}
\email[]{yanggang@zjnu.edu.cn}
\affiliation{Department of Physics, Zhejiang Normal University, Jinhua 321004, China}

\author{Jialun Ping}
\email[]{jlping@njnu.edu.cn}
\affiliation{Department of Physics, Nanjing Normal University, Nanjing 210023, China}

\author{Jorge Segovia}
\email[]{jsegovia@upo.es}
\affiliation{Dpto. Sistemas F\'isicos, Qu\'imicos y Naturales, Univ. Pablo de Olavide, 41013 Sevilla, Spain}



\begin{abstract}
Fully-heavy tetraquark states, \emph{i.e.} $cc\bar{c}\bar{c}$, $bb\bar{b}\bar{b}$, $bb\bar{c}\bar{c}$ ($cc\bar{b}\bar{b}$), $cb\bar{c}\bar{c}$, $cb\bar{b}\bar{b}$, and $cb\bar{c}\bar{b}$, are systematically investigated by means of a non-relativistic quark model based on lattice-QCD studies of the two-body $Q\bar{Q}$ interaction, which exhibits a spin-independent \emph{Cornell} potential along with a spin-spin term.
The four-body problem is solved using the Gaussian expansion method; additionally, the so-called complex scaling technique is employed so that bound, resonance, and scattering states can be treated on the same footing. Moreover, a complete set of four-body configurations, including meson-meson, diquark-antidiquark, and K-type configurations, as well as their couplings, are considered for spin-parity quantum numbers $J^{P(C)}=0^{+(+)}$, $1^{+(\pm)}$, and $2^{+(+)}$ in the $S$-wave channel. 
Several narrow resonances, with two-meson strong decay widths less than 30 MeV, are found in all of the tetraquark systems studied. Particularly, the fully-charm resonances recently reported by the LHCb Collaboration, at the energy range between 6.2 and 7.2 GeV in the di-$J/\psi$ invariant spectrum, can be well identified in our calculation. Focusing on the fully-bottom tetraquark spectrum, resonances with masses between 18.9 and 19.6 GeV are found. For the remaining charm-bottom cases, the masses are obtained within a energy region from 9.8 GeV to 16.4 GeV. All these predicted resonances can be further examined in future experiments.
\end{abstract}


\maketitle


\section{Introduction}

Many efforts have been made in the past twenty years in order to understand exotic tetra-, penta- and even hexa-quark systems, constituted mostly by heavy quarks. For example, in 2015, the hidden-charm pentaquark states $P_c(4380)$, $P_c(4312)$, $P_c(4440)$ and $P_c(4457)$ were discovered by the LHCb collaboration~\cite{lhcb:2019pc, Aaij:2015tga}, adding them to a large number of unconventional heavy mesons called jointly XYZ states, \emph{e.g.} the $X(3872)$ discovered by the Belle Collaboration in 2003~\cite{SKC2003}, the $Y(4260)$ discovered by the BaBar Collaboration in 2005~\cite{ba:2005prl}, $Z_c(3900)$ discovered by the BESIII Collaboration in 2013~\cite{ta:2013prl}, etc. 

The fully-heavy tetraquark states $QQ\bar{Q}\bar{Q}$ ($Q=c,\,b$) have recently attracted much attention. In 2017, the CMS Collaboration reported a benchmark measurement of $\Upsilon(1S)$-pair production in $pp$ collisions at $\sqrt{s}$=8 TeV~\cite{vkams:2017jhep}. A preliminary analysis of the CMS data shows an excess at 18.4 GeV in the $\Upsilon(1S) \ell^+ \ell^-$ decay channels~\cite{SD2018PHD, SD2018II, KY2018}. This excess, if confirmed by future experiments, may indicate a fully-bottom tetraquark state ($bb\bar{b}\bar{b}$). Besides, a significant peak at $18.2\,\text{GeV}$ was observed in Cu+Au collisions at RHIC~\cite{LCBland2019} but the LHCb and CMS collaborations~\cite{raba:2018jhep, cms4b2020} were not able to confirm it from the $\Upsilon(1S)\mu^+ \mu^-$ invariant mass spectrum. Recently, in the di-$J/\psi$ invariant mass spectrum, a narrow peak at 6.9 GeV, a broad one between 6.2 and 6.8 GeV, and a hint for a possible structure around 7.2 GeV were reported by the LHCb collaboration, which could indicate the existence of fully-charm tetraquarks~\cite{LA2020CERNINDICO}. It is then expected that more investigations on the existence of fully-heavy tetraquark states will be performed in the future in, for instance, the LHCb experiment at CERN.

The search of fully-heavy tetraquark states is valuable from the experimental side but also from the theoretical view point~\cite{KY2013, JR200801962, KCSZ200807670}. The debates on the existence of $QQ\bar{Q}\bar{Q}$ $(Q=c,\,b)$ states were quite intense before the LHCb's findings in the di-$J/\psi$ invariant mass spectrum. A $bb\bar{b}\bar{b}$ bound state was supported by various models~\cite{avb:2012prd, mna:2018epjc, aeap:2018epjc, mabjfcdres2019}, QCD sum rules studies~\cite{zgwqqqq:2017epjc, wchxc:2017plb}, and a naive diffusion Monte Carlo calculation~\cite{yb:2019plb}. The study of its decay properties concluded that a fully-bottom tetraquark state with $J^P=2^+$ is possible~\cite{CBAGLM2020}. As for the potential $cc\bar{c}\bar{c}$ state, it was suggested that its mass is located between $5$ and $6$ GeV through various phenomenological models~\cite{avb:2012prd, vrdfsn:2019cpc, avbakl:2011prd, mksnjl:2017prd} and the Bethe-Salpeter equation approach~\cite{whge:2012plb}.  Additionally, narrow $bb\bar{b}\bar{c}$ and $bc\bar{b}\bar{c}$ tetraquark states were predicted~\cite{jmrav:2017prd, jwyrl:2018prd}. On the contrary, there were counter examples on the existence of $QQ\bar{Q}\bar{Q}$ tetraquark states. For example, it was shown that bound states of $cc\bar{c}\bar{c}$ and $bb\bar{b}\bar{b}$ systems are impossible through model investigations of Refs.~\cite{jmrav:2017prd, jwyrl:2018prd, xc:2019epja, mslqfl:2019prd, gjw:2019arx, jmravjv2018} but also in preliminary lattice-QCD computations~\cite{cheec:2018prd}.

The interpretation of the newly reported structures in the di-$J/\psi$ invariant mass spectrum have been carried out by various theoretical approaches. In a non-relativistic model with a compact diquark-antidiquark configuration, masses of the $S$-wave fully-charm tetraquark states are predicted to be between 5.96 and 6.32 GeV~\cite{PLTO2020}. In a potential model including the linear-confinement and the one-gluon-exchange contributions, fully-charm tetraquark masses are predicted at 6.5 GeV and 6.9 GeV, which can be identified with the observed structures as $S$- and $P$-wave fully-charm tetraquarks~\cite{MSLFXLXHZQZ2020}. This conclusion differs from a study with a dynamical diquark model, where the two resonances around 6.7 GeV and 6.9 GeV are identified with the 1P and 2S multiplets~\cite{JFGRFL200801631}. Furthermore, the narrow and broad structures observed in the di-$J/\psi$ invariant mass spectrum are explained as radial excitations of the fully-charm state within the QCD sum rules approach~\cite{ZGW2020FHT}, the string-junction picture~\cite{MKJLR200904429}, and the extended relativized quark model~\cite{QLDYCYBD}. On the other hand, based on the perturbative QCD approach~\cite{RMASNARDRGR200801569}, it was predicted that both the narrow and broad structures prefer molecular-like configurations. On the contrary, a holography inspired model suggests a compact picture for the structure around 6.9 GeV~\cite{JSDW200801095}. Concerning the quantum numbers of the di-$J/\psi$ structures, the spin-parity of the $X(6900)$ resonance is suggested to be $J^P=2^+$ within the constituent quark model of Ref.~\cite{XJYXHHJP200613745} and the relativistic quark model of Ref.~\cite{RNFVOG200913237}, while it is predicted to be $0^+$ or $1^+$ by an effective potential model~\cite{JZSSPZ200910319} and $0^+$ by a diffusion Monte Carlo calculation~\cite{MCGFDSJS200911889}.

In addition, a unitary coupled-channels approach seems to describe well di-J/$\psi$ invariant mass spectrum finding hints of a near-threshold molecular state $X(6200)$ with quantum numbers $J^{PC}=0^{++}$ or $2^{++}$~\cite{XDVBFGCHAN2009}. Meanwhile, an ab initio perturbative QCD investigation suggested that there exists another state near the peak at 6.9 GeV and thus the nature of $X(6900)$ can be uncovered by calculating the cross section of this undiscovered state~\cite{YMHZ200908376}. On the other hand, strong decay properties of the fully-charm tetraquark states were investigated in Ref.~\cite{HCWCXLSZ2020}, predicting that the broad structure is a $S$-wave state with either $J^{PC}=0^{++}$ or $2^{++}$ and the narrow peak can be identified as a $P$-wave state with quantum numbers $J^{PC}=0^{-+}$ or $1^{-+}$. The production of the fully-charm tetraquark state at 6.9 GeV was also studied by a $p\bar{p}$ annihilation process~\cite{XYQLHXYXYHXC2020}, a dynamic simulation~\cite{JZDCXLTM200807430}, and a model-independent fragmentation mechanism~\cite{FFYHYJWSXXJZ2009}.

In this work, we explore the possibility of having bound, resonance and scattering states of fully-heavy quark systems, \emph{viz.} $QQ\bar{Q}\bar{Q}$ $(Q=c,\,b)$, with spin-parity $J^{P(C)}=0^{+(+)}$, $1^{+(\pm)}$ and $2^{+(+)}$ in the $S$-wave channel.\footnote{Note here that an exploratory analysis of this study for the $cc\bar c \bar c$ and $bb\bar b\bar b$ tetraquark systems was recently published in Ref.~\cite{gy:symTP}.} We employ a non-relativistic quark model for the two-body interaction between heavy quarks according to the Lattice-QCD study of Ref.~\cite{TKSSPRD2012}. The four-body problem is solved by using the Gaussian expansion method~\cite{Hiyama:2003cu}, which has been demonstrated to be as accurate as a Faddeev calculation. The complete set of four-body configurations: meson-meson, diquark-antidiquark, and K-type arrangements, as well as their couplings, are considered in the calculation. Meanwhile, a coupled-channels calculation which treats the bound, resonance, and scattering states on the same footing is performed by employing the complex scaling method~\cite{JA22269, EB22280, Bsimon1972, YKHo1983, NMoieyev1998} according to the so-called ABC theorem~\cite{JA22269, EB22280}. This tool has been already used in previous studies of nuclear~\cite{AMKK2006, MKMK2014, HIKMM2015, HLCK2016} and hadron~\cite{gy:2020dht, gy:2020dhts,gy:2020dcp, gy:2021x2900} physics.

This manuscript is organized as follows. The theoretical framework is presented in Sec.~\ref{sec:model}, including the lattice-QCD inspired potential model and the general structure of the four-body wave function. Section~\ref{sec:results} is devoted to the analysis and discussion of the obtained results. And we summarize our theoretical work in Sec.~\ref{sec:summary}.


\section{Theoretical framework}
\label{sec:model}

The kinetic motion for heavy quarks (charm and bottom) can be treated in a non-relativistic way, and their interactions can be described via a potential model. Here, we employ a potential model inspired by the Lattice-QCD investigation of Ref.~\cite{TKSSPRD2012}, \emph{viz.} the interaction between a heavy quark and a heavy antiquark can be well approximated by the spin-independent Cornell potential along with a spin-spin term. For the 
four-body system $QQ\bar{Q}\bar{Q}$ $(Q=c,\,b)$, the Hamiltonian results to be
\begin{equation}
H = \sum_{i=1}^{4}\left( M_Q+\frac{\hat{\bf p}^2_i}{2M_Q}\right) + \sum_{j>i=1}^{4} V_{ij}({\bf r}_{ij}) \,,
\label{eq:Hamiltonian}
\end{equation}
where $M_Q$ is the mass of the heavy quarks and antiquarks. The two-body interaction potential can be written as
\begin{equation}\label{CQMV}
V_{ij}({\bf r}_{ij}) = -\frac{3}{16}({\bf \lambda}^a_i\cdot{\bf \lambda}^a_j) \left[-\frac{\alpha}{|{\bf r}_{ij}|}+\sigma |{\bf r}_{ij}|+\beta e^{-\gamma|{\bf r}_{ij}|}({\bf s}_{i}\cdot{\bf s}_{j}) \right]  \,,
\end{equation}
which includes the coulomb, linear-confining and spin-spin interactions. The color-dependence of the interaction is encoded in the SU(3) Gell-Mann matrices, $\lambda^a_i$ $(a=1,2,...,8)$. Since the two-body interaction depends only on the distance between the two quarks/antiquarks labeled by $i$ and $j$, $|{\bf r}_{ij}|=|{\bf r}_{i}-{\bf r}_{j}|$, the four-body problem can be factorized into a center-of-mass motion and relative motions. In the following, we subtract the center-of-mass motion and focus on the relative motions of the four-body system. The model parameters $\alpha$, $\beta$, $\gamma$, and $\sigma$ can be determined via a calculation of the mass spectrum of the $S$-wave $Q\bar{Q}$ mesons. Table~\ref{model} lists the values of the model parameters, which are the ones collected in Ref.~\cite{HFUECHENC2020} except the spin-spin coefficient of the charm-bottom case, $\beta_{cb}$, which is determined herein. The calculated masses of the $S$-wave $Q\bar{Q}$ mesons along with their experimental values are listed in Table~\ref{Mmeson}. The theoretical results are consistent with the experimental data, providing a solid ground to study possible bound and resonance states in the four-body system $QQ\bar{Q}\bar{Q}$.

\begin{figure}[ht]
\epsfxsize=3.4in \epsfbox{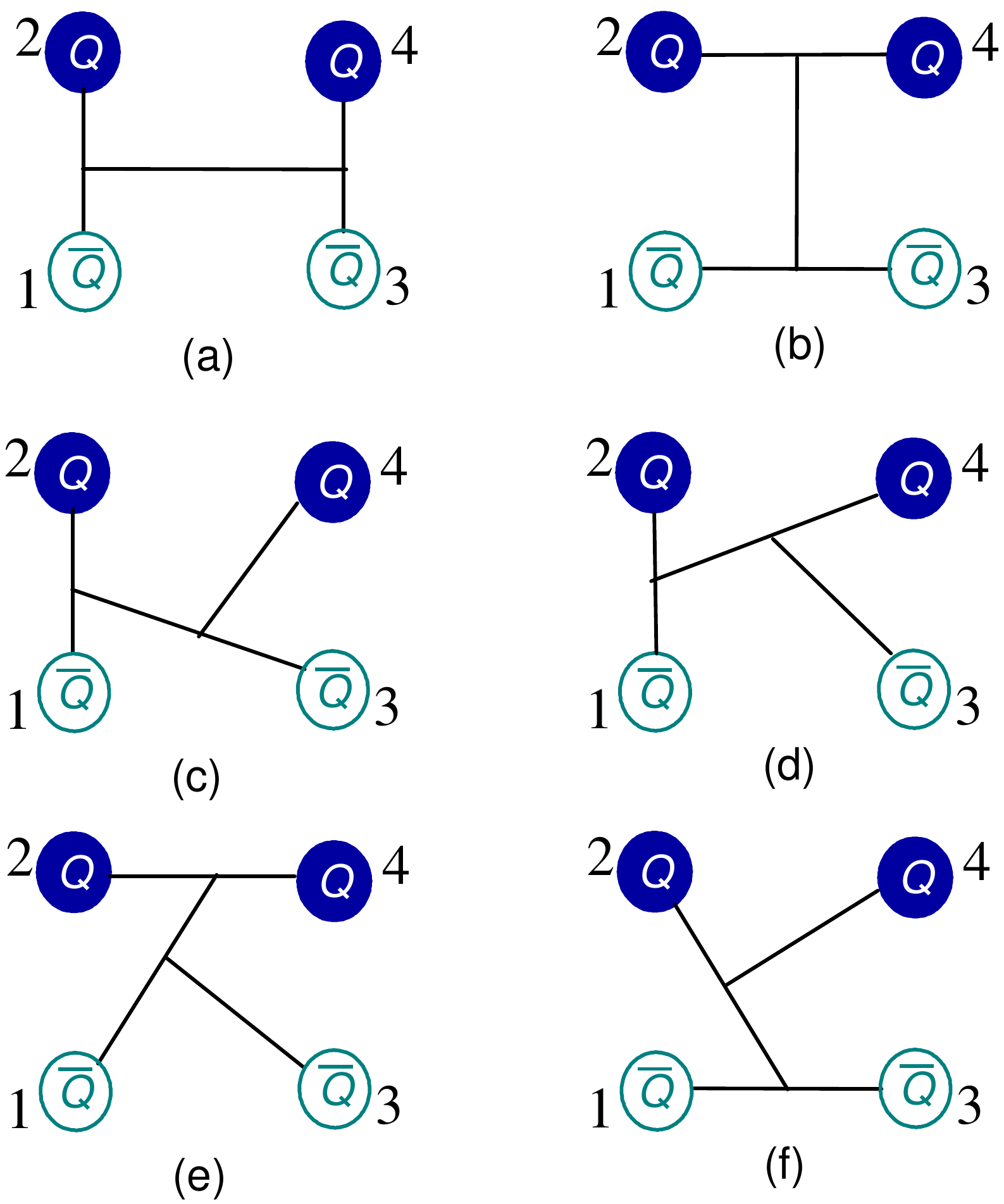}
\caption{All possible configurations in the fully-heavy tetraquarks. Panel $(a)$ is the meson-meson structure, panel $(b)$ is the diquark-antidiquark one and the other panels are of K-type. $(Q=c,\,b)$} \label{QQqq}
\end{figure}

Figure~\ref{QQqq} shows a complete set of configurations for the fully-heavy tetraquark system $QQ\bar{Q}\bar{Q}$ $(Q=c,\,b)$. Figs.~\ref{QQqq}(a) and \ref{QQqq}(b) correspond to the meson-meson structure and diquark-antidiquark one, respectively.  Figs.~\ref{QQqq}(c)-(f) are the so-called K-type configurations. In general, all these configurations are coupled to each other; on the other hand, it was shown~\cite{Harvey:1980rva, Vijande:2009kj} that if all possible excited states are included, it is enough to consider only the color-singlet channel. Since it is difficult to include all excited states and we are interested in the low-lying ones, it is more convenient to consider all possible configurations and the couplings among them. The price to pay following this procedure is that the overlap matrix becomes singular durinf the process of diagonalization because the exchange of identical particles. This can be fixed by diagonalizing the overlap matrix and drop the eigenvectors which correspond to vanishing eigenvalues. Then, a new Hamiltonian matrix can be generated using the rest of eigenvectors and the energy levels of the four-body system can be obtained from the reconstructed Hamiltonian matrix.

\begin{table}[!t]
\caption{\label{model} Model parameters.}
\begin{ruledtabular}
\begin{tabular}{lll}
Quark masses & $M_c$ $({\rm MeV})$ & 1290 \\
             & $M_b$ $({\rm MeV})$ & 4700 \\[2ex]
Coulomb & $\alpha$ & 0.4105 \\[2ex]
Confinement & $\sigma$ $({\rm GeV}^2$) & 0.2 \\[2ex]
Spin-Spin & $\gamma$ $({\rm GeV})$     & 1.982 \\
          & $\beta_{cc}$ $({\rm GeV})$ & 2.06 \\
          & $\beta_{cb}$ $({\rm GeV})$ & 0.789 \\
          & $\beta_{bb}$ $({\rm GeV})$ & 0.318 \\
\end{tabular}
\end{ruledtabular}
\end{table}

\begin{table}[!t]
\caption{\label{Mmeson} Theoretical and experimental values of the masses for the $S$-wave $Q\bar{Q}$ mesons (in units of MeV).}
\begin{ruledtabular}
\begin{tabular}{ccc}
  ~~State & $M_{\rm the.}$    &  $M_{\rm exp.}$~~ \\[1ex]
  ~~$\eta_c(1S)$ & 2968  & 2981~~ \\
  ~~$\eta_c(2S)$ & 3655  & 3639~~ \\
  ~~$J/\psi(1S)$ & 3102  & 3097~~ \\
  ~~$\psi(2S)$   & 3720  & 3686~~ \\
  ~~$B_c(1S)$    & 6275  & 6275~~ \\
  ~~$B_c(2S)$    & 6885  & 6872~~ \\
  ~~$B^*_c(1S)$  & 6349  & - ~~ \\
  ~~$B^*_c(2S)$  & 6918  & - ~~ \\    
  ~~$\eta_b(1S)$  & 9401  & 9398~~ \\
  ~~$\eta_b(2S)$  & 9961  & 9999~~ \\
  ~~$\Upsilon(1S)$ & 9463  & 9460~~ \\
  ~~$\Upsilon(2S)$ & 9981  & 10023~~ \\
\end{tabular}
\end{ruledtabular}
\end{table}

\subsection{Color, flavor, and spin structure}

There are four fundamental degrees of freedom for heavy quarks, \emph{i.e.} color, flavor, spin, and spatial degrees of freedom. 

The color-space of a tetraquark system should be spanned by $12$ basis functions. In our description, they correspond to the complete configurations shown in Fig.~\ref{QQqq} and each one gives two basis wave functions or color channels. That is to say, the meson-meson configuration, Fig.~\ref{QQqq}(a), gives two color channels, $\varphi_1$ and $\varphi_2$, corresponding to the product of two color-singlets (${\bf 1_{\rm c}\otimes 1_{\rm c}}$) and the product of two color-octets (${\bf 8_{\rm c}\otimes 8_{\rm c}}$). They are explicitly given by
\begin{eqnarray}
\varphi_1 &=& \frac{1}{3}(\rm \bar{r}r+\bar{g}g+\bar{b}b)\times (\bar{r}r+\bar{g}g+\bar{b}b) ,\nonumber\\
\varphi_2 &=& \frac{\sqrt{2}}{12}(\rm 3\bar{b}r\bar{r}b+3\bar{g}r\bar{r}g+3\bar{b}g\bar{g}b+3\bar{g}b\bar{b}g+3\bar{r}g\bar{g}r
\nonumber\\
&&\rm +\ 3\bar{r}b\bar{b}r+2\bar{r}r\bar{r}r+2\bar{g}g\bar{g}g+2\bar{b}b\bar{b}b-\bar{r}r\bar{g}g
\nonumber\\
&&\rm -\ \bar{g}g\bar{r}r-\bar{b}b\bar{g}g-\bar{b}b\bar{r}r-\bar{g}g\bar{b}b-\bar{r}r\bar{b}b) \,.
\end{eqnarray}
The diquark-antidiquark configuration, Fig.~\ref{QQqq}(b), gives again two color basis,  $\varphi_3$ and $\varphi_4$, corresponding to the product of a color triplet and a color antitriplet (${\bf 3_{\rm c}\otimes \bar{3}_{\rm c}}$) and the product of a color sextet and a color antisextet (${\bf 6_{\rm c}\otimes \bar{6}_{\rm c}}$). They are explicitly given by
\begin{eqnarray}
\varphi_3 &=& \frac{\sqrt{3}}{6}(\rm \bar{r}r\bar{g}g-\bar{g}r\bar{r}g+\bar{g}g\bar{r}r-\bar{r}g\bar{g}r+\bar{r}r\bar{b}b-\bar{b}r\bar{r}b \nonumber\\
&&\rm+\ \bar{b}b\bar{r}r-\bar{r}b\bar{b}r+\bar{g}g\bar{b}b-\bar{b}g\bar{g}b+\bar{b}b\bar{g}g-\bar{g}b\bar{b}g), \nonumber\\
\varphi_4 &=& \frac{\sqrt{6}}{12}(\rm 2\bar{r}r\bar{r}r+2\bar{g}g\bar{g}g+2\bar{b}b\bar{b}b+\bar{r}r\bar{g}g+\bar{g}r\bar{r}g\nonumber\\
&&\rm +\ \bar{g}g\bar{r}r+\bar{r}g\bar{g}r+\bar{r}r\bar{b}b+\bar{b}r\bar{r}b+\bar{b}b\bar{r}r \nonumber\\
&&\rm +\ \bar{r}b\bar{b}r+\bar{g}g\bar{b}b+\bar{b}g\bar{g}b+\bar{b}b\bar{g}g+\bar{g}b\bar{b}g) .
\end{eqnarray}

\begin{table}[!t]
\caption{\label{SpinIndex} Indices of the spin wave functions in Eq.~(\ref{spinwave}). Their specific numbers are listed in the columns corresponding to the configurations shown in Fig.~\ref{QQqq}.
}
\begin{ruledtabular}
\begin{tabular}{lcccccc}
    & Dimeson & Diquark-antidiquark & $K_1$ & $K_2$ & $K_3$ & $K_4$\\[2ex]
$u_1$ & 1 & 3 & & & & \\
$u_2$ & 2 & 4 & & & & \\
$u_3$ &   &   & 5 & 7 & 9 & 11 \\
$u_4$ &   &   & 6 & 8 & 10 & 12 \\[2ex]
$v_1$ & 1 & 4 & & & & \\
$v_2$ & 2 & 5 & & & & \\
$v_3$ & 3 & 6 & & & & \\
$v_4$ &   &   & 7 & 10 & 13 & 16 \\
$v_5$ &   &   & 8 & 11 & 14 & 17 \\
$v_6$ &   &   & 9 & 12 & 15 & 18 \\
%
\end{tabular}
\end{ruledtabular}
\end{table}

Each K-type configuration, Figs.~\ref{QQqq}(c)-(f), corresponds with two color channels. For convenience, we denote the K-type configurations from (c) to (f) in Fig.~\ref{QQqq} as K$_1$, K$_2$, K$_3$, and K$_4$, respectively. The corresponding color wave functions have been constructed in Ref.~\cite{gy:2020dhts}. The color wave functions $\varphi_5$ and $\varphi_6$, corresponding to the K$_1$ configuration [Fig.~\ref{QQqq}(c)], are given by 
\begin{align}
\varphi_5 = \varphi_2,\ \ \ \ \varphi_6 =\varphi_1.
\end{align}
The color wave functions $\varphi_7$ and $\varphi_8$ from the K$_2$ configuration [Fig.~\ref{QQqq}(d)] are similar. We have
\begin{align}
\varphi_7 = \varphi_1,\ \ \ \ \varphi_8 =\varphi_2.
\end{align}
The color wave functions $\varphi_9$ and $\varphi_{10}$ from the K$_3$ configuration [Fig.~\ref{QQqq}(e)] can be shown to be
\begin{eqnarray}
\varphi_9 &=& \frac{1}{2\sqrt{6}}(\rm \bar{r}b\bar{b}r+\bar{r}r\bar{b}b+\bar{g}b\bar{b}g+\bar{g}g\bar{b}b+\bar{r}g\bar{g}r+\bar{r}r\bar{g}g
\nonumber\\
&&\rm +\ \bar{b}b\bar{g}g+\bar{b}g\bar{g}b+\bar{g}g\bar{r}r+\bar{g}r\bar{r}g+\bar{b}b\bar{r}r+\bar{b}r\bar{r}b)
\nonumber\\
&&\rm +\ \frac{1}{\sqrt{6}}(\bar{r}r\bar{r}r+\bar{g}g\bar{g}g+\bar{b}b\bar{b}b),\nonumber\\
\varphi_{10} &=& \frac{1}{2\sqrt{3}}(\rm \bar{r}b\bar{b}r-\bar{r}r\bar{b}b+\bar{g}b\bar{b}g-\bar{g}g\bar{b}b+\bar{r}g\bar{g}r-\bar{r}r\bar{g}g
\nonumber\\
&&\rm -\ \bar{b}b\bar{g}g+\bar{b}g\bar{g}b-\bar{g}g\bar{r}r+\bar{g}r\bar{r}g-\bar{b}b\bar{r}r+\bar{b}r\bar{r}b).
\end{eqnarray}
And, finally, the color wave functions $\varphi_{11}$ and $\varphi_{12}$ from the K$_4$ configuration [Fig.~\ref{QQqq}(f)] read
\begin{align}
\varphi_{11} = \varphi_9,\ \ \ \ \varphi_{12} =-\varphi_{10}.
\end{align}

The flavor structure of the $QQ\bar{Q}\bar{Q}$ ($Q=c,\,b$) tetraquark system is trivial and will be neglected. The total spin quantum number can be $S=0$, $1$, and $2$. Since there is no spin-orbital interaction in the Hamiltonian, the third component of the total spin $(M_S)$ can be set to be equal to $S$ without loss of generality. The four-body spin wave functions, $\xi_{S}^{\alpha}$, for the configurations shown in Fig.~\ref{QQqq} can be listed as
\begin{widetext}
\begin{eqnarray}
\label{spinwave}
\xi_0^{u_1} &=& \chi_{0,0}(1,2)\otimes\chi_{0,0}(3,4), \nonumber\\
\xi_0^{u_2} &=& \frac{1}{\sqrt{3}}\Big[\chi_{1,1}(1,2)\otimes\chi_{1,-1}(3,4)-\chi_{1,0}(1,2)\otimes\chi_{1,0}(3,4)+\chi_{1,-1}(1,2)\otimes\chi_{1,1}(3,4)\Big], \nonumber\\
\xi_0^{u_3} &=& \frac{1}{\sqrt{3}}\Big[\chi_{1,1}(1,2)\otimes |\downarrow_3\downarrow_4\rangle+\chi_{1,-1}(1,2)\otimes |\uparrow_3\uparrow_4\rangle-\chi_{1,0}(1,2)\otimes\chi_{0,0}(3,4)\Big], \nonumber\\
\xi_0^{u_4} &=& \frac{1}{\sqrt{2}}\chi_{0,0}(1,2)\otimes \Big(|\uparrow_3\downarrow_4\rangle-|\downarrow_3\uparrow_4\rangle\Big), \nonumber\\
\xi_1^{v_1} &=& \chi_{0,0}(1,2)\otimes\chi_{1,1}(3,4), \nonumber\\
\xi_1^{v_2} &=& \chi_{1,1}(1,2)\otimes\chi_{0,0}(3,4), \nonumber\\
\xi_1^{v_3} &=& \frac{1}{\sqrt{2}} \Big[\chi_{1,1}(1,2)\otimes\chi_{1,0}(3,4)-\chi_{1,0}(1,2)\otimes\chi_{1,1}(3,4)\Big], \nonumber\\
\xi_1^{v_4} &=& \sqrt{\frac{3}{4}}\chi_{1,1}(1,2)\otimes |\uparrow_3\downarrow_4\rangle-\sqrt{\frac{1}{12}}\chi_{1,1}(1,2)\otimes|\downarrow_3\uparrow_4\rangle
-\sqrt{\frac{1}{6}}\chi_{1,0}(1,2)\otimes |\uparrow_3\uparrow_4\rangle, \nonumber\\
\xi_1^{v_5} &=& \sqrt{\frac{2}{3}}\chi_{1,1}(1,2)\otimes |\uparrow_3\downarrow_4\rangle-\sqrt{\frac{1}{3}}\chi_{1,0}(1,2)\otimes |\uparrow_3\uparrow_4\rangle, \nonumber\\
\xi_1^{v_6} &=& \chi_{0,0}(1,2)\otimes|\uparrow_3\uparrow_4\rangle, \nonumber\\
\xi_2 &=&  \chi_{1,1}(1,2)\otimes\chi_{1,1}(3,4).
\end{eqnarray}
\end{widetext}
The indices $u_1,...,u_4$, $v_1,...,v_6$, and $w_1$ number the spin channels corresponding to the configurations shown in Fig.~\ref{QQqq}. Their specific values are listed in Table~\ref{SpinIndex}.
These spin wave functions are obtained by considering the coupling of sub-clusters with the SU(2) algebra. The necessary bases are given by
\begin{eqnarray}
\chi_{1,1}(i,j) &=& |\uparrow_i\uparrow_j\rangle, \ \ \chi_{1,-1}(i,j) = |\downarrow_i\downarrow_j\rangle,\nonumber\\
\chi_{1,0}(i,j) &=& \frac{1}{\sqrt{2}}\Big( |\uparrow_i\downarrow_j\rangle+ |\downarrow_i\uparrow_j\rangle\Big), \nonumber\\
\chi_{0,0}(i,j) &=& \frac{1}{\sqrt{2}}\Big( |\uparrow_i\downarrow_j\rangle- |\downarrow_i\uparrow_j\rangle\Big).
\end{eqnarray}


\subsection{Computational method}

We solve the four-body problem by means of an exact and high-efficiency numerical approach, the Gaussian expansion method~\cite{Hiyama:2003cu}. Three relative orbital motions of the four-body system are all expanded by using the Gaussian bases,  
\begin{equation}
\phi_{nlm}({\bf r}) = N_{nl} |{\bf r}|^{l} e^{-\nu_{n} |{\bf r}|^2} Y_{lm}(\hat{\bf r}),
\end{equation}
where the widths $\nu_n$ are taken as the sizes of the geometric progression\footnote{The details of the Gaussian parameters can be found in Refs.~\cite{Yang:2015bmv}.} and the normalization factor $N_{nl}$ is not shown explicitly. The general spatial wave function of the four-body system can be formally expressed as
\begin{equation}
\label{eq:WFexp}
\psi_{LM_L}= \Big[ \left[ \phi_{n_1l_1m_1}(\mbox{\boldmath{$\rho$}}) \phi_{n_2l_2m_2}(\mbox{\boldmath{$\lambda$}})\right]_{lm} \phi_{n_3l_3m_3}( {\bf R}) \Big]_{L M_L},
\end{equation}
where $\mbox{\boldmath{$\rho$}}$, $\mbox{\boldmath{$\lambda$}}$, and ${\bf R}$ are the internal Jacobi coordinates for the four-body configurations in Fig.~\ref{QQqq}. For the meson-meson configuration, Fig.~\ref{QQqq}(a), they are given by
\begin{eqnarray}
\mbox{\boldmath{$\rho$}}&= & {\bf r}_1-{\bf r}_2, \ \ \ 
\mbox{\boldmath{$\lambda$}} ={\bf r}_3 - {\bf r}_4,\nonumber\\
{\bf R} &=& \frac{m_1{\bf r}_1 + m_2{\bf r}_2}{m_1+m_2}- \frac{ m_3{\bf r}_3 + m_4{\bf r}_4}{m_3+m_4}.
\end{eqnarray}
For the diquark-antdiquark configuration, Fig.~\ref{QQqq}(b), they read
\begin{eqnarray}
\mbox{\boldmath{$\rho$}}&= & {\bf r}_1-{\bf r}_3,\ \ \ 
\mbox{\boldmath{$\lambda$}} = {\bf r}_2 - {\bf r}_4,\nonumber\\
{\bf R} &=& \frac{m_1{\bf r}_1 + m_3{\bf r}_3}{m_1+m_3}- \frac{ m_2{\bf r}_2 + m_4{\bf r}_4}{m_2+m_4}.
\end{eqnarray}
For the K-type configurations in Figs.~\ref{QQqq}(c)-(f), we have
\begin{eqnarray}
\mbox{\boldmath{$\rho$}}&=& {\bf r}_i-{\bf r}_j, \nonumber\\
\mbox{\boldmath{$\lambda$}}&=& {\bf r}_k- \frac{ m_i{\bf r}_i + m_j{\bf r}_j}{m_i+m_j},\nonumber\\
{\bf R} &=& {\bf r}_l- \frac{ m_i{\bf r}_i +  m_j{\bf r}_j+ m_k{\bf r}_k}{m_i+m_j+m_k},
\end{eqnarray}
where the indices $i, j, k, l$ should be properly assigned according to different K-type configurations. Furthermore, with these coordinates the center-of-mass motion can be completely eliminated.

The complete basis set of the four-body wave functions that fulfill the Pauli principle can be expressed as
\begin{equation}
\label{TPs}
\Psi_{JM_J,\alpha\beta}={\cal A} \left[ \left( \psi_{LM_L}\otimes \xi_{SM_S}^{\alpha} \right)_{JM_J} \otimes \varphi_\beta^{\phantom{\dag}} \right] \,,
\end{equation}
where $J$ is the quantum number of the total angular momentum. Here the antisymmetrizing operator $\cal{A}$ is added considering the exchange of identical particles.  Particularly, for the fully-charm, -bottom and $\bar{c}b\bar{c}b$ tetraquark states, it is explicitly given by
\begin{equation}
{\cal{A}} = 1-(13)-(24)+(13)(24).
\end{equation}
Meanwhile, for the $\bar{c}c\bar{c}b$ and $\bar{b}b\bar{b}c$ tetraquark states, it reads as
\begin{equation}
{\cal{A}} = 1-(13),
\end{equation}
and $\cal{A}$=1 for the $\bar{c}c\bar{b}b$ case. This antisymmetrizing procedure is necessary since the basis set is constructed from sub-clusters, \emph{i.e.} meson-meson, diquark-antidiquark, and K-type structures. The Hamiltonian matrix can be constructed using this complete basis set of the four-body wave functions.

We employ the complex scaling method (CSM)~\cite{JA22269, EB22280, Bsimon1972, YKHo1983, NMoieyev1998, AMKK2006, MKMK2014} according to the ABC theorem~\cite{JA22269, EB22280} in order to search for possible resonant states of the fully-heavy tetraquark system. Using the CSM, the resonance energy (its position and width) is obtained as a stable eigenvalue of the complex scaled Schr\"{o}dinger euqation:
\begin{equation}
\label{CSMSE}
\left[ H(\theta)-E(\theta) \right] \Psi_{JM,\alpha\beta}(\theta)=0,	
\end{equation}
where $H(\theta)$ is obtained from the original Hamiltonian by transforming the four-body Jacobi coordinates with respect to a common complex scaling angle $\theta$:
\begin{equation}
|\mbox{\boldmath{$\rho$}}|\rightarrow|\mbox{\boldmath{$\rho$}}|e^{i\theta}, \ \ \ 
|\mbox{\boldmath{$\lambda$}}|\rightarrow|\mbox{\boldmath{$\lambda$}}|e^{i\theta},\ \ \ 
|{\bf R}|\rightarrow|{\bf R}|e^{i\theta}.
\end{equation}
According to the ABC theorem~\cite{JA22269, EB22280}, the eigenvalues of Eq. (\ref{CSMSE}) can be separated into three groups:
\begin{itemize}
\item[(i)] A bound state corresponds to a pole on the real axis whose energy is smaller than the meson-meson threshold for a specific $J^{P(C)}$ state. It remains unchanged under the complex scaling transformation and such energy is identified as the mass of the four-body bound state.
\item[(ii)] The discretized continuum states, associated with the cuts, are rotated downward by an angle of $2\theta$ with respect the real axis. 
\item[(iii)] A resonant state corresponds to a complex pole, $E_{\rm res}$, that is independent of the angle $\theta$. Moreover, it is isolated from the discretized continuum spectrum, lying along the $2\theta$-rotated line when the relation $\tan 2\theta > - \text{Im}(E_{\text{res}})/\text{Re}(E_{\text{res}})$ is satisfied. The mass $M$ and width $\Gamma$ of the four-body resonance are given by
\begin{equation}
M={\rm Re}(E_{\rm res}),\ \ \ \ \Gamma=- 2 {\rm Im}(E_{\rm res}).
\end{equation}

\end{itemize}


\section{Results and discussion}
\label{sec:results}

We present herein our results for the low-lying $S$-wave states ($L=0$ and hence $J=S$) of the fully-heavy tetraquark systems $QQ\bar{Q}\bar{Q}$ $(Q=c,b)$. They are organized as $cc\bar{c}\bar{c}$, $bb\bar{b}\bar{b}$, $cb\bar{c}\bar{c}$, $cb\bar{b}\bar{b}$, $bb\bar{c}\bar{c}$ and $cb\bar{c}\bar{b}$ tetraquark states. Since the wave function of the tetraquark system is constructed at the fundamental quark level, some forbidden di-meson channels at hadron level, such as the di-$J/\psi$ structure $J/\psi J/\psi$ in the $1^{+(-)}$ state, are included; however, they will exhibit self-consistently scattering nature in our numerical calculations. These forbidden di-meson channels, interpreted as thresholds, are actually useful for us to identify possible tetraquark states.

\subsection{Fully-charm system $cc\bar{c}\bar{c}$}

We find no bound state for all $S$-wave channels $J^{P(C)}=0^{+(+)}$, $1^{+(-)}$, and $2^{+(+)}$ of fully-charm system $cc\bar{c}\bar{c}$. However, using the CSM, we identify several narrow resonances. Table~\ref{Rsum-cccc} summarizes our findings, the results are discussed in the following.

\begin{table}[!t]
\caption{\label{Rsum-cccc} Predicted resonances of the fully-charm tetraquark system from the complete coupled-channel calculation with CSM. Their masses and widths are summarized in the third and fourth column, respectively, in units of MeV.}
\begin{ruledtabular}
\begin{tabular}{llcc}
~~$J^{P(C)}$ & Resonance & Mass &  Width~~ \\[2ex]
~~$0^{+(+)}$ & $\eta_c (1S)\eta_c (2S)$ & 6640 & 0.66~~ \\
             & $\eta_c (1S)\eta_c (2S)$ & 6762 & 1.14~~ \\
             & $J/\psi (1S)\psi (2S)$   & 6875 & 3.74~~ \\
             & $J/\psi (1S)\psi (2S)$   & 6987 & 4.00~~ \\
             & $J/\psi (1S)\psi (2S)$   & 7195 & 8.00~~ \\[2ex]
~~$1^{+(-)}$ & $J/\psi (1S)J/\psi (1S)$ & 6274 & 0.06~~ \\
             & $J/\psi (1S)J/\psi (1S)$   & 6653 & 1.20~~ \\
             & $\eta_c (2S)J/\psi (1S)$   & 6885 & 4.10~~ \\[2ex]
~~$2^{+(+)}$ & $J/\psi (1S)\psi (2S)$     & 7007 & 1.02~~ \\[2ex]
\end{tabular}
\end{ruledtabular}
\end{table}

\begin{table}[!t]
\caption{\label{GresultCC1} Calculation of the $J^{P(C)}=0^{+(+)}$ state of the fully-charm system in the real case ($\theta=0^\circ$). The first and second columns show all possible structures and channels. They are obtained by the necessary bases in spin and color degrees of freedom summarized in the third column. The values in the parentheses after the first two meson-meson structures give the theoretical results of the noninteracting meson-meson thresholds. The lowest energies of all channels without considering any channel-channel coupling are listed in the fourth column. The last column shows the lowest energies in a calculation taking into account the couplings among the channels belonging to each structure (configuration). The last row gives the lowest energy from the complete coupled-channel calculation. All results are in units of MeV.}
\begin{ruledtabular} 
\begin{tabular}{lcccc}
~~Structure  & Channel & $\xi_0^{\alpha}$; $\varphi_\beta^{\phantom{\dag}}$ & $M$ & Mixed~~ \\
 & & $[\alpha;~\beta]$ &  &  \\[2ex]
~~$\eta_c \eta_c(5936)$      & 1 & [1; 1]  & $5936$ &~~ \\
~~$J/\psi J/\psi(6204)$       & 2  & [2; 1]  & $6204$ &$5936$~~  \\[2ex]
~~$\eta^8_c \eta^8_c$           & 3  & [1; 2]  & $6403$ &~~ \\
~~$J/\psi^8 J/\psi^8$       & 4  & [2; 2]  & $6346$ &$6268$~~  \\[2ex]
~~$[cc]^{\bf 6_{\rm c}}[\bar{c}\bar{c}]^{\bf \bar{6}_{\rm c}}$   & 5  & [3; 4]   & $6404$ &~~ \\
~~$[cc]^{\bf 3_{\rm c}}[\bar{c}\bar{c}]^{\bf \bar{3}_{\rm c}}$   & 6  & [4; 3]  & $6421$  &$6343$~~\\[2ex]
~~K$_1$   & 7   & [5; 5] & $6349$ & \\
                 & 8   & [5; 6] & $6341$ & \\
                 & 9   & [6; 5] & $6416$ &  \\
                 & 10  & [6; 6] & $6160$ & $6150$~~ \\[2ex]
~~K$_2$   & 11   & [7; 7] & $6341$ & \\
                 & 12   & [7; 8] & $6349$ & \\
                 & 13   & [8; 7] & $6160$ &  \\
                 & 14  & [8; 8] & $6416$ & $6150$~~ \\[2ex]
~~K$_3$   & 15   & [9; 10] & $6420$ & \\
                 & 16   & [10; 9] & $6402$ &$6340$~~  \\[2ex]
~~K$_4$   & 17   & [11; 12] & $6420$ & \\
                 & 18   & [12; 11] & $6402$ &$6340$~~  \\[2ex]
\multicolumn{4}{l}{Lowest energy of the fully-coupled result:}  & $5936$~~ \\
\end{tabular}
\end{ruledtabular}
\end{table}

\begin{figure}[ht]
\includegraphics[clip, trim={3.0cm, 1.0cm, 2.0cm, 1.0cm}, width=0.50\textwidth]{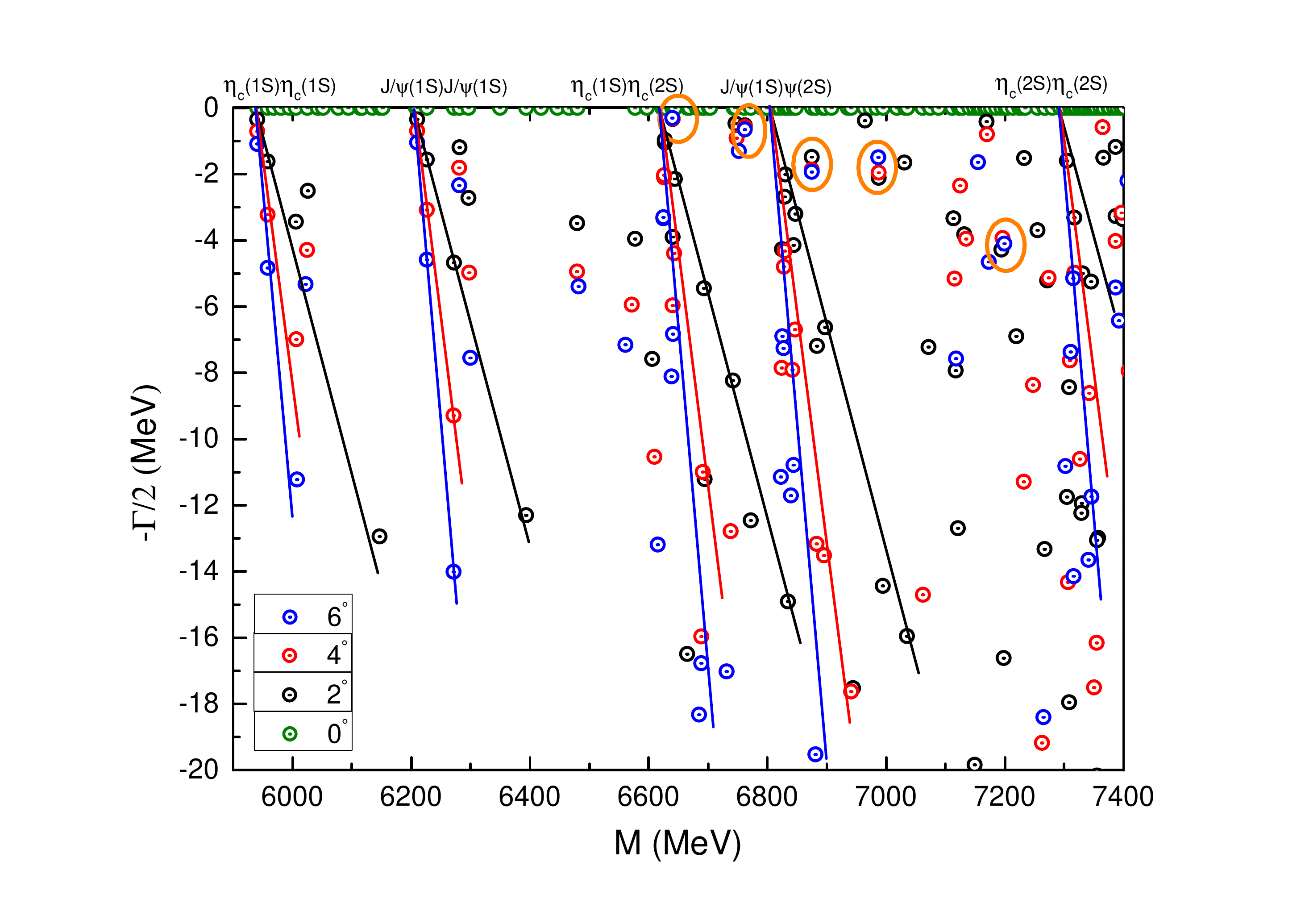}
\caption{Complex energy spectrum of the fully-charm tetraquark system with $J^{P(C)}=0^{+(+)}$ from the complete coupled-channel calculation with CSM. The parameter $\theta$ varies from $0^\circ$ to $6^\circ$.} \label{PP1}
\end{figure}

{\bf The $\bm{J^{P(C)}=0^{+(+)}}$ channel:} To seek for possible bound states of the fully-charm tetraquark system, we first consider the real case ($\theta=0^\circ$), shown in Table~\ref{GresultCC1}. Therein, the lowest energies of all channels from a fully-uncoupled calculation (without considering any channel-channel coupling), the lowest energies of all structures from a partially-coupled calculation (considering only the couplings among the channels belonging to each structure), and the lowest energy from a fully-coupled calculation (considering all channel-channel couplings), are listed. The energy (mass) spectrum from the fully-coupled calculation is also shown in Fig.~\ref{PP1} (green circles on the real axis).

The meson-meson configuration includes two channels, $\eta_c \eta_c$ and $J/\psi J/\psi$, and the diquark-antidiquark configuration also includes two channels, $[cc]^{\bf 3_{\rm c}}[\bar{c}\bar{c}]^{\bf \bar{3}_{\rm c}}$ and $[cc]^{\bf 6_{\rm c}}[\bar{c}\bar{c}]^{\bf \bar{6}_{\rm c}}$. In Table~\ref{GresultCC1} we notice that the theoretical values of the di-meson threshold, 5936 MeV for $\eta_c \eta_c$ and 6204 MeV for $J/\psi J/\psi$, are quite comparable with the experimental values 5962 MeV and 6194 MeV, respectively. Then in the fully-uncoupled calculation, bound state is unavailable in these channels. Particularly, the masses of the hidden-color channels $\eta_c \eta_c$ and $J/\psi J/\psi$ are both around 6.4 GeV, which is also very close to those of the diquark-antidiquark channels. Meanwhile, the masses of the K-type channels, which are all above the di-meson thresholds, generally range from 6.2 to 6.4 GeV.

In the partially-coupled calculations, which are performed for each kind of configuration, the calculated masses are still all above the lowest threshold value, 5936 MeV for $\eta_c \eta_c$. From Table~\ref{GresultCC1} we find that, the masses of the excited states for the diquark-antidiquark and K-type configurations range from 6.15 to 6.35 GeV. Since we are treating a tetraquark system in which all quarks and antiquarks have the same flavor, no interaction in the Hamiltonian (\ref{eq:Hamiltonian}) can distinguish between the configurations K$_1$ and K$_2$, as well as between K$_3$ and K$_4$; Thus, the results are degenerate and this is also true for other $J^{P(C)}$ states.

Finally, in the fully-coupled calculation, the lowest mass of the spectrum is 5936 MeV, which is still slightly above the $\eta_c\eta_c$ threshold. Therefore, we can conclude that no bound state exists with $J^{P(C)}=0^{+(+)}$. However, as it will be shown next, several resonances are found in the complex-range and complete coupled-channel calculation. 

\begin{table}[!t]
\caption{\label{GresultCC2} Calculation of the $J^{P(C)}=1^{+(-)}$ state of the fully-charm system in the real case ($\theta=0^\circ$). The first and second columns show all possible structures and channels. They are obtained by the necessary bases in spin and color degrees of freedom summarized in the third column. The values in the parentheses after the first two meson-meson structures give the theoretical results of the noninteracting meson-meson thresholds. The lowest energies of all channels without considering any channel-channel coupling are listed in the fourth column. The last column shows the lowest energies in a calculation taking into account the couplings among the channels belonging to each structure (configuration). The last row gives the lowest energy from the complete coupled-channel calculation. All results are in units of MeV.}
\begin{ruledtabular}
\begin{tabular}{lcccc}
~~Structure  & Channel & $\xi_1^{\alpha}$; $\varphi_\beta^{\phantom{\dag}}$ & $M$ & Mixed~~ \\
 & & $[\alpha;~\beta]$ &  &  \\[2ex]
~~$\eta_c J/\psi(6070)$      & 1  & [1; 1] & $6070$ &~~ \\
~~$J/\psi J/\psi(6204)$      & 2  & [3; 1]  & $6204$ &$6070$~~ \\[2ex]
~~$\eta^8_c J/\psi^8$                 & 3 & [1; 2]   & $6325$ &~~ \\
~~$J/\psi^8 J/\psi^8$                 & 4  & [3; 2]  & $6340$ & $6325$~~ \\[2ex]
~~$[cc]^{\bf 3_{\rm c}}[\bar{c}\bar{c}]^{\bf \bar{3}_{\rm c}}$   & 5 & [6; 3]   & $6439$ & $6439$~~ \\[2ex]
~~K$_1$   & 6  & [7; 5]   & $6428$ & \\
                 & 7  & [8; 5]  & $6407$ & \\
                 & 8  & [9; 5]  & $6382$ & \\
                 & 9  & [7; 6]  & $6429$ & \\
                 & 10  & [8; 6]  & $6431$ & \\
                 & 11  & [9; 6]  & $6354$ & $6271$~~ \\[2ex]
~~K$_2$   & 12 & [10; 7]   & $6429$ & \\
                 & 13  & [11; 7]  & $6431$ & \\
                 & 14  & [12; 7]  & $6354$ & \\
                 & 15  & [10; 8]  & $6428$ & \\
                 & 16  & [11; 8]  & $6407$ & \\
                 & 17  & [12; 8]  & $6382$ & $6271$~~ \\[2ex]
~~K$_3$   & 18  & [13; 10]  & $6442$ & \\
                 & 19   & [14; 10]  & $6442$ & \\
                 & 20   & [15; 9]  & $6931$ & $6436$~~ \\[2ex]
~~K$_4$   & 21   & [16; 12]  & $6442$ & \\
                 & 22   & [17; 12]  & $6442$ & \\
                 & 23   & [18; 11]  & $6931$ & $6436$~~ \\[2ex]
\multicolumn{4}{l}{Lowest mass of the fully-coupled result:}    & $6070$~~  \\
\end{tabular}
\end{ruledtabular}
\end{table}

\begin{figure}[ht]
\includegraphics[clip, trim={3.0cm, 1.0cm, 2.0cm, 1.0cm}, width=0.50\textwidth]{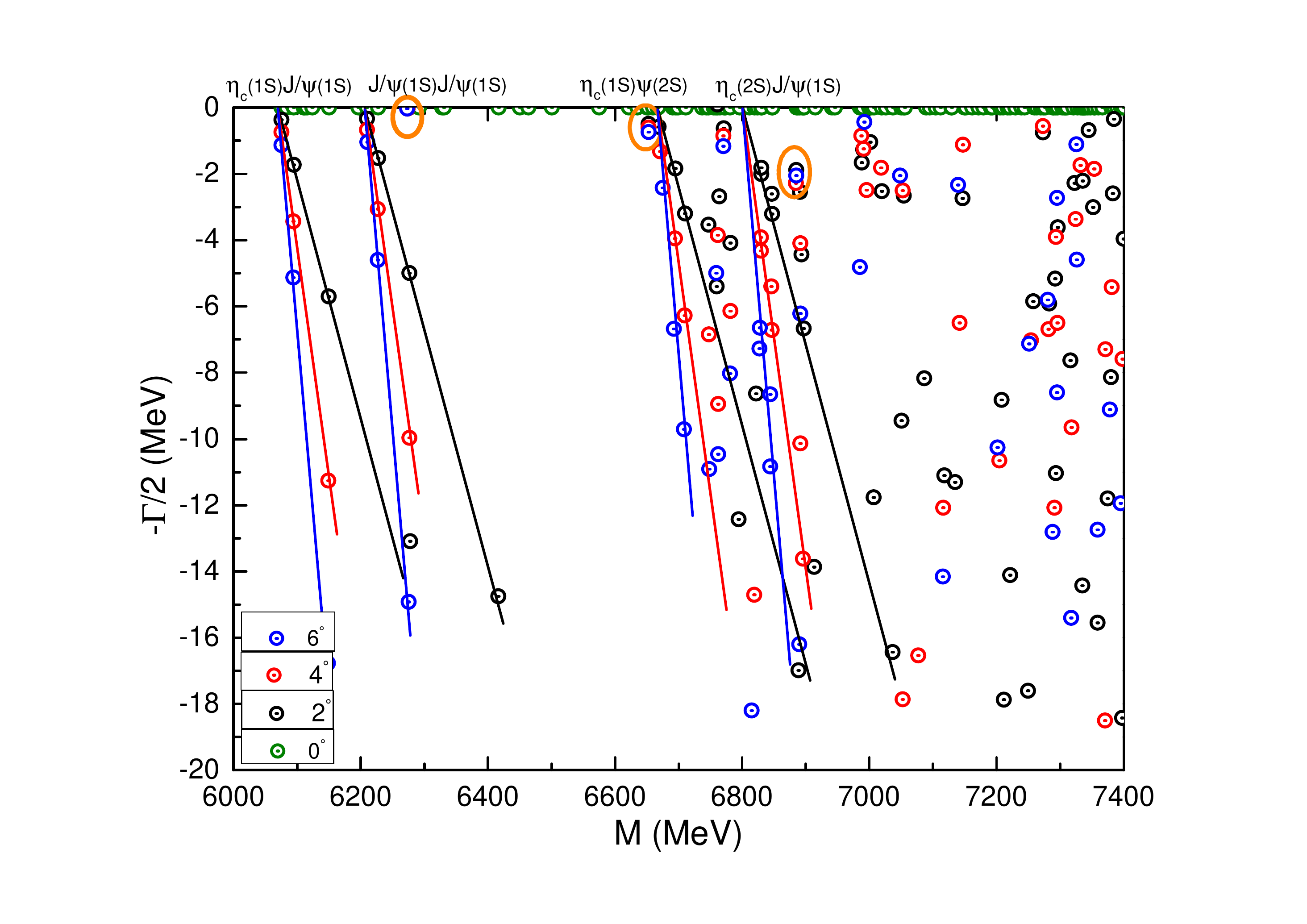}
\caption{Complex energy spectrum of the fully-charm tetraquark system with $J^{P(C)}=1^{+(-)}$ from the complete coupled-channel calculation with CSM. The parameter $\theta$ varies from $0^\circ$ to $6^\circ$.} \label{PP2}
\end{figure}

Figure~\ref{PP1} presents the complex energy spectrum with the angle $\theta$ ranging from $0^\circ$ to $6^\circ$. In the mass range of $5.9-7.4$ GeV, there are five meson-meson thresholds, $\eta_c (1S)\eta_c (1S)$, $J/\psi (1S)J/\psi (1S)$, $\eta_c (1S)\eta_c (2S)$, $J/\psi (1S)\psi (2S)$, and $\eta_c (2S)\eta_c (2S)$. The calculated complex eigenenergies are generally aligned along the threshold lines, which are rotated downward making an angle of $2\theta$ with the real axis, indicating the nature of the scattering states. Being cautious due to the calculation noise in the highly excited region, we can identify five fixed poles that are independent of $\theta$, as circled in Fig.~\ref{PP1}. Apparently, they can be identified as the resonance states of the fully-charm system $cc\bar{c}\bar{c}$.

The first two poles, with masses and widths (6640 MeV, 0.66 MeV) and (6762 MeV, 1.14 MeV), are just above the $\eta_c (1S)\eta_c (2S)$ threshold lines. Therefore, they can be identified as di-$\eta_c$ resonances, with a radial excitation. Similarly, the other three poles seems to be resonances of the type $J/\psi (1S)\psi (2S)$, and their masses and widths are (6875 MeV, 3.74 MeV), (6987 MeV, 4.00 MeV) and (7195 MeV, 8.00 MeV), respectively.

It is worth highlighting that the obtained $J/\psi (1S)\psi (2S)$ resonances with masses 6875 MeV and 6987 MeV are both quite close to the narrow structure around 6.9 GeV claimed by the LHCb collaboration~\cite{LA2020CERNINDICO} and hence they could be candidates for this. Meanwhile, the remaining $\eta_c (1S)\eta_c (2S)$ and $J/\psi (1S)\psi (2S)$ resonances are consistent with the broad structure between 6.2 and 6.8 GeV, and the possible peak shown around 7.2 GeV~\cite{LA2020CERNINDICO}. Their predicted small widths indicate that these resonances are stable against two-meson strong decays, indicating that they could be confirmed in future experiments.

\begin{table}[!t]
\caption{\label{GresultCC3} Calculation of the $J^{P(C)}=2^{+(+)}$ state of the fully-charm system in the real case ($\theta=0^\circ$). The first and second columns show all possible structures and channels. They are obtained by the necessary bases in spin and color degrees of freedom summarized in the third column. The value in the parenthesis after the first meson-meson structure gives the theoretical result of the noninteracting meson-meson threshold. The lowest energies of all channels without considering any channel-channel coupling are listed in the fourth column. The last column shows the lowest energies in a calculation taking into account the couplings among the channels belonging to each structure (configuration). The last row gives the lowest energy from the complete coupled-channel calculation. All results are in units of MeV.}
\begin{ruledtabular}
\begin{tabular}{lcccc}
~~Structure  & Channel & $\xi_2$; $\varphi_\beta^{\phantom{\dag}}$ & $M$ & Mixed~~ \\
 & & $[\beta]$ &  &  \\[2ex]
~~$J/\psi J/\psi(6204)$       & 1  & 1 & $6204$ &$6204$~~  \\[2ex]
~~$J/\psi^8 J/\psi^8$       & 2  & 2  & $6388$ &$6388$~~  \\[2ex]
~~$[cc]^{\bf 3_{\rm c}}[\bar{c}\bar{c}]^{\bf \bar{3}_{\rm c}}$   & 3  & 3  & $6472$  &$6472$~~\\[2ex]
~~K$_1$   & 4  & 5  & $6378$ & \\
                 & 5  & 6 & $6374$ & $6358$~~ \\[2ex]
~~K$_2$   & 6  & 7  & $6374$ & \\
                 & 7 & 8 & $6378$ & $6358$~~ \\[2ex]
~~K$_3$   & 8 & 10  & $6473$ & $6473$~~ \\[2ex]
~~K$_4$   & 9  & 12 & $6473$ & $6473$~~ \\[2ex]
\multicolumn{4}{l}{Lowest mass of the fully-coupled result:}  & $6204$~~ \\
\end{tabular}
\end{ruledtabular}
\end{table}

{\bf The $\bm{J^{P(C)}=1^{+(-)}}$ channel:} There are 23 channels in this case as listed in Table~\ref{GresultCC2}. We have four meson-meson channels, including the color-singlet and hidden-color states of $\eta_c J/\psi$ and $J/\psi J/\psi$, one diquark-antidiquark channel $[cc]^{\bf 6_{\rm c}}[\bar{c}\bar{c}]^{\bf \bar{6}_{\rm c}}$, and 18 K-type ones. A bound state is still unavailable in each single-channel calculation. In particular, the theoretical masses for the two lowest di-meson states, $\eta_c J/\psi$ and $J/\psi J/\psi$, in the color-singlet channel, are 6070 MeV and 6204 MeV, respectively. Moreover, the coupled-channel effect is too weak to acquire binding energy within these two channels. In the partially-coupled calculation, the masses of K$_1$ and K$_2$ configurations are remarkably low due to the strong coupled-channel effect. However, these masses (6271 MeV) are still above the theoretical thresholds of $\eta_c J/\psi$ and $J/\psi J/\psi$.  In the fully-coupled calculation, the lowest mass of the spectrum is 6070 MeV, which is still slightly above the $\eta_c J/\psi$ threshold. Therefore, we conclude that no bound state exists with $J^{P(C)}=1^{+(-)}$.

Within the CSM, the complete coupled-channel calculation for $J^{P(C)}=1^{+(-)}$ $cc\bar c\bar c$ tetraquark system has been performed and the complex energy spectrum is shown in Fig.~\ref{PP2}. While most of the complex eigenenergies decline with increasing angle $\theta$ indicating that they are scattering states of $\eta_c J/\psi$ and $J/\psi J/\psi$, we find three unchanged poles in the complex plane. Their masses and widths are (6274 MeV, 0.06 MeV), (6653 MeV, 1.20 MeV), and (6885 MeV, 4.10 MeV), respectively.

\begin{figure}[ht]
\includegraphics[clip, trim={3.0cm, 1.0cm, 2.0cm, 1.0cm}, width=0.50\textwidth]{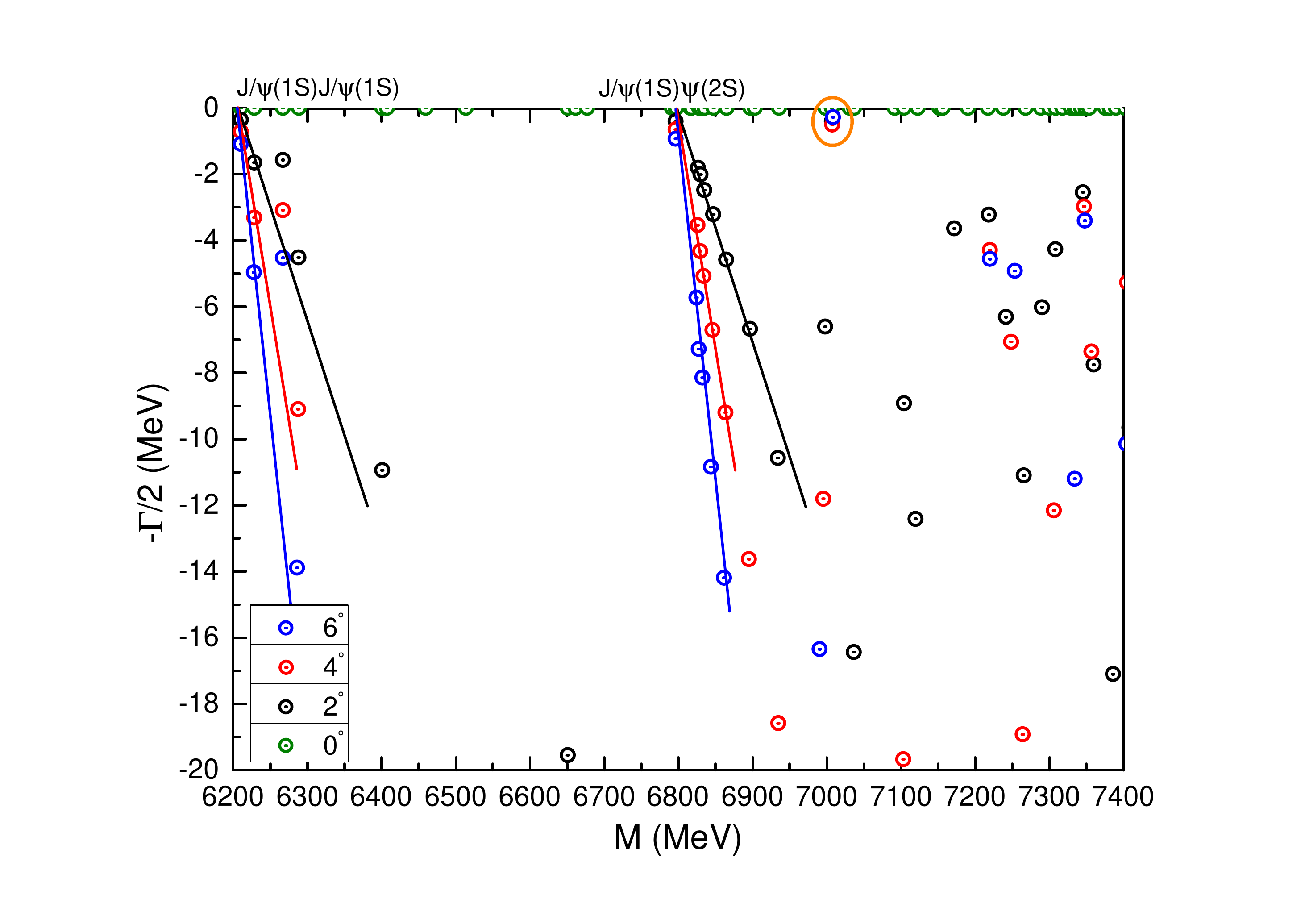}
\caption{Complex energy spectrum of the fully-charm tetraquark system with $J^{P(C)}=2^{+(+)}$ from the complete coupled-channel calculation with CSM. The parameter $\theta$ varies from $0^\circ$ to $6^\circ$.} \label{PP3}
\end{figure}

The resonance pole (6885 MeV, 4.10 MeV) can be identified as an $\eta_c (2S)J/\psi (1S)$ resonance. Its mass is quite close to that of the X(6900) structure seen by the LHCb collaboration. The other two resonances, (6274 MeV, 0.06 MeV) and (6653 MeV, 1.20 MeV), can be identified with a $J/\psi (1S)J/\psi (1S)$ nature. They are located in the energy range where the broad structure is seen by the LHCb collaboration.

{\bf The $\bm{J^{P(C)}=2^{+(+)}}$ channel:} We have two di-meson channels, $J/\psi J/\psi$ and $J/\psi^8 J/\psi^8$, one diquark-antidiquark channel, $[cc]^{\bf 6_{\rm c}}[\bar{c}\bar{c}]^{\bf \bar{6}_{\rm c}}$, and six K-type ones, as listed in Table~\ref{GresultCC3}. Again, no bound state is found in this highest total-spin case by all kinds of computations, including the partially-coupled and fully-coupled ones. The lowest mass of the spectrum from the fully-coupled calculation is 6204 MeV, which is slightly above the theoretical threshold of di-$J/\psi$.

Using the CSM, we obtain the complex energy spectrum from a complete coupled-channel calculation in Fig.~\ref{PP3}. Therein, the nature of the scattering states, $J/\psi(1S) J/\psi(1S)$ and $J/\psi(1S) \psi(2S)$, are clearly shown by varying the angle $\theta$. One fixed pole at (7007 MeV, 1.02 MeV) can be identified, and it can be regarded as a $J/\psi(1S) \psi(2S)$ resonance. Since the mass of this resonance is close to 6.9 GeV, the first radial excitation state of di-$J/\psi$, \emph{i.e.} $J/\psi  (1S)\psi (2S)$, in $J^{P(C)}=2^{+(+)}$ state is also a possible candidate for the X(6900).

\begin{table}[!t]
\caption{\label{Rsum-bbbb} Predicted resonances of the fully-bottom tetraquark system from the complete coupled-channel calculation with CSM. Their masses and widths are summarized in the third and fourth column, respectively, in units of MeV.}
\begin{ruledtabular}
\begin{tabular}{cccc}
~~$J^{P(C)}$ & Resonance & Mass &  Width~~ \\
~~$0^{+(+)}$ 
              & $\eta_b (1S)\eta_b (1S)$  & 18882   & 1.42~~ \\
              & $\Upsilon (1S)\Upsilon (1S)$  & 19019  &1.28~~\\
              & $\eta_b (1S)\eta_b (2S)$  & 19394   & 0.16~~ \\
              & $\Upsilon (1S)\Upsilon (2S)$  & 19454   & 0.46~~ \\[2ex]
~~$1^{+(-)}$ 
              & $\Upsilon (1S)\Upsilon (1S)$  & 19198   & 0.22~~ \\
              & $\eta_b (1S)\Upsilon (2S)$  & 19402   & 0.22~~ \\[2ex]
~~$2^{+(+)}$ 
              & $\Upsilon (1S)\Upsilon (2S)$  & 19473   & 0.42~~ \\
              & $\Upsilon (1S)\Upsilon (2S)$  & 19633   & 0.40~~ \\   
\end{tabular}
\end{ruledtabular}
\end{table}

\begin{figure}[ht]
\includegraphics[clip, trim={3.0cm, 1.0cm, 2.0cm, 1.0cm}, width=0.50\textwidth]{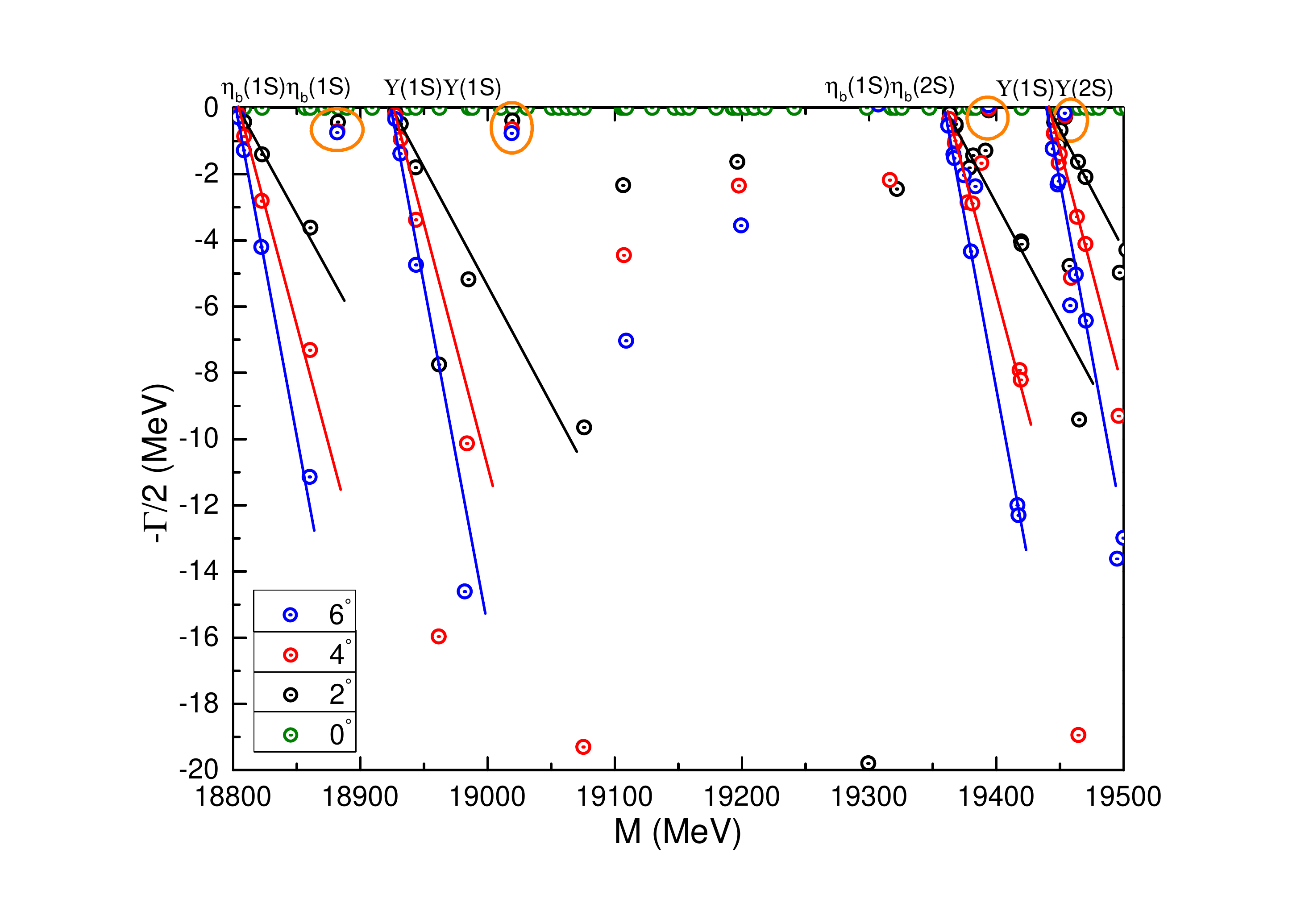}
\caption{Complex energy spectrum of the fully-bottom tetraquark system with $J^{P(C)}=0^{+(+)}$ from the complete coupled-channel calculation with CSM. The parameter $\theta$ varies from $0^\circ$ to $6^\circ$.} \label{PP4}
\end{figure}


\subsection{Fully-bottom system $bb\bar{b}\bar{b}$}

We do not find bound states for all $S$-wave cases: $J^{P(C)}=0^{+(+)}$, $1^{+(-)}$, and $2^{+(+)}$, of the fully-bottom system $bb\bar{b}\bar{b}$. However, performing a complete coupled-channel calculation within the CSM formalism, several narrow resonances are found and they are summarized in Table~\ref{Rsum-bbbb}. The results are discussed in the following.

{\bf The $\bm{J^{P(C)}=0^{+(+)}}$ channel:} Table~\ref{GresultCC4} lists the four di-meson channels, including the color-singlet and hidden-color configurations of $\eta_b \eta_b$ and $\Upsilon \Upsilon$, the two diquark-antidiquark channels, $[bb]^{\bf 3_{\rm c}}[\bar{b}\bar{b}]^{\bf \bar{3}_{\rm c}}$ and $[bb]^{\bf 6_{\rm c}}[\bar{b}\bar{b}]^{\bf \bar{6}_{\rm c}}$, and twelve K-type allowed structures. For the color-singlet channels of di-$\eta_b$ and di-$\Upsilon$ configurations, the lowest masses are just their corresponding theoretical thresholds, 18802 MeV and 18926 MeV, respectively. For other exotic structures, most of their masses are above 19.1 GeV, while two channels in the K$_1$ and K$_2$ configurations possess masses around 19 GeV, close to the threshold of the $\Upsilon \Upsilon$ channel. When a partially-coupled calculation is performed for each structure, masses are ranging from 18.8 to 19.1 GeV. The lowest mass of the spectrum from the fully-coupled calculation is 18802 MeV, which is slightly above the threshold of di-$\eta_b$. Therefore, no bound state exists for $J^{P(C)}=0^{+(+)}$.

\begin{table}[!t]
\caption{\label{GresultCC4} Calculation of the $J^{P(C)}=0^{+(+)}$ state of the fully-bottom system in the real case ($\theta=0^\circ$). The first and second columns show all possible structures and channels. They are obtained by the necessary bases in spin and color degrees of freedom summarized in the third column. The values in the parentheses after the first two meson-meson structures give the theoretical results of the noninteracting meson-meson thresholds. The lowest energies of all channels without considering any channel-channel coupling are listed in the fourth column. The last column shows the lowest energies in a calculation taking into account the couplings among the channels belonging to each structure (configuration). The last row gives the lowest energy from the complete coupled-channel calculation. All results are in units of MeV.}
\begin{ruledtabular}
\begin{tabular}{lcccc}
~~Structure  & Channel & $\xi_0^{\alpha}$; $\varphi_\beta^{\phantom{\dag}}$ & $M$ & Mixed~~ \\
 & & $[\alpha;~\beta]$ &  &  \\[2ex]
~~$\eta_b \eta_b (18802)$      & 1  & [1; 1]  & $18802$ &~~ \\
~~$\Upsilon \Upsilon (18926)$   & 2  & [2; 1]  & $18926$ &$18802$~~  \\[2ex]
~~$\eta^8_b \eta^8_b$           & 3  & [1; 2]  & $19243$ &~~ \\
~~$\Upsilon^8 \Upsilon^8$       & 4  & [2; 2]  & $19237$ &$19144$~~  \\[2ex]
~~$[bb]^{\bf 6_{\rm c}}[\bar{b}\bar{b}]^{\bf \bar{6}_{\rm c}}$   & 5  & [3; 4] & $19173$ &~~ \\
~~$[bb]^{\bf 3_{\rm c}}[\bar{b}\bar{b}]^{\bf \bar{3}_{\rm c}}$   & 6  & [4; 3]  & $19196$  &$19148$~~\\[2ex]
~~K$_1$   & 7 & [5; 5]   & $19237$ & \\
                 & 8  & [5; 6]  & $19058$ & \\
                 & 9  & [6; 5]  & $19241$ &  \\
                 & 10 & [6; 6]  & $18992$ & $18977$~~ \\[2ex]
~~K$_2$   & 11  & [7; 7]  & $19058$ & \\
                 & 12  & [7; 8]  & $19237$ & \\
                 & 13  & [8; 7]  & $18992$ &  \\
                 & 14  & [8; 8]  & $19241$ & $18977$~~ \\[2ex]
~~K$_3$   & 15  & [9; 10]  & $19197$ & \\
                 & 16  & [10; 9]  & $19167$ &$19143$~~  \\[2ex]
~~K$_4$   & 17  & [11; 12]  & $19197$ & \\
                 & 18   & [12; 11]  & $19167$ &$19143$~~  \\[2ex]
\multicolumn{4}{l}{Lowest mass of the fully-coupled result:}  & $18802$~~ \\
\end{tabular}
\end{ruledtabular}
\end{table}

We now look for possible resonances by performing a complete coupled-channel calculation within CSM. The complex energy spectrum is presented in Fig.~\ref{PP4}, with energy ranging from 18.8 to 19.5 GeV. There are four scattering states, $\eta_b (1S) \eta_b (1S)$, $\Upsilon (1S) \Upsilon (1S)$, $\eta_b (1S)\eta_b (2S)$, and $\Upsilon (1S)\Upsilon (2S)$; with varying angle $\theta$, the complex eigenenergies are generally aligned along the cut lines. In addition to the scattering states, four resonance poles are found whose structure and mass can be identified with $\eta_b (1S)\eta_b (1S) (18882)$, $\Upsilon (1S)\Upsilon (1S) (19019)$, $\eta_b (1S)\eta_b (2S) (19394)$ and $\Upsilon (1S)\Upsilon (2S) (19454)$. Their widths are small, they are 1.42 MeV, 1.28 MeV, 0.16 MeV, and 0.46 MeV, respectively.

\begin{table}[!t]
\caption{\label{GresultCC5} Calculation of the $J^{P(C)}=1^{+(-)}$ state of the fully-bottom system in the real case ($\theta=0^\circ$). The first and second columns show all possible structures and channels. They are obtained by the necessary bases in spin and color degrees of freedom summarized in the third column. The values in the parentheses after the first two meson-meson structures give the theoretical results of the noninteracting meson-meson thresholds. The lowest energies of all channels without considering any channel-channel coupling are listed in the fourth column. The last column shows the lowest energies in a calculation taking into account the couplings among the channels belonging to each structure (configuration). The last row gives the lowest energy from the complete coupled-channel calculation. All results are in units of MeV.}
\begin{ruledtabular}
\begin{tabular}{lcccc}
~~Structure  & Channel & $\xi_1^{\alpha}$; $\varphi_\beta^{\phantom{\dag}}$ & $M$ & Mixed~~ \\
 & & $[\alpha;~\beta]$ &  &  \\[2ex]
~~$\eta_b \Upsilon (18864)$      & 1  & [1; 1] & $18864$ &~~ \\
~~$\Upsilon \Upsilon (18926)$      & 2 & [3; 1]   & $18926$ &$18864$~~ \\[2ex]
~~$\eta^8_b \Upsilon^8$                 & 3 & [1; 2]   & $19126$ &~~ \\
~~$\Upsilon^8 \Upsilon^8$                 & 4  & [3; 2]  & $19227$ & $19126$~~ \\[2ex]
~~$[bb]^{\bf 3_{\rm c}}[\bar{b}\bar{b}]^{\bf \bar{3}_{\rm c}}$   & 5  & [6; 3]  & $19205$ & $19205$~~ \\[2ex]
~~K$_1$   & 6  & [7; 5]  & $19197$ & \\
                 & 7  & [8; 5]  & $19191$ & \\
                 & 8  & [9; 5]  & $19170$ & \\
                 & 9  & [7; 6]  & $19186$ & \\
                 & 10  & [8; 6]  & $19197$ & \\
                 & 11  & [9; 6]  & $19139$ & $19053$~~ \\[2ex]
~~K$_2$   & 12  & [10; 7]  & $19186$ & \\
                 & 13  & [11; 7]  & $19197$ & \\
                 & 14  & [12; 7]  & $19139$ & \\
                 & 15  & [10; 8]  & $19197$ & \\
                 & 16  & [11; 8]  & $19191$ & \\
                 & 17  & [12; 8]  & $19170$ & $19053$~~ \\[2ex]
~~K$_3$   & 18  & [13; 10]  & $19212$ & \\
                 & 19   & [14; 10] & $19211$ & \\
                 & 20  & [15; 9]  & $19603$ & $19206$~~ \\[2ex]
~~K$_4$   & 21  & [16; 12]  & $19212$ & \\
                 & 22  & [17; 12]  & $19211$ & \\
                 & 23  & [18; 11]  & $19603$ & $19206$~~ \\[2ex]
\multicolumn{4}{l}{Lowest mass of the fully-coupled result:}    & $18864$~~  \\
\end{tabular}
\end{ruledtabular}
\end{table}

Although the fully-bottom tetraquark state has not yet been reported experimentally, our theoretical findings could be valuable for future experimental studies. In fact, the $\Upsilon (1S)\Upsilon(1S) (19019)$ resonance is also supported in other theoretical works as, for instance, Ref.~\cite{JZDCXLTM200807430}.

\begin{figure}[ht]
\includegraphics[clip, trim={3.0cm, 1.0cm, 2.0cm, 1.0cm}, width=0.50\textwidth]{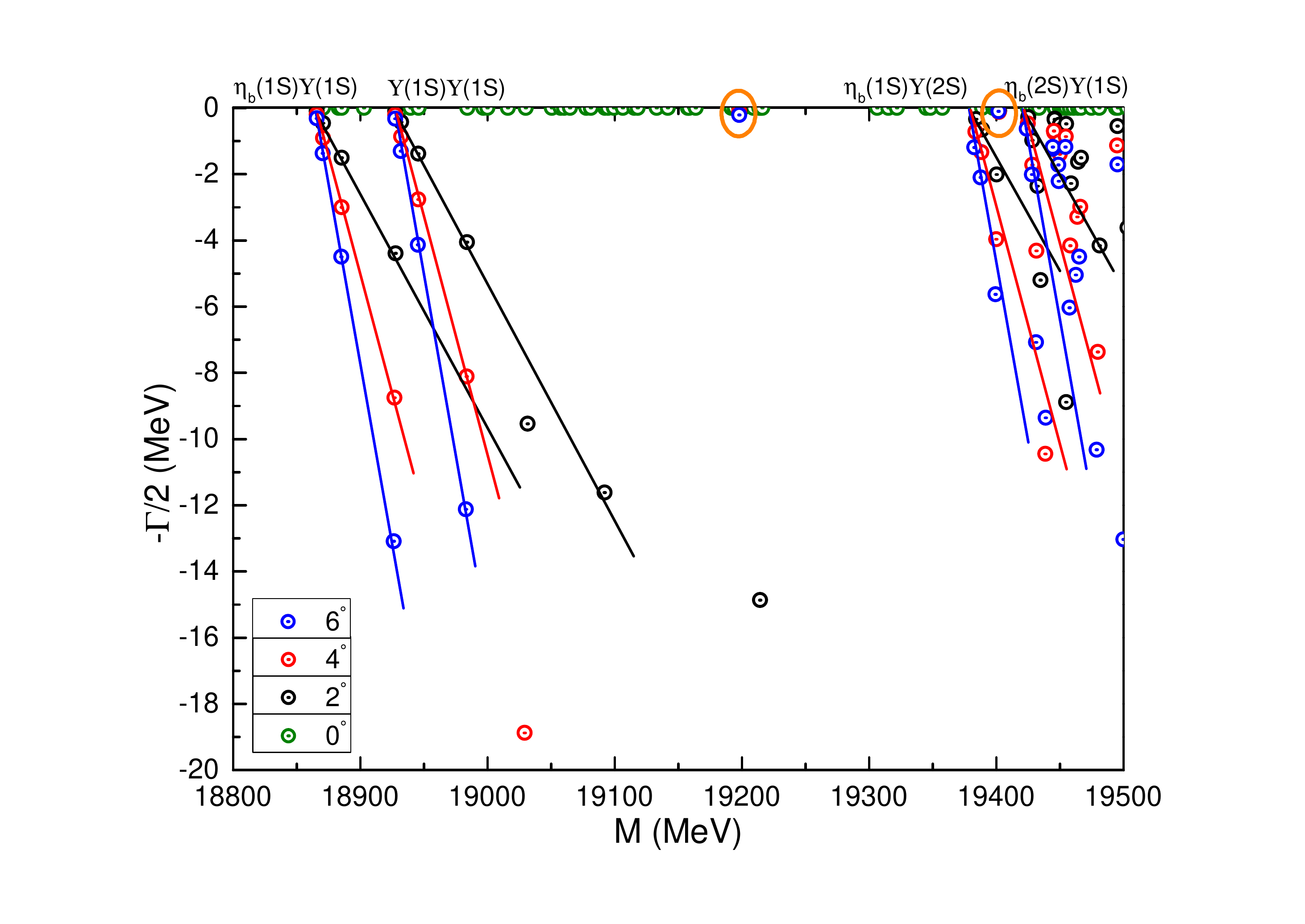}
\caption{Complex energy spectrum of the fully-bottom tetraquark system with $J^{P(C)}=1^{+(-)}$ from the complete coupled-channel calculation with CSM. The parameter $\theta$ varies from $0^\circ$ to $6^\circ$.} \label{PP5}
\end{figure}

{\bf The $\bm{J^{P(C)}=1^{+(-)}}$ channel:} there are 23 channels listed in Table~\ref{GresultCC5}, including four di-meson channels with $\eta_b \Upsilon$ and $\Upsilon \Upsilon$ di-meson configurations in the singlet- and hidden-color states, one diquark-antidiquark channel, $[bb]^{\bf 6_{\rm c}}[\bar{b}\bar{b}]^{\bf \bar{6}_{\rm c}}$, and 18 K-type ones. First of all, the lowest channel $\eta_b \Upsilon$ is unbound, with a mass of 18864 MeV. The other color-singlet di-meson channel, $\Upsilon \Upsilon$, is also unbound with the lowest mass 18926 MeV. The masses of the remaining configurations populate densely a small energy range, $19.1-19.2$ GeV; except for two channels from K$_3$ and K$_4$ structures whose masses are located at $\sim$19.6 GeV. The lowest mass of the spectrum from the fully-coupled calculation is 18864 MeV, which is slightly above the threshold of $\eta_b \Upsilon$. Therefore, no bound state exists for $J^{P(C)}=1^{+(-)}$.

Again, we look for possible resonances by performing a complete coupled-channel calculation with CSM. The complex energy spectrum is shown in Fig.~\ref{PP5}. Apart from four continuum states, $\eta_b (1S)\Upsilon (1S)$, $\Upsilon (1S)\Upsilon (1S)$, $\eta_b (1S)\Upsilon (2S)$, and $\eta_b (2S)\Upsilon (1S)$, we find two resonance poles in the complex energy plane. Attending to their components and masses, we can identify these two resonance poles as $\Upsilon (1S)\Upsilon (1S) (19198)$ and $\eta_b (1S)\Upsilon (2S) (19402)$. Both have extremely small two-meson decay widths: 0.22 MeV.

{\bf The $\bm{J^{P(C)}=2^{+(+)}}$ channel:} we have 9 channels in this case,  including two di-meson channels, $\Upsilon \Upsilon$ configurations in the singlet- and hidden-color states, one diquark-antidiquark channel
$[bb]^{\bf 6_{\rm c}}[\bar{b}\bar{b}]^{\bf \bar{6}_{\rm c}}$, and six K-type ones. The calculated lowest mass for $\Upsilon \Upsilon$ is 18926 MeV, which is equal to its theoretical threshold. The masses of the other channels are all above 19.1 GeV, and there is an approximate degeneracy among the channels $[bb]^{\bf 6_{\rm c}}[\bar{b}\bar{b}]^{\bf \bar{6}_{\rm c}}$, K$_3$, and K$_4$. The lowest mass of the spectrum from the fully-coupled calculation is 18926 MeV, which is slightly above the threshold of $\Upsilon \Upsilon$. Therefore, no bound state exists for $J^{P(C)}=2^{+(+)}$. 

\begin{figure}[ht]
\includegraphics[clip, trim={3.0cm, 1.0cm, 2.0cm, 1.0cm}, width=0.50\textwidth]{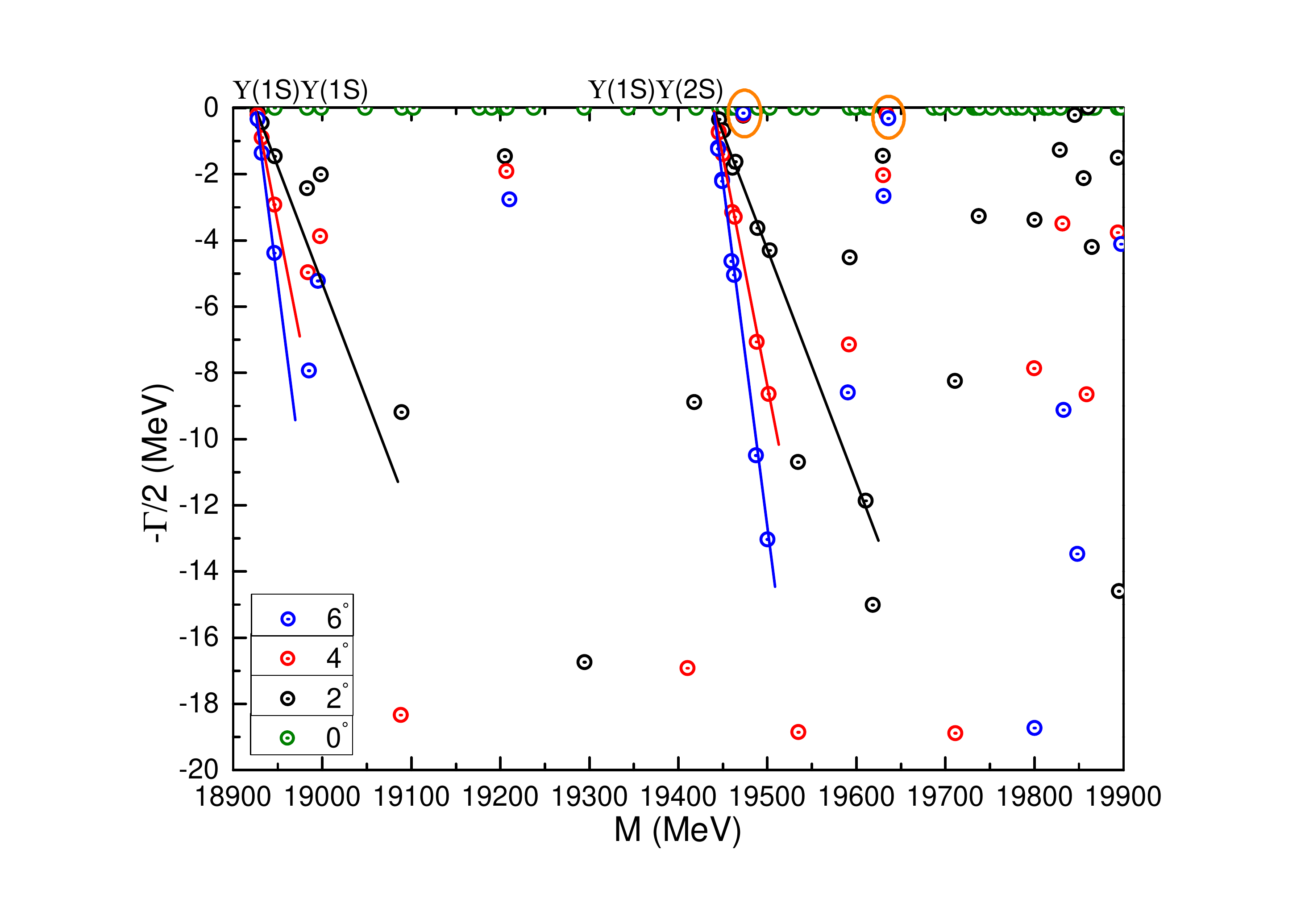}
\caption{Complex energy spectrum of the fully-bottom tetraquark system with $J^{P(C)}=2^{+(+)}$ from the complete coupled-channel calculation with CSM. The parameter $\theta$ varies from $0^\circ$ to $6^\circ$.} \label{PP6}
\end{figure}

Fig.~\ref{PP6} shows the complex energy spectrum for $J^{P(C)}=2^{+(+)}$, obtained by using CSM. We find two resonance poles above the $\Upsilon(1S) \Upsilon(2S)$ threshold. They can be identified as $\Upsilon(1S) \Upsilon(2S) (19473)$ and $\Upsilon(1S) \Upsilon(2S) (19633)$, with their widths 0.42 MeV and 0.40 MeV, respectively. 

\begin{table}[!t]
\caption{\label{GresultCC6} Calculation of the $J^{P(C)}=2^{+(+)}$ state of the fully-bottom system in the real case ($\theta=0^\circ$). The first and second columns show all possible structures and channels. They are obtained by the necessary bases in spin and color degrees of freedom summarized in the third column. The value in the parenthesis after the first meson-meson structure gives the theoretical result of the noninteracting meson-meson threshold. The lowest energies of all channels without considering any channel-channel coupling are listed in the fourth column. The last column shows the lowest energies in a calculation taking into account the couplings among the channels belonging to each structure (configuration). The last row gives the lowest energy from the complete coupled-channel calculation. All results are in units of MeV.}
\begin{ruledtabular}
\begin{tabular}{lcccc}
~~Structure  & Channel & $\xi_2$; $\varphi_\beta^{\phantom{\dag}}$ & $M$ & Mixed~~ \\
 & & $[\beta]$ &  &  \\[2ex]
~~$\Upsilon \Upsilon (18926)$       & 1  & 1  & $18926$ &$18926$~~  \\[2ex]
~~$\Upsilon^8 \Upsilon^8$       & 2  & 2  & $19197$ &$19197$~~  \\[2ex]
~~$[bb]^{\bf 3_{\rm c}}[\bar{b}\bar{b}]^{\bf \bar{3}_{\rm c}}$   & 3  & 3   & $19223$  &$19223$~~\\[2ex]
~~K$_1$   & 4  & 5  & $19147$ & \\
                 & 5  & 6  & $19101$ & $19093$~~ \\[2ex]
~~K$_2$   & 6 & 7  & $19101$ & \\
                 & 7 & 8  & $19147$ & $19093$~~ \\[2ex]
~~K$_3$   & 8  & 10  & $19225$ & $19225$~~ \\[2ex]
~~K$_4$   & 9  & 12  & $19225$ & $19225$~~ \\[2ex]
\multicolumn{4}{l}{Lowest mass of the fully-coupled result:}  & $18926$~~ \\
\end{tabular}
\end{ruledtabular}
\end{table}

\subsection{Charm-bottom system $\bar{c}c\bar{c}b$}

For the charm-bottom system $\bar{c}c\bar{c}b$, narrow resonances are found in $J^{P}=0^{+}$, $1^{+}$, and $2^{+}$ states, respectively. Table~\ref{Rsum-cccb} summarizes the results and the details are discussed in the following.

\begin{table}[!t]
\caption{\label{Rsum-cccb} Predicted resonances of the $\bar{c}c\bar{c}b$ tetraquark system from the complete coupled-channel calculation with CSM. Their masses and widths are summarized in the third and fourth column, respectively, in units of MeV.}
\begin{ruledtabular}
\begin{tabular}{cccc}
~~$J^{P}$ & Resonance & Mass &  Width~~ \\
~~$0^{+}$ 
              & $J/\psi (1S) B^*_c (1S)$  & 9813   & 4.02~~ \\[2ex]
~~$1^{+}$ 
              & $J/\psi (1S) B^*_c (1S)$  & 9823   & 2.41~~ \\
              & $\eta_c (1S) B^*_c (2S)$  & 9898   & 1.33~~ \\
              & $\eta_c (1S) B^*_c (2S)$  & 9928   & 3.80~~ \\
              & $\eta_c (1S) B^*_c (2S)$  & 9974   & 8.21~~ \\
              & $\eta_c (2S) B^*_c (1S)$  & 10284   & 3.00~~ \\[2ex]
~~$2^{+}$ 
              & $\psi (2S) B^*_c (1S)$  & 10563   & 6.61~~ \\  
\end{tabular}
\end{ruledtabular}
\end{table}

{\bf The $\bm{J^{P}=0^{+}}$ channel:} Table~\ref{GresultCC7} lists the lowest masses of each four configurations of $\bar{c}c\bar{c}b$ tetraquark in $0^{+(+)}$ states. In the real-range calculation, $\theta=0^\circ$, the two dimeson color-singlet channels $\eta_c B_c$ and $J/\psi B^*_c$ are just scattering states, and their hidden-color cases present masses about 9.6 GeV. The two diquark-antidiquark structures are almost degenerate at 9.67 GeV. As for the other K-type configurations, which are all excited channels, the calculated single-channel masses are generally in a mass region of $9.45-9.69$ GeV. Then, in the coupled-channel investigations on each configuration, bound states are still not found, the lowest mass is just the theoretical threshold value of $\eta_c B_c$ at 9243 MeV, and the other excited configurations are generally at 9.5 GeV. This fact remains in a complete coupled-channels calculation, whose lowest mass at 9243 MeV is shown in the last row of Table~\ref{GresultCC7}.

\begin{figure}[ht]
\includegraphics[clip, trim={3.0cm, 1.0cm, 2.0cm, 1.0cm}, width=0.50\textwidth]{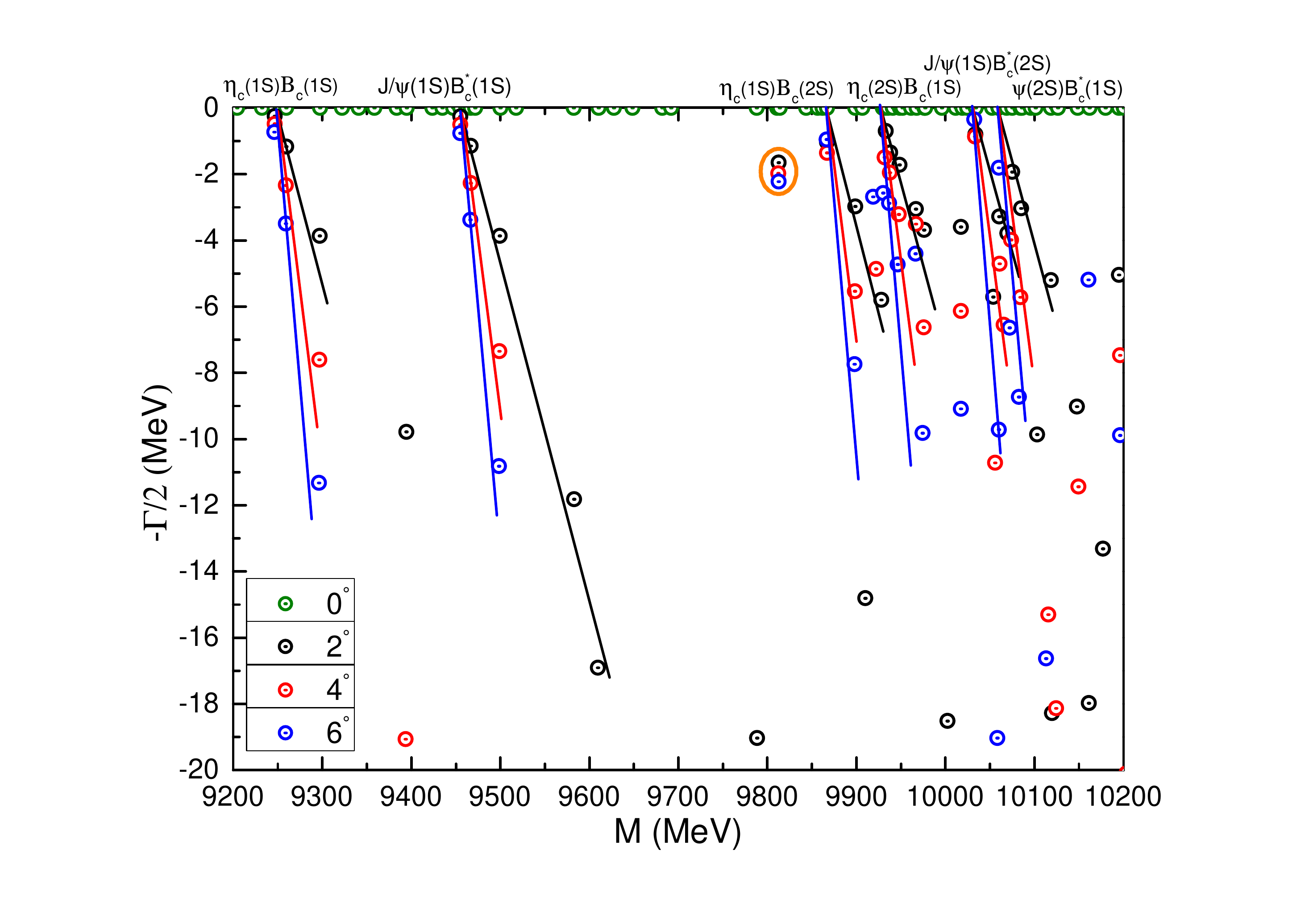}
\caption{Complex energy spectrum of the $\bar{c}c\bar{c}b$ system with $J^{P}=0^{+}$ from the complete coupled-channel calculation with CSM. The parameter $\theta$ varies from $0^\circ$ to $6^\circ$.} \label{PP7}
\end{figure}

In a further step, when a complex-scaling study is considered in a fully-coupled calculation, one narrow resonance state is found. Fig.~\ref{PP7} presents the distributions of the complex energies of $\bar{c}c\bar{c}b$ tetraquarks with rotated angle $\theta$ varied from $0^\circ$ to $6^\circ$. In a mass region from 9.2 GeV to 10.2 GeV, the scattering nature of $\eta_c B_c$ and $J/\psi B^*_c$ both in the ground and first radial excitation states are well established. One can also see in Fig.~\ref{PP7} a stable resonance pole, whose mass and width are 9813 MeV and 4.02 MeV, respectively. This resonance is below the threshold lines of $\eta_c(1S)B_c(2S)$, and the dominant channel, therefore, can be identified as the $J/\psi(1S)B^*_c(1S)$ state.

\begin{table}[!t]
\caption{\label{GresultCC7} Calculation of the $J^{P}=0^{+}$ state of the $\bar{c}c\bar{c}b$ system in the real case ($\theta=0^\circ$). The first and second columns show all possible structures and channels. They are obtained by the necessary bases in spin and color degrees of freedom summarized in the third column. The values in the parentheses after the first two meson-meson structures give the theoretical results of the noninteracting meson-meson thresholds. The lowest energies of all channels without considering any channel-channel coupling are listed in the fourth column. The last column shows the lowest energies in a calculation taking into account the couplings among the channels belonging to each structure (configuration). The last row gives the lowest energy from the complete coupled-channel calculation. All results are in units of MeV.}
\begin{ruledtabular}
\begin{tabular}{lcccc}
~~Structure  & Channel & $\xi_0^{\alpha}$; $\varphi_\beta^{\phantom{\dag}}$ & $M$ & Mixed~~ \\
 & & $[\alpha;~\beta]$ &  &  \\[2ex]
~~$\eta_c B_c (9243)$      & 1  & [1; 1]  & $9243$ &~~ \\
~~$J/\psi B^*_c (9451)$   & 2  & [2; 1]  & $9451$ &$9243$~~  \\[2ex]
~~$\eta^8_c B^8_c$           & 3  & [1; 2]  & $9683$ &~~ \\
~~$J/\psi^8 B^{*8}_c$       & 4  & [2; 2]  & $9625$ &$9559$~~  \\[2ex]
~~$[cb]^{\bf 6_{\rm c}}[\bar{c}\bar{c}]^{\bf \bar{6}_{\rm c}}$   & 5  & [3; 4] & $9672$ &~~ \\
~~$[cb]^{\bf 3_{\rm c}}[\bar{c}\bar{c}]^{\bf \bar{3}_{\rm c}}$   & 6  & [4; 3]  & $9673$  &$9620$~~\\[2ex]
~~K$_1$   & 7 & [5; 5]   & $9617$ & \\
                 & 8  & [5; 6]  & $9696$ & \\
                 & 9  & [6; 5]  & $9690$ &  \\
                 & 10 & [6; 6]  & $9558$ & $9551$~~ \\[2ex]
~~K$_2$   & 11  & [7; 7]  & $9602$ & \\
                 & 12  & [7; 8]  & $9624$ & \\
                 & 13  & [8; 7]  & $9447$ &  \\
                 & 14  & [8; 8]  & $9699$ & $9310$~~ \\[2ex]
~~K$_3$   & 15  & [9; 9]  & $10105$ & \\
                 & 16  & [9; 10]  & $9673$ &  \\
                 & 17  & [10; 9]  & $9666$ &  \\
                 & 18  & [10; 10]  & $10172$ &$9599$~~  \\[2ex]
~~K$_4$   & 19  & [11; 12]  & $9672$ & \\
                 & 20   & [12; 11]  & $9689$ &$9621$~~  \\[2ex]
\multicolumn{4}{l}{Lowest mass of the fully-coupled result:}  & $9243$~~ \\
\end{tabular}
\end{ruledtabular}
\end{table}

{\bf The $\bm{J^{P}=1^{+}}$ channel:} There are in all 30 channels listed in Table~\ref{GresultCC8} to be considered for the $\bar{c}c\bar{c}b$ tetraquark in $1^{+}$ state. In particular, six channels devote to the meson-meson structures, three channels belong to the diquark-antidiquark ones, and 21 channels correspond to the K-type configurations. When they are considered as isolated channels or when a coupled-channels calculation in each configurations is performed, no bound state is found. The lowest mass, 9317 MeV, is the theoretical threshold value of $\eta_c B^*_c$; the other two dimeson channels, $J/\psi B_c$ and $J/\psi B^*_c$, are at 9377 MeV and 9451 MeV, respectively; the exotic configurations, \emph{i.e.} hidden-color, diquark-antidiquark and K-types structures, are generally located in a energy range from 9.62 GeV to 9.68 GeV; and the lowest mass in each of the conficuration's coupled-channels calculation is around 9.60 GeV.

\begin{figure}[!t]
\includegraphics[clip, trim={3.0cm 1.0cm 2.0cm 1.0cm}, width=0.45\textwidth]{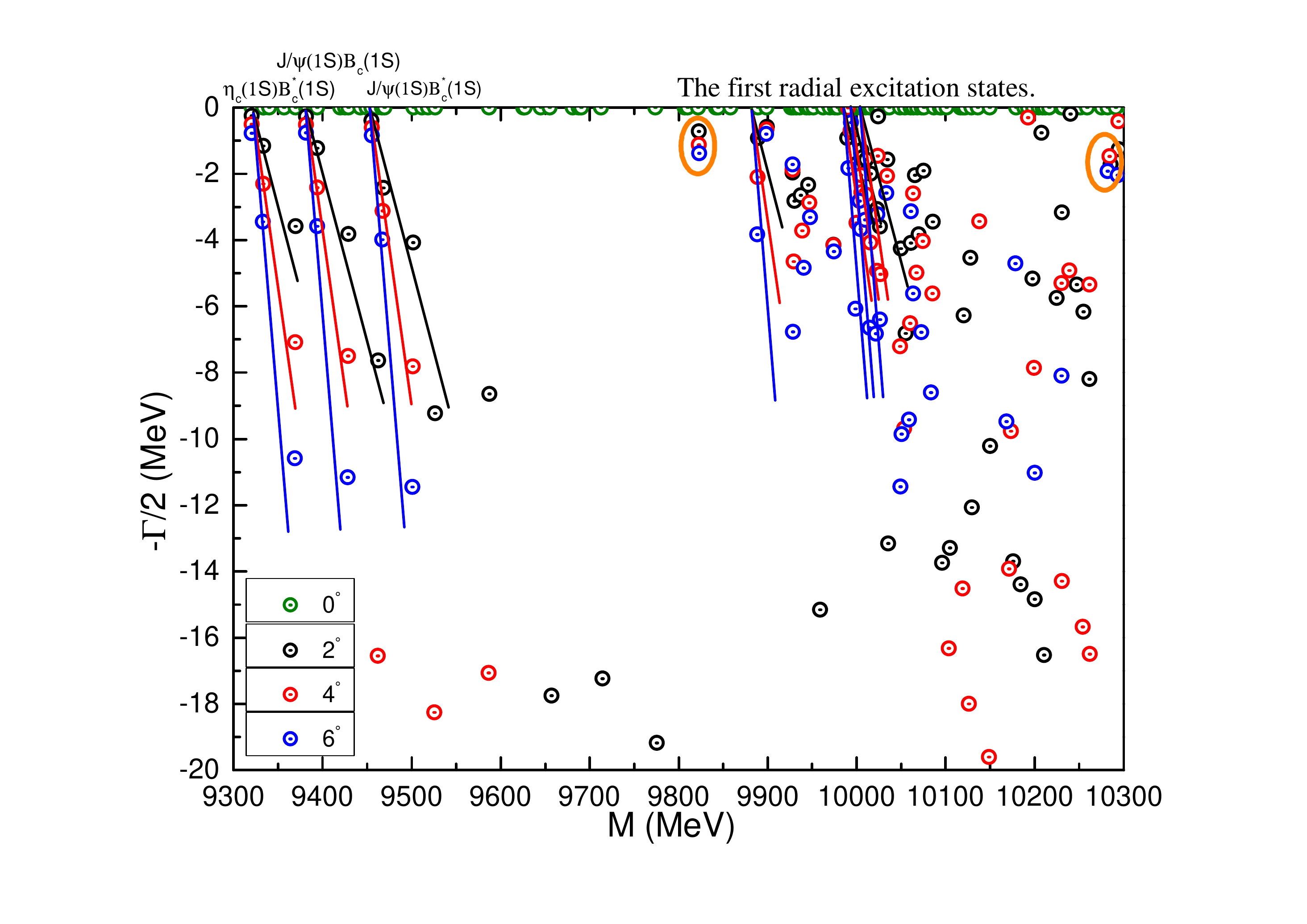} \\
\includegraphics[clip, trim={3.0cm 1.0cm 2.0cm 1.0cm}, width=0.45\textwidth]{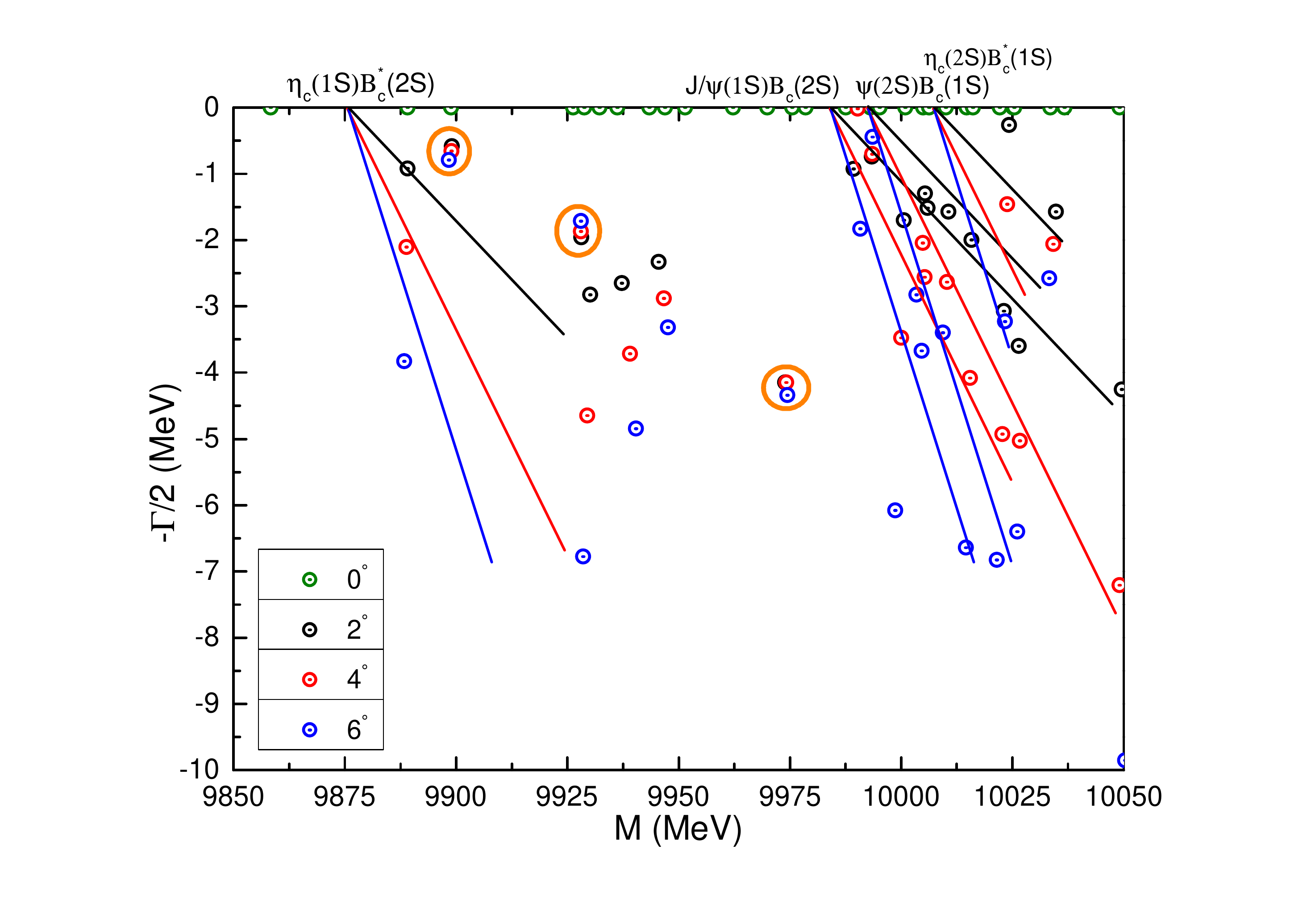}
\caption{\label{PP8} {\it Top panel:} Complex energy spectrum of the $\bar{c}c\bar{c}b$ system with $J^{P}=1^{+}$ from the complete coupled-channel calculation with CSM. The parameter $\theta$ varies from $0^\circ$ to $6^\circ$. {\it Bottom panel:} Enlarged case of the top panel, with real values of energy ranging from $9.85\,\text{GeV}$ to $10.05\,\text{GeV}$.} 
\end{figure}

However, several resonances are obtained when the complex scaling method is employed in a complete coupled-channels calculation. Figure~\ref{PP8} presents our results; in particular, the top panel shows the distributions of $\eta_c B_c^*$, $J/\psi B_c$ and $J/\psi B^*_c$ continuum states, their ground and first radial excitation states basically align along the corresponding threshold lines. Nevertheless, two stable resonance poles are circled therein and they can be identified as $J/\psi(1S)B^*_c(1S)(9823)$ and $\eta_c(2S)B^*_c(1S)(10284)$ states, with resonance's widths 2.41 MeV and 3.00 MeV, respectively.

Since the first radial excitation of $\eta_c B_c^*$, $J/\psi B_c$ and $J/\psi B^*_c$ states are densely distribute in the top panel of Fig.~\ref{PP8}, an enlarged zoom, with energy range from 9.85 GeV to 10.05 GeV, is presented as bottom panel. Therein, the first radial excitation states of $\eta_c(1S)B^*_c(2S)$, $J/\psi(1S)B_c(2S)$, $\psi(2S)B_c(1S)$ and $\eta_c(2S)B^*_c(1S)$ are clearly shown. Additionally, three more $\eta_c(1S)B^*_c(2S)$ resonances are found, \emph{i.e.} $\eta_c(1S)B^*_c(2S)(9898)$, $\eta_c(1S)B^*_c(2S)(9928)$ and $\eta_c(1S)B^*_c(2S)(9974)$, their resonance widths are 1.33 MeV, 3.80 MeV and 8.21 MeV, respectively.

\begin{table}[!t]
\caption{\label{GresultCC8} Calculation of the $J^{P}=1^{+}$ state of the $\bar{c}c\bar{c}b$ system in the real case ($\theta=0^\circ$). All results are in units of MeV.}
\begin{ruledtabular}
\begin{tabular}{lcccc}
~~Structure  & Channel & $\xi_1^{\alpha}$; $\varphi_\beta^{\phantom{\dag}}$ & $M$ & Mixed~~ \\
 & & $[\alpha;~\beta]$ &  &  \\[2ex]
~~$\eta_c B^*_c (9317)$         & 1  & [1; 1] & $9317$ &~~ \\
~~$J/\psi B_c (9377)$  & 2 & [2; 1]   & $9377$ & \\
~~$J/\psi B^*_c (9451)$        & 3 & [3; 1]   & $9451$ &$9317$~~ \\[2ex]
~~$\eta^8_c B^{*8}_c$         & 4  & [1; 2] & $9659$ &~~ \\
~~$J/\psi^8 B^8_c$  & 5  & [2; 2]   & $9651$ & \\
~~$J/\psi^8 B^{*8}_c$         & 6  & [3; 2]   & $9622$ &$9587$~~ \\[2ex]
~~$[cb]^{\bf 6_{\rm c}}[\bar{c}\bar{c}]^{\bf \bar{6}_{\rm c}}$   & 7  & [4; 4]  & $9665$ & \\
~~$[cb]^{\bf 3_{\rm c}}[\bar{c}\bar{c}]^{\bf \bar{3}_{\rm c}}$   & 8  & [5; 3]  & $9675$ & \\
~~$[cb]^{\bf 3_{\rm c}}[\bar{c}\bar{c}]^{\bf \bar{3}_{\rm c}}$   & 9  & [6; 3]  & $9687$ & $9638$~~ \\[2ex]
~~K$_1$   & 10  & [7; 5]  & $9671$ & \\
                 & 11  & [8; 5]  & $9619$ & \\
                 & 12  & [9; 5]  & $9668$ & \\
                 & 13  & [7; 6]  & $9673$ & \\
                 & 14  & [8; 6]  & $9676$ & \\
                 & 15  & [9; 6]  & $9577$ & $9568$~~ \\[2ex]
~~K$_2$   & 16  & [10; 7]  & $9574$ & \\
                 & 17  & [11; 7]  & $9599$ & \\
                 & 18  & [12; 7]  & $9650$ & \\
                 & 19  & [10; 8]  & $9636$ & \\
                 & 20  & [11; 8]  & $9633$ & \\
                 & 21  & [12; 8]  & $9674$ & $9546$~~ \\[2ex]
~~K$_3$   & 22  & [13; 9]  & $9681$ & \\
                 & 23   & [14; 9] & $9678$ & \\
                 & 24  & [15; 9]  & $10165$ & \\
                 & 25  & [13; 10]  & $9696$ & \\
                 & 26   & [14; 10] & $9696$ & \\
                 & 27  & [15; 10]  & $9677$ & $9619$~~ \\[2ex]
~~K$_4$   & 28  & [16; 12]  & $9689$ & \\
                 & 29  & [17; 12]  & $9679$ & \\
                 & 30  & [18; 11]  & $9678$ & $9640$~~ \\[2ex]
\multicolumn{4}{l}{Lowest mass of the fully-coupled result:}    & $9317$~~  \\
\end{tabular}
\end{ruledtabular}
\end{table}

{\bf The $\bm{J^{P}=2^{+}}$ channel:} We have ten channels listed in Table~\ref{GresultCC9}. There is only one channel for the dimeson structure, $J/\psi B^*_c$, another one for the diquark-antidiquark case, $[cb]^{{\bf 3}c}[\bar{c}\bar{c}]^{\bar{\bf 3}c}$, and the rest belong to the K-type configurations. We obtain a lowest mass of 9451 MeV, which is just the theoretical threshold value of $J/\psi B^*_c$. Generally, the unconventional structures have masses which lie within the range 9.65$-$9.71 GeV. Moreover, no bound state is found when coupled-channel effects are taken into account in each tetraquark configuration separately.

\begin{figure}[ht]
\includegraphics[clip, trim={3.0cm, 1.0cm, 2.0cm, 1.0cm}, width=0.50\textwidth]{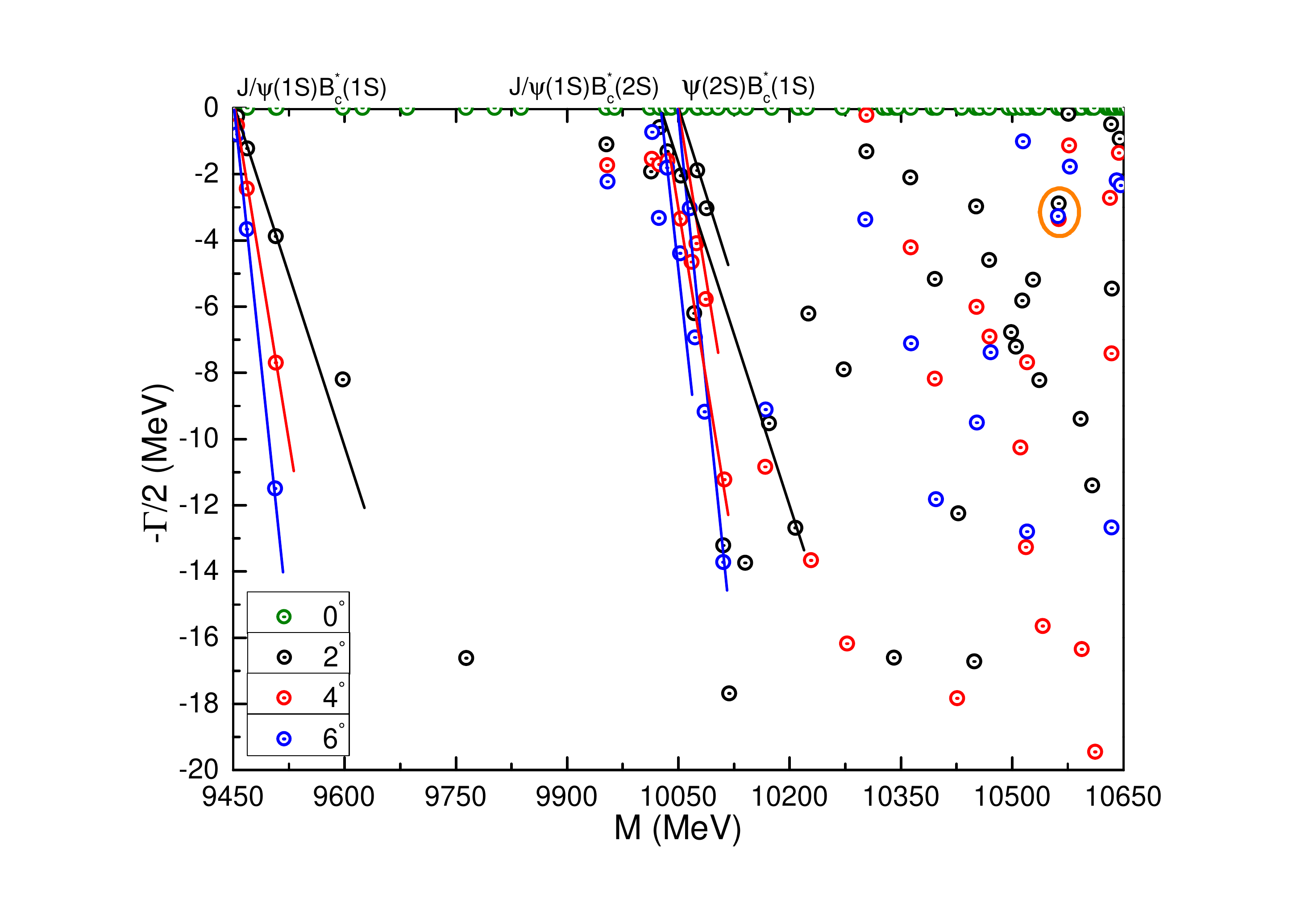}
\caption{Complex energy spectrum of the $\bar{c}c\bar{c}b$ system with $J^{P}=2^{+}$ from the complete coupled-channel calculation with CSM. The parameter $\theta$ varies from $0^\circ$ to $6^\circ$.} \label{PP9}
\end{figure}

Figure~\ref{PP9} shows the distribution of the complex energies obtained for the $\bar{c}c\bar{c}b$ tetraquark system. The nature as scattering states of the $J/\psi B^*_c$ system in ground and first radial excitation are well shown. However, from 9.45 GeV to 10.65 GeV energy region, a narrow resonance pole is obtained at 10.56 GeV, and its width is 6.61 MeV. Due to the location in the complex plane, we can identify it as a $\psi(2S)B^*_c(1S)(10563)$ resonance.

\begin{table}[!t]
\caption{\label{GresultCC9} Calculation of the $J^{P}=2^{+}$ state of the $\bar{c}c\bar{c}b$ system in the real case ($\theta=0^\circ$).
The first and second columns show all possible structures and channels. They are obtained by the necessary bases in spin and color degrees of freedom summarized in the third column. The value in the parenthesis after the first meson-meson structure gives the theoretical result of the noninteracting meson-meson threshold. The lowest energies of all channels without considering any channel-channel coupling are listed in the fourth column. The last column shows the lowest energies in a calculation taking into account the couplings among the channels belonging to each structure (configuration). The last row gives the lowest energy from the complete coupled-channel calculation. All results are in units of MeV.}
\begin{ruledtabular}
\begin{tabular}{lcccc}
~~Structure  & Channel & $\xi_2$; $\varphi_\beta^{\phantom{\dag}}$ & $M$ & Mixed~~ \\
 & & $[\beta]$ &  &  \\[2ex]
~~$J/\psi B^*_c (9451)$       & 1  & 1  & $9451$ &$9451$~~  \\[2ex]
~~$J/\psi^8 B^{*8}_c$       & 2  & 2  & $9650$ &$9650$~~  \\[2ex]
~~$[cb]^{\bf 3_{\rm c}}[\bar{c}\bar{c}]^{\bf \bar{3}_{\rm c}}$   & 3  & 3   & $9714$  &$9714$~~\\[2ex]
~~K$_1$   & 4  & 5  & $9695$ & \\
                 & 5  & 6  & $9697$ & $9688$~~ \\[2ex]
~~K$_2$   & 6 & 7  & $9613$ & \\
                 & 7 & 8  & $9682$ & $9593$~~ \\[2ex]
~~K$_3$   & 8  & 9  & $10184$ & ~~ \\
                  & 9  & 10  & $9716$ & $9704$~~ \\[2ex]
~~K$_4$   & 10  & 12  & $9717$ & $9717$~~ \\[2ex]
\multicolumn{4}{l}{Lowest mass of the fully-coupled result:}  & $9451$~~ \\
\end{tabular}
\end{ruledtabular}
\end{table}

\subsection{Charm-bottom system $\bar{b}b\bar{b}c$}

We find some resonance structures with quantum numbers $J^{P}=1^{+}$ and $2^{+}$ states for the charm-bottom system $\bar{b}b\bar{b}c$. Table~\ref{Rsum-bbbc} summarizes our findings, the details of the calculation are discussed in the following.

\begin{table}[!t]
\caption{\label{Rsum-bbbc} Predicted resonances of the $\bar{b}b\bar{b}c$ tetraquark system from the complete coupled-channel calculation with CSM. Their masses and widths are summarized in the third and fourth column, respectively, in units of MeV.}
\begin{ruledtabular}
\begin{tabular}{cccc}
~~$J^{P}$ & Resonance & Mass &  Width~~ \\
~~$1^{+}$ 
              & $\Upsilon (1S) \bar{B}^*_c (1S)$  & 16031   & 24.61~~ \\
              & $\Upsilon (2S) \bar{B}_c (1S)$  & 16303   & 1.80~~ \\[2ex]
~~$2^{+}$ 
              & $\Upsilon (1S) \bar{B}^*_c (1S)$  & 16312   & 1.12~~ \\  
\end{tabular}
\end{ruledtabular}
\end{table}

{\bf The $\bm{J^{P}=0^{+}}$ channel:} There are 20 channels listed in Table~\ref{GresultCC10}. In particular, there are two meson-meson channels, $\eta_b \bar{B}_c$ and $\Upsilon \bar{B}^*_c$, two diquark-antidiaquark ones, $[cb]^{{\bf 6}c}[\bar{b}\bar{b}]^{\bar{\bf 6}c}$ and $[cb]^{{\bf 3}c}[\bar{b}\bar{b}]^{\bar{\bf 3}c}$, and 14 channels of K-type configurations. Firstly, single-channel calculations deliver a lowest mass of 15676 MeV for the $\eta_b \bar{B}_c$, confirming its nature as a scattering state. The color-singlet channel of $\Upsilon \bar{B}^*_c$ configuration is also a scattering state with a mass of 15812 MeV. The remaining excited channels are located within 15.92$-$16.48 GeV. Although the coupled-channels effect helps in pushing down the lowest masses of each configuration, they are still above the threshold values of $\eta_b \bar{B}_c$ and $\Upsilon \bar{B}^*_c$. Moreover, when a complete coupled-bases calculation is performed, the coupling is still too weak to have a bound state.

\begin{figure}[ht]
\includegraphics[clip, trim={3.0cm, 1.0cm, 2.0cm, 1.0cm}, width=0.50\textwidth]{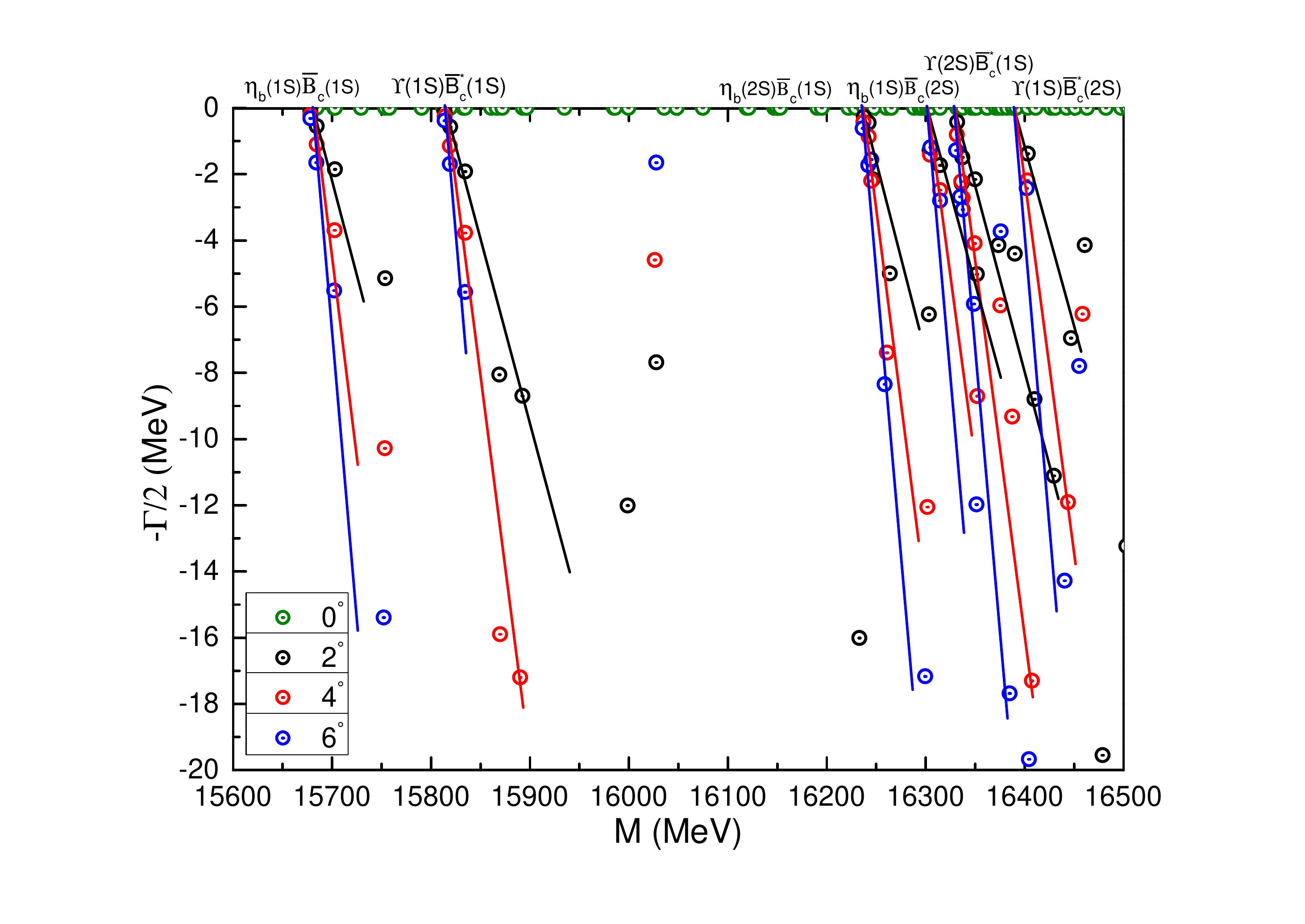}
\caption{Complex energy spectrum of the $\bar{b}b\bar{b}c$ system with $J^{P}=0^{+}$ from the complete coupled-channel calculation with CSM. The parameter $\theta$ varies from $0^\circ$ to $6^\circ$.} \label{PP10}
\end{figure}

A complex range analysis on the fully coupled-channels calculation is done and the results are presented in Fig.~\ref{PP10}. One can notice that, within a mass region of 15.6$-$16.5 GeV, six continuum states are clearly identified: $\eta_b(1S)\bar{B}_c(1S)$, $\Upsilon(1S)\bar{B}^*_c(1S)$, $\eta_b(2S)\bar{B}_c(1S)$, $\eta_b(1S)\bar{B}_c(2S)$, $\Upsilon(2S)\bar{B}^*_c(1S)$ and $\Upsilon(1S)\bar{B}^*_c(2S)$. However, neither bound nor resonance states are obtained in this case.

\begin{table}[!t]
\caption{\label{GresultCC10} Calculation of the $J^{P}=0^{+}$ state of the $\bar{b}b\bar{b}c$ system in the real case ($\theta=0^\circ$). The first and second columns show all possible structures and channels. They are obtained by the necessary bases in spin and color degrees of freedom summarized in the third column. The values in the parentheses after the first two meson-meson structures give the theoretical results of the non-interacting meson-meson thresholds. The lowest energies of all channels without considering any channel-channel coupling are listed in the fourth column. The last column shows the lowest energies in a calculation taking into account the couplings among the channels belonging to each structure (configuration). The last row gives the lowest energy from the complete coupled-channel calculation. All results are in units of MeV.}
\begin{ruledtabular}
\begin{tabular}{lcccc}
~~Structure  & Channel & $\xi_0^{\alpha}$; $\varphi_\beta^{\phantom{\dag}}$ & $M$ & Mixed~~ \\
 & & $[\alpha;~\beta]$ &  &  \\[2ex]
~~$\eta_b \bar{B}_c (15676)$      & 1  & [1; 1]  & $15676$ &~~ \\
~~$\Upsilon \bar{B}^*_c (15812)$   & 2  & [2; 1]  & $15812$ &$15676$~~  \\[2ex]
~~$\eta^8_b \bar{B}^8_c$           & 3  & [1; 2]  & $16091$ &~~ \\
~~$\Upsilon^8 \bar{B}^{*8}_c$       & 4  & [2; 2]  & $16064$ &$15976$~~  \\[2ex]
~~$[cb]^{\bf 6_{\rm c}}[\bar{b}\bar{b}]^{\bf \bar{6}_{\rm c}}$   & 5  & [3; 4] & $16042$ &~~ \\
~~$[cb]^{\bf 3_{\rm c}}[\bar{b}\bar{b}]^{\bf \bar{3}_{\rm c}}$   & 6  & [4; 3]  & $16058$  &$16012$~~\\[2ex]
~~K$_1$   & 7 & [5; 5]   & $16059$ & \\
                 & 8  & [5; 6]  & $15988$ & \\
                 & 9  & [6; 5]  & $16089$ &  \\
                 & 10 & [6; 6]  & $15919$ & $15906$~~ \\[2ex]
~~K$_2$   & 11  & [7; 7]  & $16032$ & \\
                 & 12  & [7; 8]  & $16059$ & \\
                 & 13  & [8; 7]  & $15967$ &  \\
                 & 14  & [8; 8]  & $16082$ & $15892$~~ \\[2ex]
~~K$_3$   & 15  & [9; 9]  & $16396$ & \\
                 & 16  & [9; 10]  & $16060$ &  \\
                 & 17  & [10; 9]  & $16037$ &  \\
                 & 18  & [10; 10]  & $16480$ &$15977$~~  \\[2ex]
~~K$_4$   & 19  & [11; 12]  & $16056$ & \\
                 & 20   & [12; 11]  & $16048$ &$16013$~~  \\[2ex]
\multicolumn{4}{l}{Lowest mass of the fully-coupled result:}  & $15676$~~ \\
\end{tabular}
\end{ruledtabular}
\end{table}

{\bf The $\bm{J^{P}=1^{+}}$ channel:} There are 30 channels when considering the $\bar{b}b\bar{b}c$ tetraquark system in $J^{P}=1^{+}$. Table~\ref{GresultCC11} shows the masses of the color-singlet channels of dimeson structures, $\eta_b\bar{B}^*_c$ (15750 MeV), $\Upsilon\bar{B}_c$ (15738 MeV) and $\Upsilon\bar{B}^*_c$ (15812 MeV); these results indicate the nature of scattering states. As for the other excited channels, their calculated masses generally locate in the range between 15.95 GeV and 16.44 GeV. In the coupled-channels studies of each configuration, the lowest mass of color singlet di-meson structures remains to be 15738 MeV, and the other exotic cases are about 16.0 GeV. Meanwhile, last row of Table~\ref{GresultCC11} indicate that the lowest channel, $\Upsilon\bar{B}_c$, is still unbound even in a complete coupled-channels calculation.

\begin{figure}[!t]
\includegraphics[clip, trim={3.0cm 1.0cm 3.0cm 1.0cm}, width=0.45\textwidth]{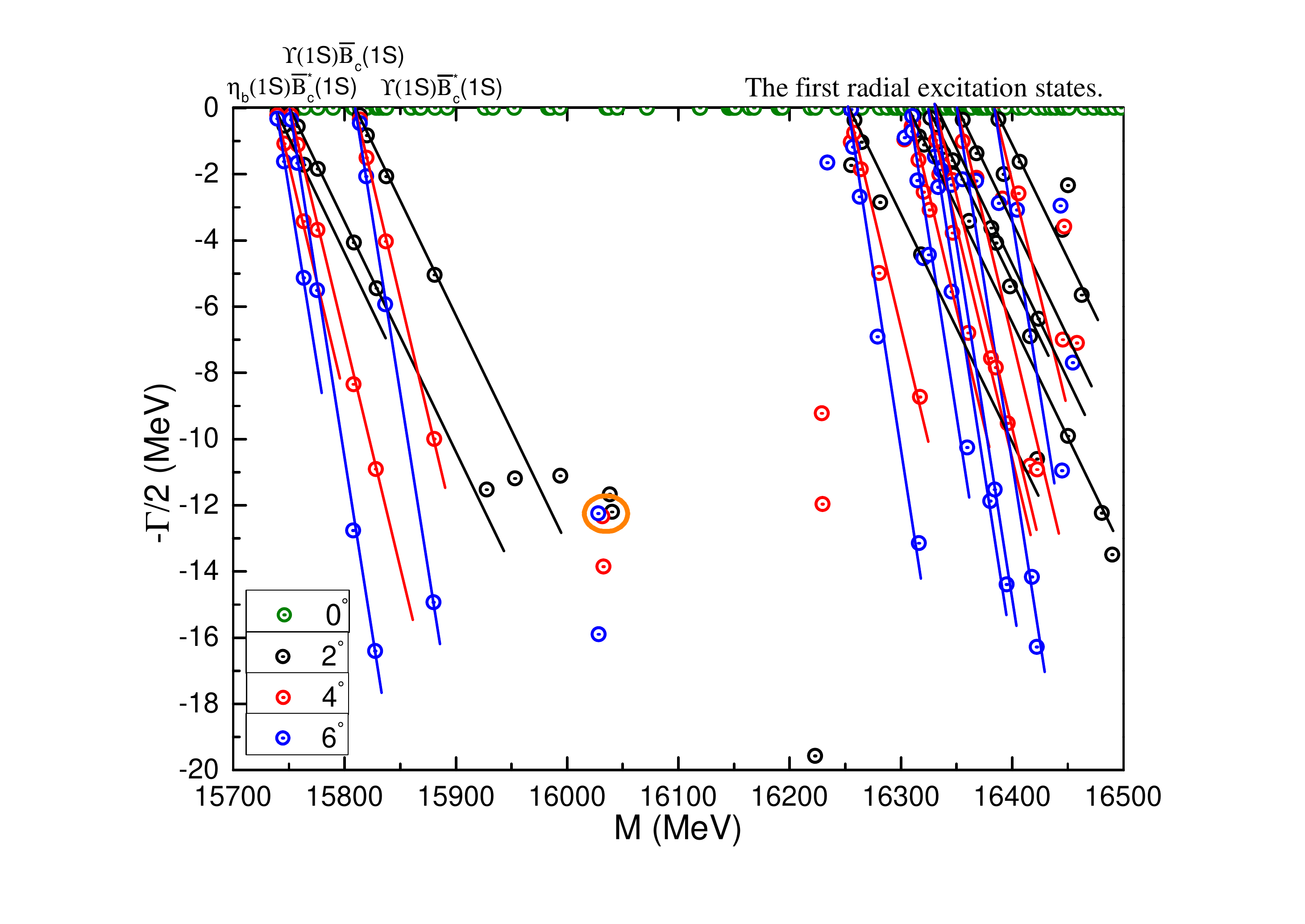} \\
\includegraphics[clip, trim={3.0cm 1.0cm 3.0cm 1.0cm}, width=0.45\textwidth]{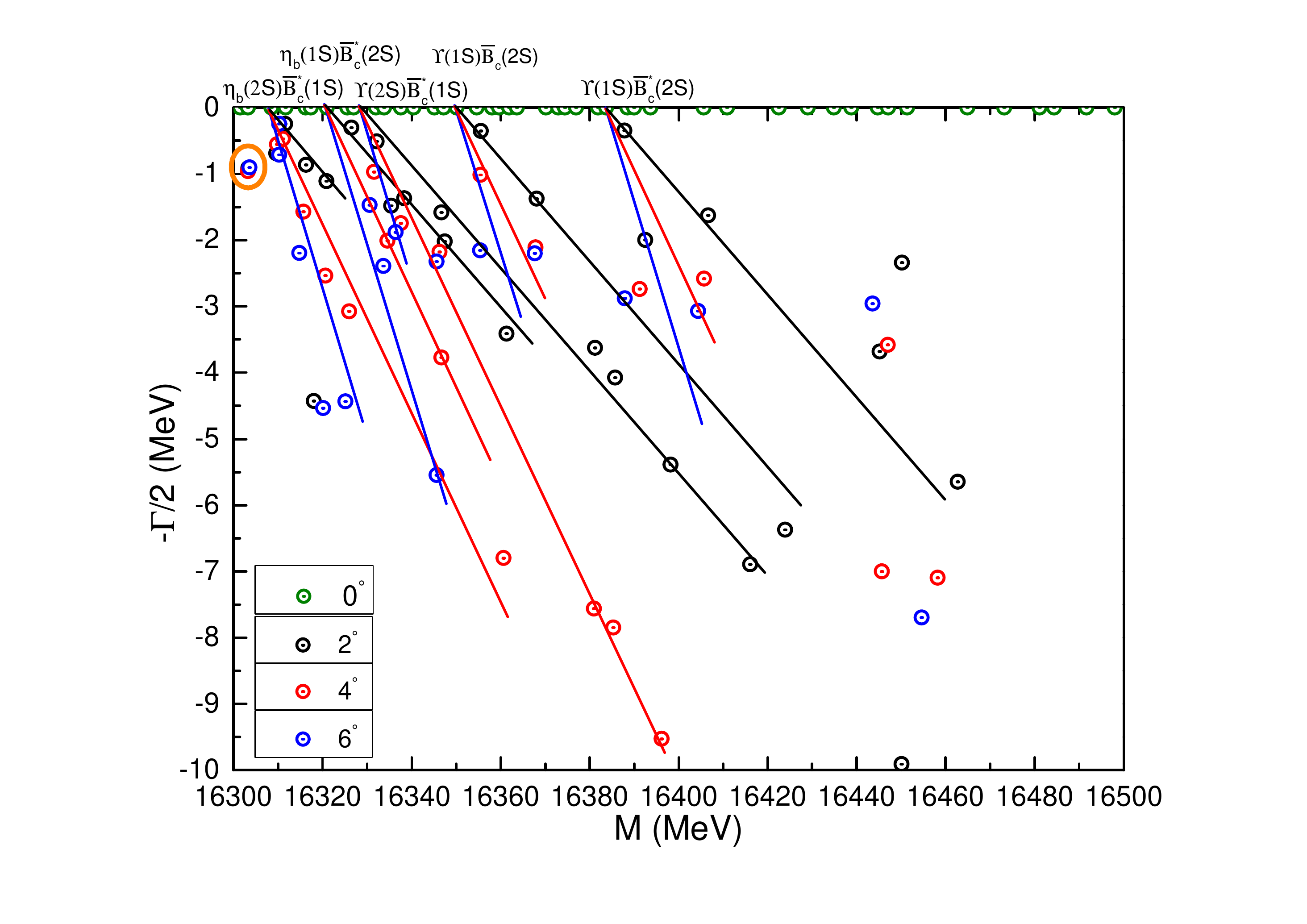}
\caption{\label{PP11} {\it Top panel:} Complex energy spectrum of the $\bar{b}b\bar{b}c$ system with $J^{P}=1^{+}$ from the complete coupled-channel calculation with CSM. The parameter $\theta$ varies from $0^\circ$ to $6^\circ$. {\it Bottom panel:} Enlarged top panel, with real values of energy ranging from $16.30\,\text{GeV}$ to $16.50\,\text{GeV}$.} 
\end{figure}

In order to find possible resonance states, as those obtained in the previous tetraquark systems, a CSM is employed in the fully coupled-channels computation, and the results are shown in Fig.~\ref{PP11}. Within a energy gap of 15.7$-$16.5 GeV, the three ground states, $\eta_b \bar{B}^*_c$, $\Upsilon\bar{B}_c$ and $\Upsilon\bar{B}^*_c$, along with their first radial excitations, are well shown in the top panel. Most of the complex-energy dots are well align along the corresponding cut lines, reflecting the nature of scattering states. However, one can notice that there is a stable resonance pole circled by an orange line, the calculated mass and width are 16031 MeV and 24.61 MeV, respectively. It can be identified as a $\Upsilon(1S)\bar{B}^*_c(1S)$ resonance.

In the bottom panel, we further enlarge the energy region between 16.3 GeV and 16.5 GeV. One can see that the five radial excited states are clearly shown as scattering states. However, one narrow pole, which is independent of the rotated angle $\theta$, is obtained. Thus, it can be identified as a $\Upsilon(2S)\bar{B}_c(1S)$ resonance state with a mass and width 16303 MeV and 1.80 MeV, respectively.

\begin{table}[!t]
\caption{\label{GresultCC11} Calculation of the $J^{P}=1^{+}$ state of the $\bar{b}b\bar{b}c$ system in the real case ($\theta=0^\circ$). All results are in units of MeV.}
\begin{ruledtabular}
\begin{tabular}{lcccc}
~~Structure  & Channel & $\xi_1^{\alpha}$; $\varphi_\beta^{\phantom{\dag}}$ & $M$ & Mixed~~ \\
 & & $[\alpha;~\beta]$ &  &  \\[2ex]
~~$\eta_b \bar{B}^*_c (15750)$         & 1  & [1; 1] & $15750$ &~~ \\
~~$\Upsilon \bar{B}_c (15738)$  & 2 & [2; 1]   & $15738$ & \\
~~$\Upsilon \bar{B}^*_c (15812)$        & 3 & [3; 1]   & $15812$ &$15738$~~ \\[2ex]
~~$\eta^8_b \bar{B}^{*8}_c$         & 4  & [1; 2] & $16067$ &~~ \\
~~$\Upsilon^8 \bar{B}^8_c$  & 5  & [2; 2]   & $16070$ & \\
~~$\Upsilon^8 \bar{B}^{*8}_c$         & 6  & [3; 2]   & $16055$ &$15993$~~ \\[2ex]
~~$[cb]^{\bf 6_{\rm c}}[\bar{b}\bar{b}]^{\bf \bar{6}_{\rm c}}$   & 7  & [4; 4]  & $16032$ & \\
~~$[cb]^{\bf 3_{\rm c}}[\bar{b}\bar{b}]^{\bf \bar{3}_{\rm c}}$   & 8  & [5; 3]  & $16049$ & \\
~~$[cb]^{\bf 3_{\rm c}}[\bar{b}\bar{b}]^{\bf \bar{3}_{\rm c}}$   & 9  & [6; 3]  & $16068$ & $16017$~~ \\[2ex]
~~K$_1$   & 10  & [7; 5]  & $16013$ & \\
                 & 11  & [8; 5]  & $16061$ & \\
                 & 12  & [9; 5]  & $16062$ & \\
                 & 13  & [7; 6]  & $15990$ & \\
                 & 14  & [8; 6]  & $15966$ & \\
                 & 15  & [9; 6]  & $15946$ & $15922$~~ \\[2ex]
~~K$_2$   & 16  & [10; 7]  & $16007$ & \\
                 & 17  & [11; 7]  & $16039$ & \\
                 & 18  & [12; 7]  & $15988$ & \\
                 & 19  & [10; 8]  & $16067$ & \\
                 & 20  & [11; 8]  & $15993$ & \\
                 & 21  & [12; 8]  & $16054$ & $15944$~~ \\[2ex]
~~K$_3$   & 22  & [13; 9]  & $16066$ & \\
                 & 23   & [14; 9] & $16074$ & \\
                 & 24  & [15; 9]  & $16442$ & \\
                 & 25  & [13; 10]  & $16074$ & \\
                 & 26   & [14; 10] & $16073$ & \\
                 & 27  & [15; 10]  & $16051$ & $15985$~~ \\[2ex]
~~K$_4$   & 28  & [16; 12]  & $16053$ & \\
                 & 29  & [17; 12]  & $16064$ & \\
                 & 30  & [18; 11]  & $16035$ & $16018$~~ \\[2ex]
\multicolumn{4}{l}{Lowest mass of the fully-coupled result:}    & $15738$~~  \\
\end{tabular}
\end{ruledtabular}
\end{table}

{\bf The $\bm{J^{P}=2^{+}}$ channel:} Twelve channels are under our investigation in the highest spin state of $\bar{b}b\bar{b}c$ tetraquark system, and the results are listed in Table~\ref{GresultCC12}. First of all, in each single channels calculations, the lowest mass is 15812 MeV for the color-singlet channel of $\Upsilon\bar{B}^*_c$ state, and this is also the theoretical threshold value. The other excited channels masses are all above 16.0 GeV. The fact of unbound state of $\bar{b}b\bar{b}c$ tetraquark system holds for different kinds of coupled-channels calculations.

\begin{figure}[ht]
\includegraphics[clip, trim={3.0cm 1.0cm 2.0cm 1.0cm}, width=0.45\textwidth]{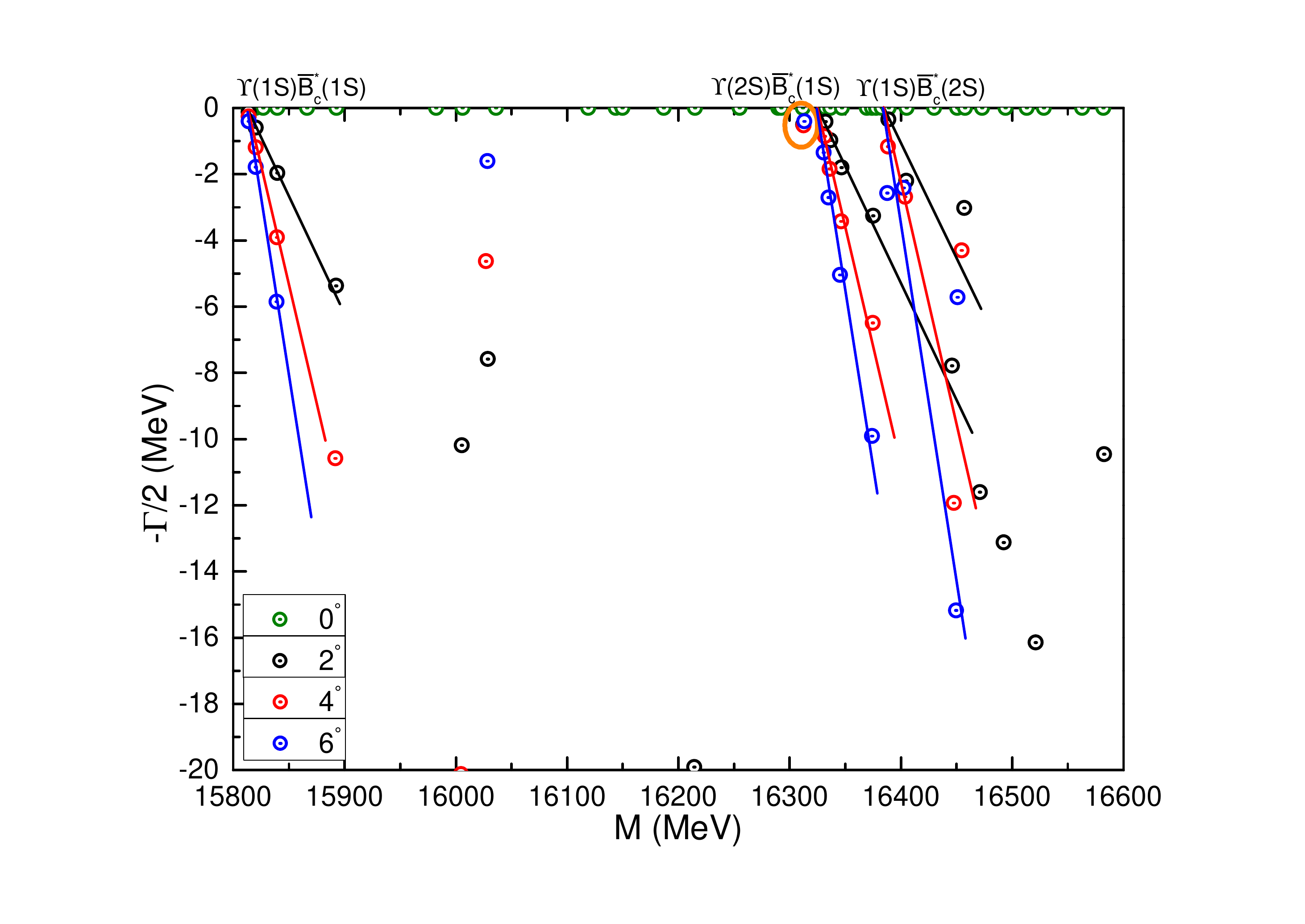} \\
\caption{Complex energy spectrum of the $\bar{b}b\bar{b}c$ system with $J^{P}=2^{+}$ from the complete coupled-channel calculation with CSM. The parameter $\theta$ varies from $0^\circ$ to $6^\circ$.} \label{PP12}
\end{figure}

In additional, Fig.~\ref{PP12} presents the distributions of complex energies of $\bar{b}b\bar{b}c$ tetraquark system in a complete coupled-channels study. In the mass region of 15.8$-$16.6 GeV, three continuum states, $\Upsilon(1S)\bar{B}^*_c(1S)$, $\Upsilon(2S)\bar{B}^*_c(1S)$ and $\Upsilon(1S)\bar{B}^*_c(2S)$, are well shown. However, one narrow resonance pole is obtained and it is quite close to (but below) the threshold of $\Upsilon(2S)\bar{B}^*_c(1S)$. Therefore, it can be identified as a $\Upsilon(1S)\bar{B}^*_c(1S)(16312)$ resonance with width is 1.12 MeV.

\begin{table}[!t]
\caption{\label{GresultCC12} Calculation of the $J^{P}=2^{+}$ state of the $\bar{b}b\bar{b}c$ system in the real case ($\theta=0^\circ$). The first and second columns show all possible structures and channels. They are obtained by the necessary bases in spin and color degrees of freedom summarized in the third column. The value in the parenthesis after the first meson-meson structure gives the theoretical result of the noninteracting meson-meson threshold. The lowest energies of all channels without considering any channel-channel coupling are listed in the fourth column. The last column shows the lowest energies in a calculation taking into account the couplings among the channels belonging to each structure (configuration). The last row gives the lowest energy from the complete coupled-channel calculation. All results are in units of MeV.}
\begin{ruledtabular}
\begin{tabular}{lcccc}
~~Structure  & Channel & $\xi_2$; $\varphi_\beta^{\phantom{\dag}}$ & $M$ & Mixed~~ \\
 & & $[\beta]$ &  &  \\[2ex]
~~$\Upsilon \bar{B}^*_c (15812)$       & 1  & 1  & $15812$ &$15812$~~  \\[2ex]
~~$\Upsilon^8 \bar{B}^{*8}_c$       & 2  & 2  & $16030$ &$16030$~~  \\[2ex]
~~$[cb]^{\bf 3_{\rm c}}[\bar{b}\bar{b}]^{\bf \bar{3}_{\rm c}}$   & 3  & 3   & $16087$  &$16087$~~\\[2ex]
~~K$_1$   & 4 & 5  & $16043$ & \\
              & 5  & 6  & $16017$ & $16008$~~ \\[2ex]
~~K$_2$   & 6 & 7  & $16055$ & \\
              & 7 & 8  & $16021$ & $15980$~~ \\[2ex]
~~K$_3$   & 8 & 9  & $16451$ & ~~ \\
              & 9  & 10  & $16090$ & $16065$~~ \\[2ex]
~~K$_4$   & 10 & 12  & $16087$ & $16087$~~ \\[2ex]
\multicolumn{4}{l}{Lowest mass of the fully-coupled result:}  & $15812$~~ \\
\end{tabular}
\end{ruledtabular}
\end{table}

\subsection{Charm-bottom system $\bar{c}b\bar{c}b$}

For the charm-bottom system $\bar{c}b\bar{c}b$, some resonance structures are found with quantum numbers $J^{P(C)}=0^{+(+)}$, $1^{+(-)}$ and $2^{+(+)}$. Table~\ref{Rsum-cbcb} summarizes the results, and the details are discussed in the following.

\begin{table}[!t]
\caption{\label{Rsum-cbcb} Predicted resonances of the $\bar{c}b\bar{c}b$ tetraquark system from the complete coupled-channel calculation with CSM. Their masses and widths are summarized in the third and fourth column, respectively, in units of MeV.}
\begin{ruledtabular}
\begin{tabular}{cccc}
~~$J^{P(C)}$ & Resonance & Mass &  Width~~ \\
~~$0^{+(+)}$ 
              & $B^*_c (1S) B^*_c (1S)$  & 12768   & 1.73~~ \\[2ex]
~~$1^{+(-)}$ 
              & $B_c (1S) B^*_c (2S)$  & 13214   & 0.66~~ \\[2ex]
~~$2^{+(+)}$ 
              & $B^*_c (1S) B^*_c (1S)$  & 12965   & 17.42~~ \\ 
              & $B^*_c (1S) B^*_c (1S)$  & 13236   & 4.40~~ \\ 
              & $B^*_c (1S) B^*_c (1S)$  & 13248   & 3.41~~ \\ 
\end{tabular}
\end{ruledtabular}
\end{table}

{\bf The $\bm{J^{P(C)}=0^{+(+)}}$ channel:} There are 18 channels, as listed in Table~\ref{GresultCC13}. The calculated masses of the two di-meson structures in color-singlet channels are 12550 MeV and 12698 MeV, respectively. These are just the theoretical threshold values of di-$B_c$ and di-$B^*_c$, and their relevant hidden-color channels masses are both around 12.91 GeV. As for the other exotic structures, diquark-antidiquark and K-types configurations, the obtained single-channels masses are generally located in the energy range 12.71$-$12.89 GeV. When the coupled-channels calculations are considered, no bound state is found.

\begin{figure}[ht]
\includegraphics[clip, trim={3.0cm 1.0cm 2.0cm 1.0cm}, width=0.45\textwidth]{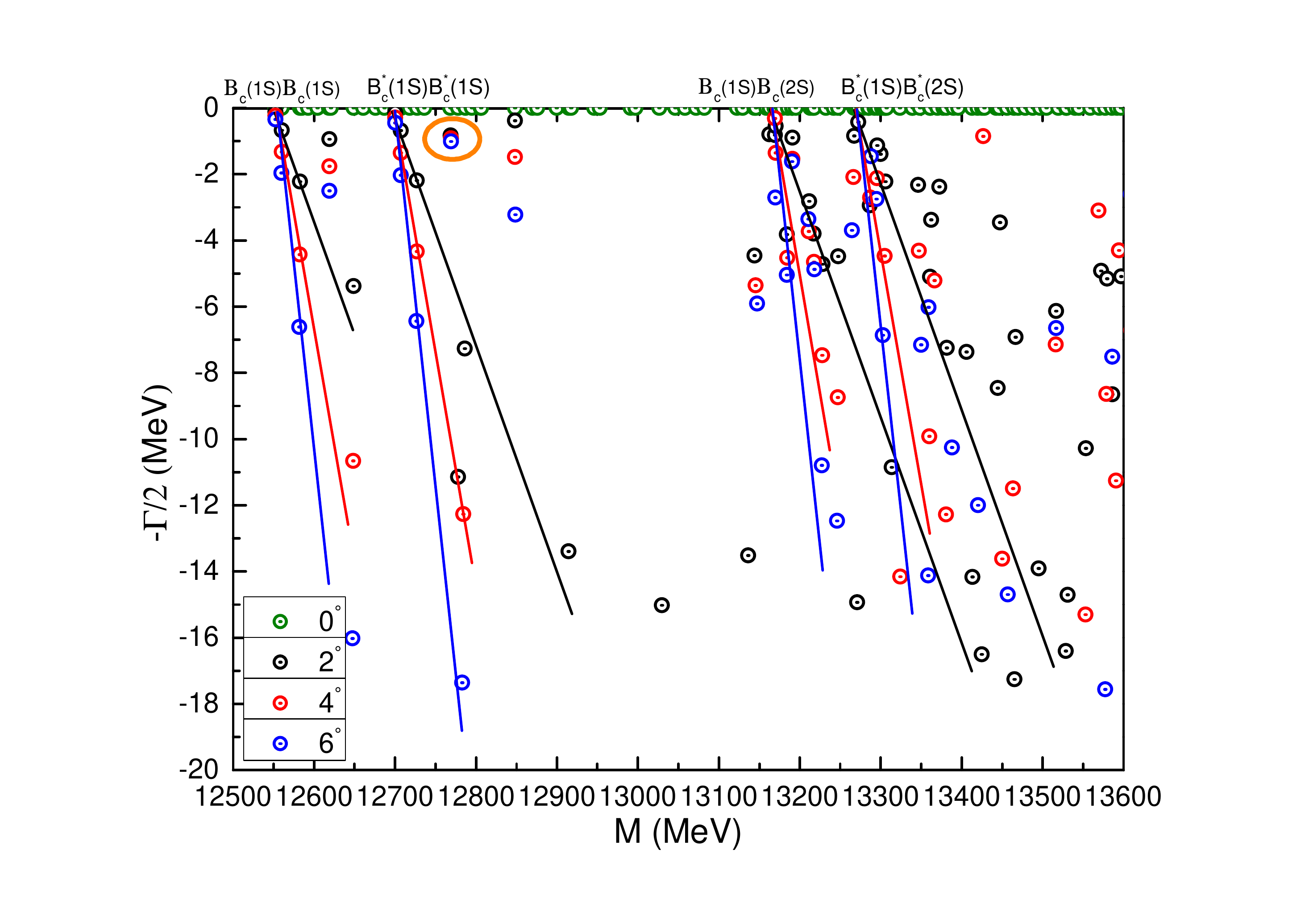} \\
\caption{Complex energy spectrum of the $\bar{c}b\bar{c}b$ system with $J^{P(C)}=0^{+(+)}$ from the complete coupled-channel calculation with CSM. The parameter $\theta$ varies from $0^\circ$ to $6^\circ$.} \label{PP13}
\end{figure}

Figure~\ref{PP13} shows our results when a complete coupled-channels calculation using the CSM formalism is performed. In an energy region of 12.5$-$13.6 GeV, the scattering states of $B_c(1S)B_c(1S)$, $B^*_c(1S)B^*_c(1S)$, $B_c(1S)B_c(2S)$ and $B^*_c(1S)B^*_c(2S)$ are well presented. One stable pole is also found and circled. It is above the $B^*_c(1S)B^*_c(1S)$ cut lines and thus it can be identified as a di-$B^*_c(1S)$ resonance with mass and width 12768 MeV and 1.73 MeV, respectively.

\begin{table}[!t]
\caption{\label{GresultCC13} Calculation of the $J^{P(C)}=0^{+(+)}$ state of the $\bar{c}b\bar{c}b$ system in the real case ($\theta=0^\circ$). The first and second columns show all possible structures and channels. They are obtained by the necessary bases in spin and color degrees of freedom summarized in the third column. The values in the parentheses after the first two meson-meson structures give the theoretical results of the noninteracting meson-meson thresholds. The lowest energies of all channels without considering any channel-channel coupling are listed in the fourth column. The last column shows the lowest energies in a calculation taking into account the couplings among the channels belonging to each structure (configuration). The last row gives the lowest energy from the complete coupled-channel calculation. All results are in units of MeV.}
\begin{ruledtabular}
\begin{tabular}{lcccc}
~~Structure  & Channel & $\xi_0^{\alpha}$; $\varphi_\beta^{\phantom{\dag}}$ & $M$ & Mixed~~ \\
 & & $[\alpha;~\beta]$ &  &  \\[2ex]
~~$B_c B_c (12550)$      & 1  & [1; 1]  & $12550$ &~~ \\
~~$B^*_c B^*_c (12698)$   & 2  & [2; 1]  & $12698$ &$12550$~~  \\[2ex]
~~$B^8_c B^8_c$           & 3  & [1; 2]  & $12949$ &~~ \\
~~$B^{*8}_c B^{*8}_c$       & 4  & [2; 2]  & $12910$ &$12853$~~  \\[2ex]
~~$[bb]^{\bf 6_{\rm c}}[\bar{c}\bar{c}]^{\bf \bar{6}_{\rm c}}$   & 5  & [3; 4] & $12888$ &~~ \\
~~$[bb]^{\bf 3_{\rm c}}[\bar{c}\bar{c}]^{\bf \bar{3}_{\rm c}}$   & 6  & [4; 3]  & $12867$  &$12836$~~\\[2ex]
~~K$_1$   & 7 & [5; 5]   & $12833$ & \\
                 & 8  & [5; 6]  & $12835$ & \\
                 & 9  & [6; 5]  & $12876$ &  \\
                 & 10 & [6; 6]  & $12742$ & $12695$~~ \\[2ex]
~~K$_2$   & 11  & [7; 7]  & $12806$ & \\
                 & 12  & [7; 8]  & $12850$ & \\
                 & 13  & [8; 7]  & $12705$ &  \\
                 & 14  & [8; 8]  & $12889$ & $12689$~~ \\[2ex]
~~K$_3$   & 15  & [9; 10]  & $12868$ & \\
                 & 16  & [10; 9]  & $12882$ &$12832$~~  \\[2ex]
~~K$_4$   & 17  & [11; 12]  & $12869$ & \\
                 & 18   & [12; 11]  & $12895$ &$12839$~~  \\[2ex]
\multicolumn{4}{l}{Lowest mass of the fully-coupled result:}  & $12550$~~ \\
\end{tabular}
\end{ruledtabular}
\end{table}

{\bf The $\bm{J^{P(C)}=1^{+(-)}}$ channel:} we have 23 channels as listed in Table~\ref{GresultCC14}. They involve two di-meson configurations, $B_c B^*_c$ and $B^*_c B^*_c$ in the singlet- as well as in the hidden-color wave functions, one diquark-antidiquark channel $[bb]^{\bf 3_{\rm c}}[\bar{c}\bar{c}]^{\bf \bar{3}_{\rm c}}$, and 18 K-type structures. Firstly, the meson-meson configurations in color-singlet channels are unbound and their calculated masses are 12624 MeV and 12698 MeV, respectively. The two hidden-color channels are characterize by a mass of $\sim$12.9 GeV, this value is also compatible for the diquark-antidiquark case. The K-type tetraquarks are at $\sim$12.87 GeV. Additionally, we do not find any bound state in coupled-channels calculations, which imply couplings in each certain tetraquarj configuration and a complete one.

\begin{figure}[!t]
\includegraphics[clip, trim={3.0cm 1.0cm 2.0cm 0.0cm}, width=0.45\textwidth]{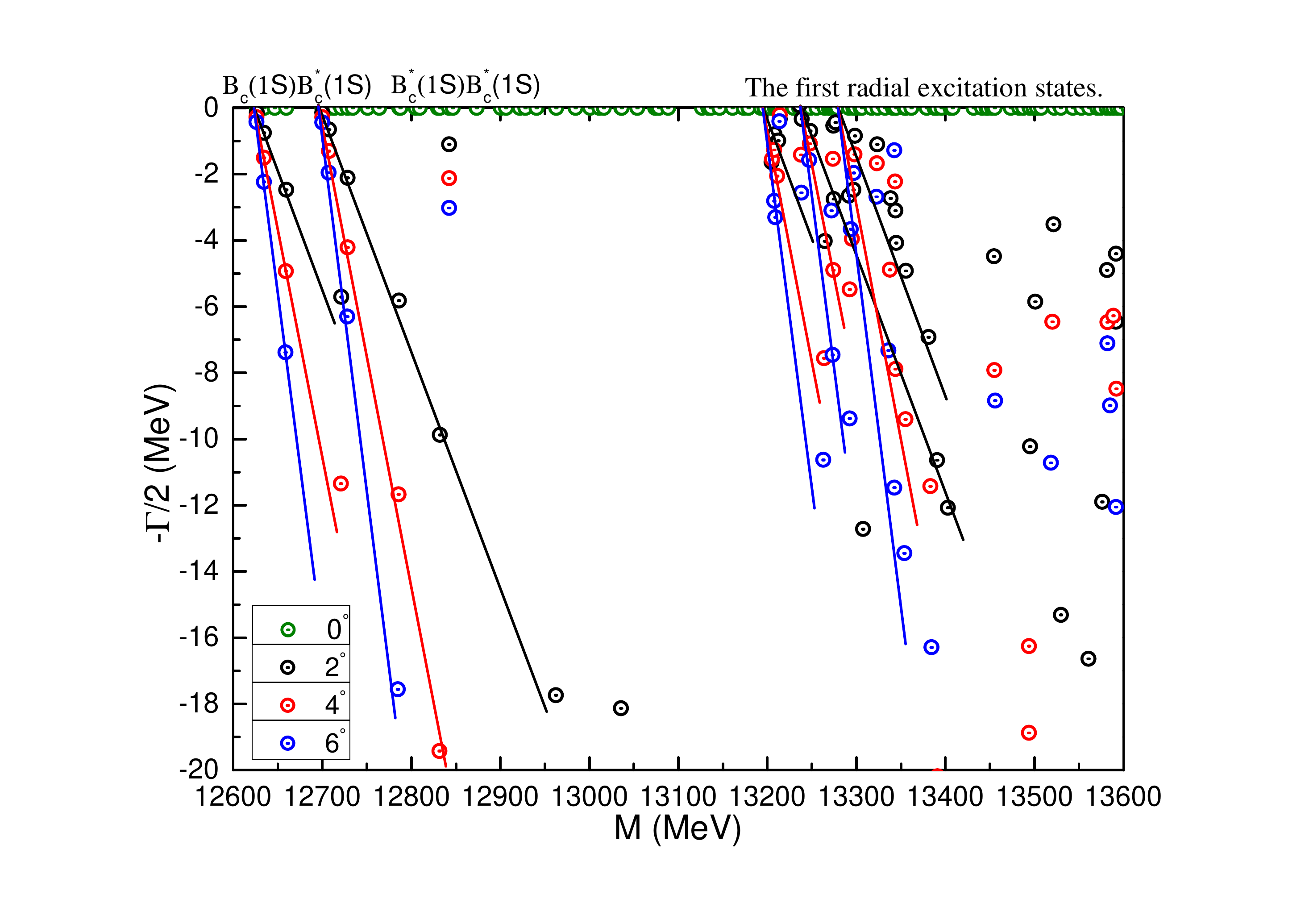} \\
\includegraphics[clip, trim={3.0cm 1.0cm 2.0cm 0.0cm}, width=0.45\textwidth]{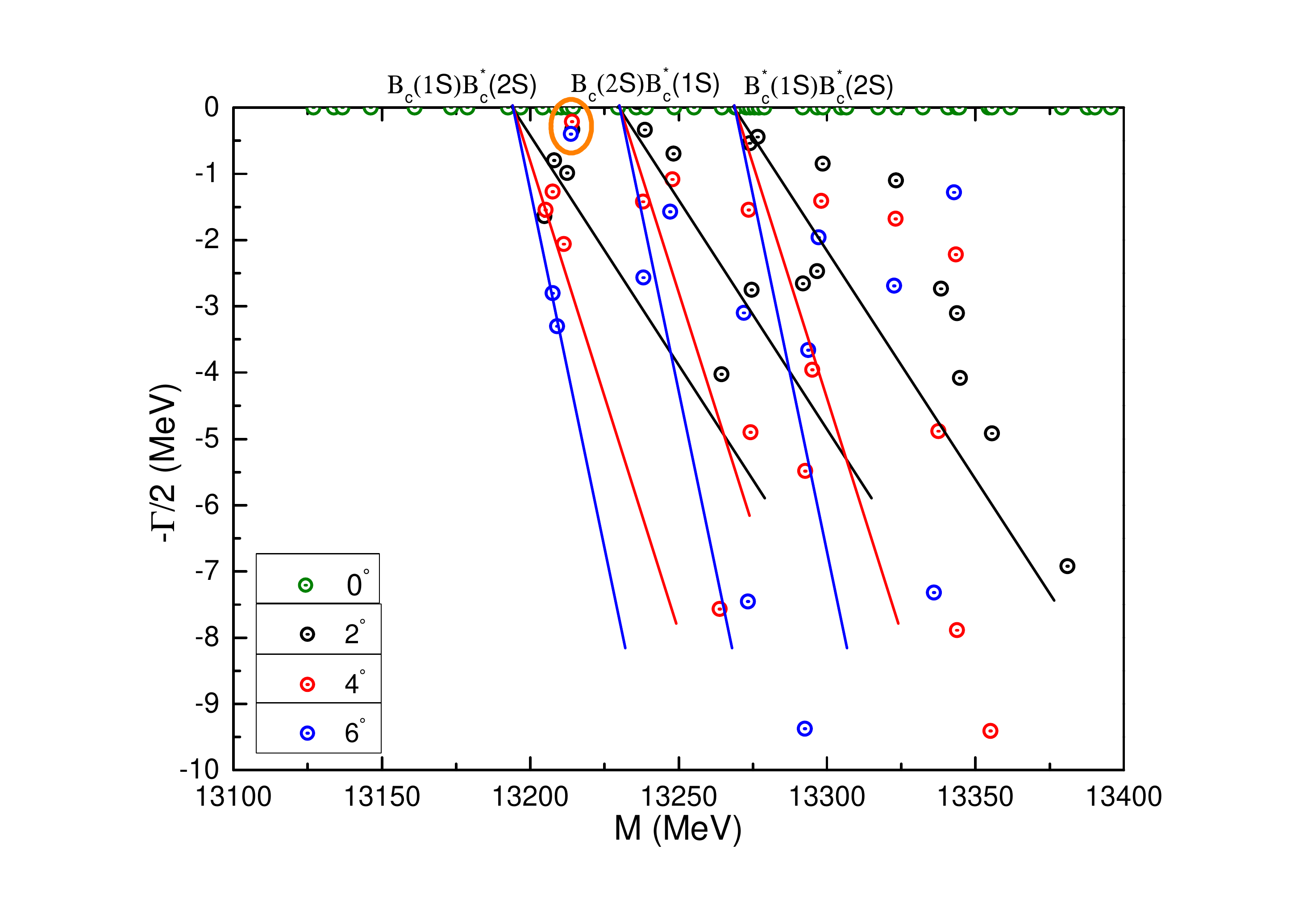}
\caption{\label{PP14} {\it Top panel:} Complex energy spectrum of the $\bar{c}b\bar{c}b$ system with $J^{P(C)}=1^{+(-)}$ from the complete coupled-channel calculation with CSM. The parameter $\theta$ varies from $0^\circ$ to $6^\circ$. {\it Bottom panel:} Enlarged top panel, with real values of energy ranging from $13.10\,\text{GeV}$ to $13.40\,\text{GeV}$.} 
\end{figure}

By employing the CSM in a complete coupled-channel investigation, the distributions of eigenenergies are shown in Fig.~\ref{PP14}. In the top panel, with an energy range from 12.6 GeV to 13.6 GeV, neither bound nor resonance states are found; the scattering states of $B_c(1S) B^*_c(1S)$, $B^*_c(1S) B^*_c(1S)$ along with their first radial excitation states are however clearly identified.

Nevertheless, when we perform an energy zoom in the bottom panel, apart from the three continuum states of $B_c(1S) B^*_c(2S)$, $B_c(2S) B^*_c(1S)$ and $B^*_c(1S) B^*_c(2S)$, a stable pole with mass and width 13214 MeV and 0.66 MeV, is found. It is located above the $B_c(1S) B^*_c(2S)$ threshold line too. 

\begin{table}[!t]
\caption{\label{GresultCC14} Calculation of the $J^{P(C)}=1^{+(-)}$ state of the $\bar{c}b\bar{c}b$ system in the real case ($\theta=0^\circ$). The first and second columns show all possible structures and channels. They are obtained by the necessary bases in spin and color degrees of freedom summarized in the third column. The values in the parentheses after the first two meson-meson structures give the theoretical results of the noninteracting meson-meson thresholds. The lowest energies of all channels without considering any channel-channel coupling are listed in the fourth column. The last column shows the lowest energies in a calculation taking into account the couplings among the channels belonging to each structure (configuration). The last row gives the lowest energy from the complete coupled-channel calculation. All results are in units of MeV.}
\begin{ruledtabular}
\begin{tabular}{lcccc}
~~Structure  & Channel & $\xi_1^{\alpha}$; $\varphi_\beta^{\phantom{\dag}}$ & $M$ & Mixed~~ \\
 & & $[\alpha;~\beta]$ &  &  \\[2ex]
~~$B_c B^*_c (12624)$         & 1  & [1; 1] & $12624$ &~~ \\
~~$B^*_c B^*_c (12698)$        & 2 & [3; 1]   & $12698$ &$12624$~~ \\[2ex]
~~$B^8_c B^{*8}_c$         & 3  & [1; 2] & $12852$ &~~ \\
~~$B^{*8}_c B^{*8}_c$         & 4  & [3; 2]   & $12910$ &$12851$~~ \\[2ex]
~~$[bb]^{\bf 3_{\rm c}}[\bar{c}\bar{c}]^{\bf \bar{3}_{\rm c}}$   & 5  & [6; 3]  & $12878$ & $12878$~~ \\[2ex]
~~K$_1$   & 6  & [7; 5]  & $12886$ & \\
                 & 7  & [8; 5]  & $12878$ & \\
                 & 8  & [9; 5]  & $12863$ & \\
                 & 9  & [7; 6]  & $12886$ & \\
                 & 10  & [8; 6]  & $12891$ & \\
                 & 11  & [9; 6]  & $12855$ & $12788$~~ \\[2ex]
~~K$_2$   & 12  & [10; 7]  & $12868$ & \\
                 & 13  & [11; 7]  & $12874$ & \\
                 & 14  & [12; 7]  & $12826$ & \\
                 & 15  & [10; 8]  & $12886$ & \\
                 & 16  & [11; 8]  & $12889$ & \\
                 & 17  & [12; 8]  & $12875$ & $12759$~~ \\[2ex]
~~K$_3$   & 18  & [15; 9]  & $13399$ & \\
                 & 19   & [13; 10] & $12888$ & \\
                 & 20  & [14; 10]  & $12887$ & $12878$~~ \\[2ex]
~~K$_4$   & 21  & [16; 12]  & $12882$ & \\
                 & 22  & [17; 12]  & $12881$ & \\
                 & 23  & [18; 11]  & $13336$ & $12880$~~ \\[2ex]
\multicolumn{4}{l}{Lowest mass of the fully-coupled result:}    & $12624$~~  \\
\end{tabular}
\end{ruledtabular}
\end{table}

{\bf The $\bm{J^{P(C)}=2^{+(+)}}$ channel:} For the highest spin state of $\bar{c}b\bar{c}b$ tetraquark system, there are 9 channels under consideration, and the results are listed in Table~\ref{GresultCC15}. Firstly, in each single-channel calculation, masses of the color-singlet, hidden-color and diquark-antidiquark structures are 12698 GeV, 12916 GeV and 12899 GeV, respectively. The other 6 K-types masses locate in the range 12.82$-$12.90 GeV. Obviously, no bound state is found and this fact does not change in the different coupled-channel calculations. The lowest mass, 12698 MeV, is just the theoretical threshold value of di-$B^*_c$. 

\begin{figure}[ht]
\includegraphics[clip, trim={3.0cm 1.0cm 2.0cm 0.0cm}, width=0.45\textwidth]{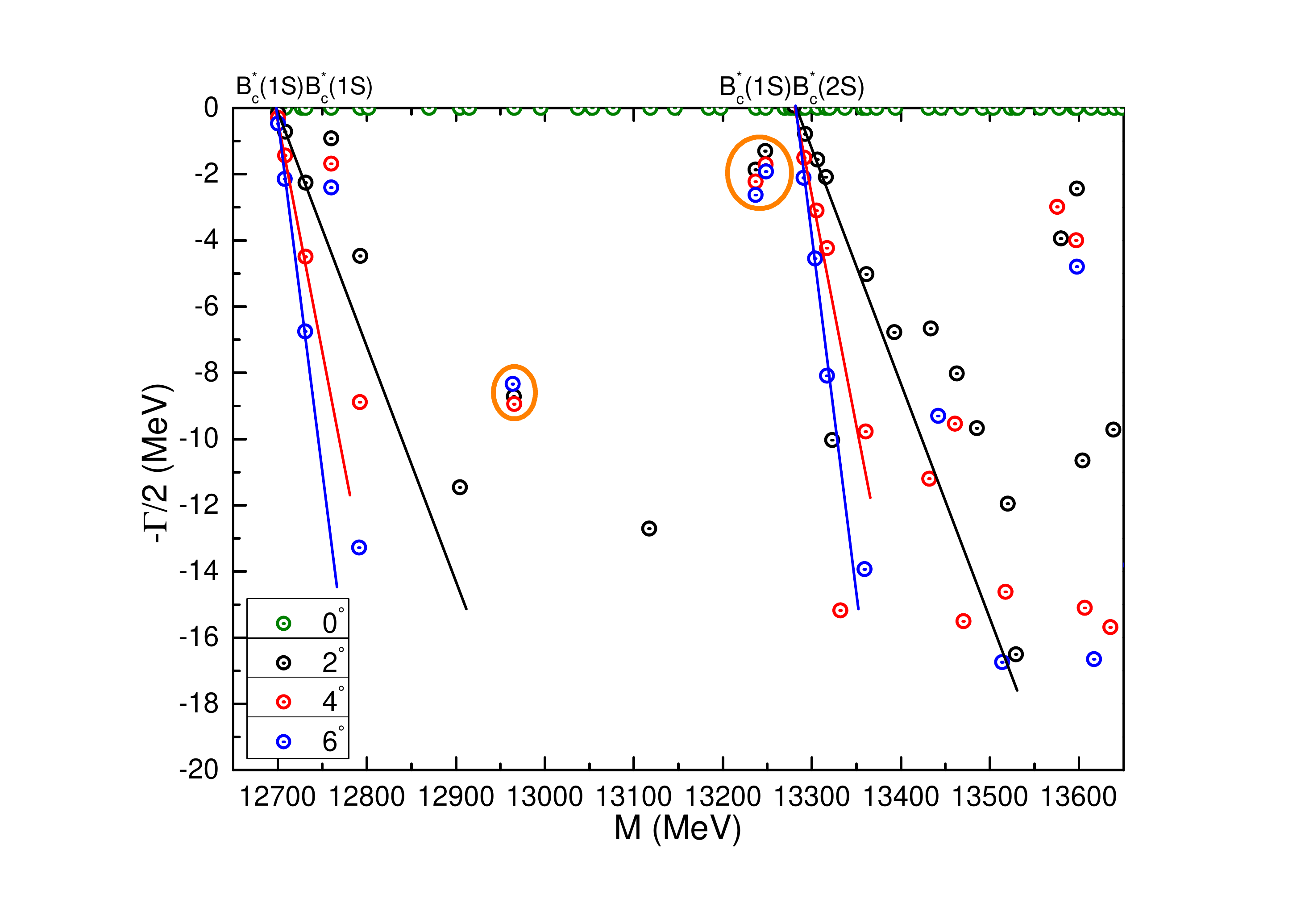} \\
\caption{Complex energy spectrum of the $\bar{c}b\bar{c}b$ system with $J^{P(C)}=2^{+(+)}$ from the complete coupled-channel calculation with CSM. The parameter $\theta$ varies from $0^\circ$ to $6^\circ$.} \label{PP15}
\end{figure}

In a further step, when a complete coupled-channels calculation is performed in the complex range, resonances are obtained and shown in Fig.~\ref{PP15}. In a mass region from 12.65$-$13.65 GeV, the scattering states of $B^*_c(1S)B^*_c(1S)$ and $B^*_c(1S)B^*_c(2S)$ are well shown. With the rotated angle varying from $0^\circ$ to $6^\circ$, there are three stable resonance poles in the complex plane and they are marked with circles. The lowest one, whose mass and width are 12965 MeV and 17.42 MeV, can be identified as a  $B^*_c(1S)B^*_c(1S)$ resonance. For the other two poles, which are quite close to each other, the calculated masses and widths are (13236 MeV, 4.40 MeV) and (13248 MeV, 3.41 MeV), respectively. Their dominant channel is again $B^*_c(1S)B^*_c(1S)$.

\begin{table}[!t]
\caption{\label{GresultCC15} Calculation of the $J^{P(C)}=2^{+(+)}$ state of the $\bar{c}b\bar{c}b$ system in the real case ($\theta=0^\circ$).
The first and second columns show all possible structures and channels. They are obtained by the necessary bases in spin and color degrees of freedom summarized in the third column. The value in the parenthesis after the first meson-meson structure gives the theoretical result of the noninteracting meson-meson threshold. The lowest energies of all channels without considering any channel-channel coupling are listed in the fourth column. The last column shows the lowest energies in a calculation taking into account the couplings among the channels belonging to each structure (configuration). The last row gives the lowest energy from the complete coupled-channel calculation. All results are in units of MeV.}
\begin{ruledtabular}
\begin{tabular}{lcccc}
~~Structure  & Channel & $\xi_2$; $\varphi_\beta^{\phantom{\dag}}$ & $M$ & Mixed~~ \\
 & & $[\beta]$ &  &  \\[2ex]
~~$B^*_c B^*_c (12698)$       & 1  & 1  & $12698$ &$12698$~~  \\[2ex]
~~$B^{*8}_c B^{*8}_c$       & 2  & 2  & $12916$ &$12916$~~  \\[2ex]
~~$[bb]^{\bf 3_{\rm c}}[\bar{c}\bar{c}]^{\bf \bar{3}_{\rm c}}$   & 3  & 3   & $12899$  &$12899$~~\\[2ex]
~~K$_1$   & 4  & 5  & $12834$ & \\
                 & 5  & 6  & $12844$ & $12828$~~ \\[2ex]
~~K$_2$   & 6 & 7  & $12819$ & \\
                 & 7 & 8  & $12842$ & $12806$~~ \\[2ex]
~~K$_3$   & 8  & 10  & $12900$ & $12900$ ~~ \\[2ex]
~~K$_4$   & 9  & 12  & $12901$ & $12901$~~ \\[2ex]
\multicolumn{4}{l}{Lowest mass of the fully-coupled result:}  & $12698$~~ \\
\end{tabular}
\end{ruledtabular}
\end{table}

\subsection{Charm-bottom system $\bar{c}c\bar{b}b$}

For the charm-bottom system $\bar{c}c\bar{b}b$, some resonance structures are found with quantum numbers $J^{P(C)}=0^{+(+)}$, $1^{+(\pm)}$ and $2^{+(+)}$. Table~\ref{Rsum-ccbb} summarizes our results and the details of the computation can be found in the following. Note herein that the $K_1$-type structure (Fig.~\ref{QQqq}(c)) is equivalent to the $K_2$-type one (Fig.~\ref{QQqq}(d)) for the $\bar{c}c\bar{b}b$ system, and this identity also holds for $K_3$- (Fig.~\ref{QQqq}(e)) and $K_4$-type (Fig.~\ref{QQqq}(f)) configurations. Hence, only $K_1$- and $K_3$-types are employed in the following Tables without losing any generality.

\begin{table}[!t]
\caption{\label{Rsum-ccbb} Predicted resonances of the $\bar{c}c\bar{b}b$ tetraquark system from the complete coupled-channel calculation with CSM. Their masses and widths are summarized in the third and fourth column, respectively, in units of MeV.}
\begin{ruledtabular}
\begin{tabular}{cccc}
~~$J^{P(C)}$ & Resonance & Mass &  Width~~ \\
~~$0^{+(+)}$ 
              & $J/\psi (1S) \Upsilon (1S)$  & 12820   & 0.62~~ \\
              & $\psi (2S) \Upsilon (1S)$  & 13449   & 2.81~~ \\[2ex]
~~$1^{+(\pm)}$ 
              & $J/\psi (1S) \Upsilon (1S)$  & 12858   & 0.60~~ \\
              & $\eta_c (1S) \Upsilon (2S)$  & 13002   & 0.94~~ \\
              & $\psi (2S) \Upsilon (1S)$  & 13290   & 1.81~~ \\[2ex]
~~$2^{+(+)}$ 
              & $J/\psi (1S) \Upsilon (1S)$  & 12826   & 1.25~~ \\
              & $\psi (2S) \Upsilon (1S)$  & 13321   & 3.40~~ \\  
\end{tabular}
\end{ruledtabular}
\end{table}

{\bf The $\bm{J^{P(C)}=0^{+(+)}}$ channel:} There are 16 channels listed in Table~\ref{GresultCC16}. In particular, the $\eta_c \eta_b$ and $J/\psi \Upsilon$ di-meson channels have masses at 12369 MeV and 12565 MeV, and so they are both scattering states. Their corresponding hidden-color channels are almost degenerate with masses close to 12.85 GeV. The diquark-antidiquark and K-type configurations are all located in the energy range 12.77$-$12.89 GeV, except the $K_1$ channel with a mass at 12.60 GeV. No bound state is available in any of the three possible calculations: single-channel, configuration-type channels calculation and full-channel computation.

\begin{figure}[ht]
\includegraphics[clip, trim={3.0cm 1.0cm 2.0cm 0.0cm}, width=0.45\textwidth]{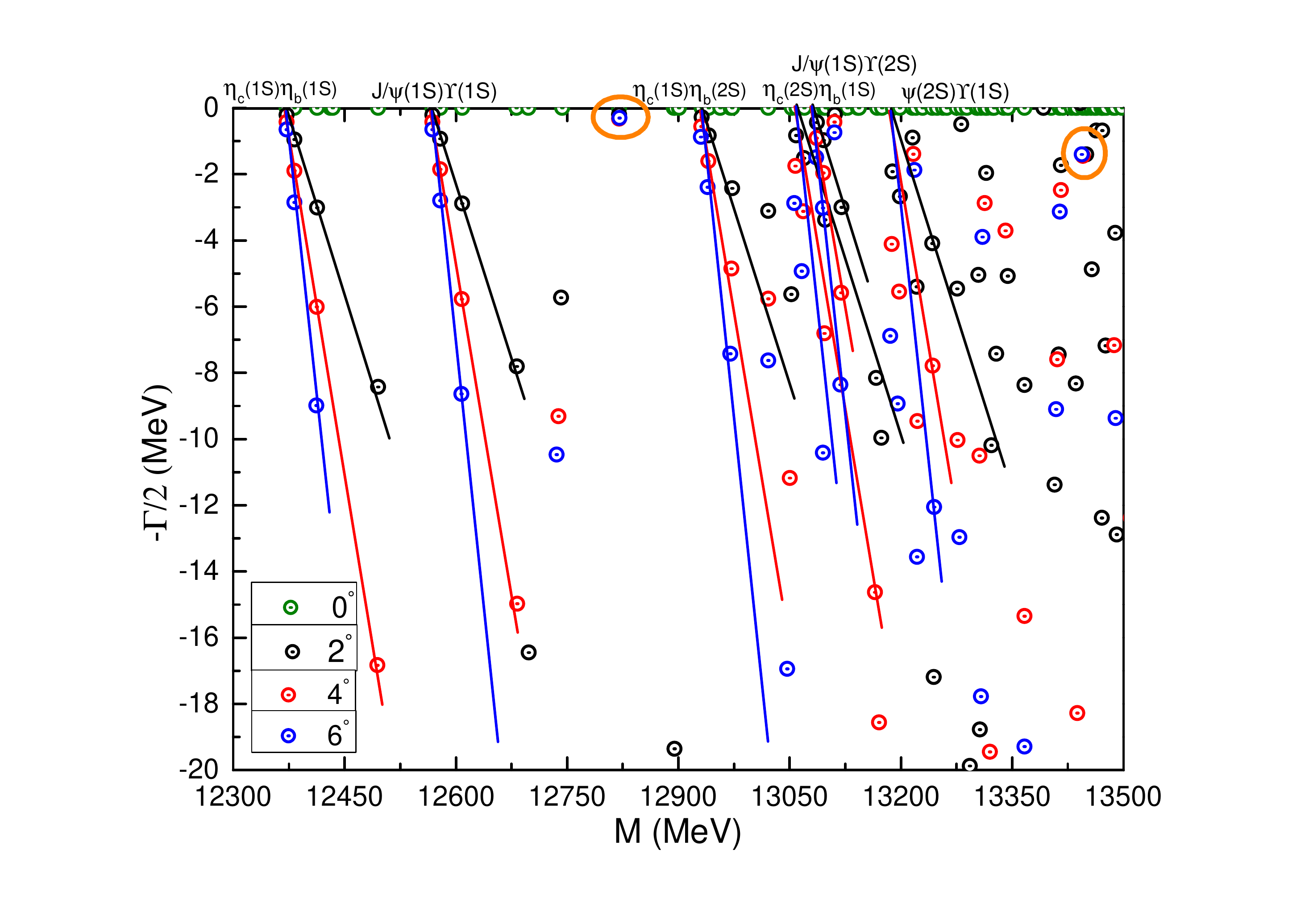} \\
\caption{Complex energy spectrum of the $\bar{c}c\bar{b}b$ system with $J^{P(C)}=0^{+(+)}$ from the complete coupled-channel calculation with CSM. The parameter $\theta$ varies from $0^\circ$ to $6^\circ$.} \label{PP16}
\end{figure}

Let us now focus on Fig.~\ref{PP16}, which presents the distribution of complex eigenenergies of $J^{P(C)}=0^{+(+)}$ $\bar{c}c\bar{b}b$ tetraquarks in a complete coupled-channels calculation by CSM, in a mass region 12.3$-$13.5 GeV. The scattering nature of $\eta_c(1S)\eta_b(1S)$, $J/\psi(1S)\Upsilon(1S)$, $\eta_c(1S)\eta_b(2S)$, $\eta_c(2S)\eta_b(1S)$, $J/\psi(1S)\Upsilon(2S)$ and $\psi(2S)\Upsilon(1S)$ is well established. However, there are two resonance poles which do not depend on the variation of the rotated angle $\theta$. One pole is identified as a $J/\psi(1S)\Upsilon(1S)(12820)$ narrow resonance with $\Gamma=0.62$ MeV, another is $\psi(2S)\Upsilon(1S)(13449)$ with $\Gamma=2.81$ MeV.

\begin{table}[!t]
\caption{\label{GresultCC16} Calculation of the $J^{P(C)}=0^{+(+)}$ state of the $\bar{c}c\bar{b}b$ system in the real case ($\theta=0^\circ$).
The first and second columns show all possible structures and channels. They are obtained by the necessary bases in spin and color degrees of freedom summarized in the third column. The values in the parentheses after the first two meson-meson structures give the theoretical results of the noninteracting meson-meson thresholds. The lowest energies of all channels without considering any channel-channel coupling are listed in the fourth column. The last column shows the lowest energies in a calculation taking into account the couplings among the channels belonging to each structure (configuration). The last row gives the lowest energy from the complete coupled-channel calculation. All results are in units of MeV.}
\begin{ruledtabular}
\begin{tabular}{lcccc}
~~Structure  & Channel & $\xi_0^{\alpha}$; $\varphi_\beta^{\phantom{\dag}}$ & $M$ & Mixed~~ \\
 & & $[\alpha;~\beta]$ &  &  \\[2ex]
~~$\eta_c \eta_b (12369)$      & 1  & [1; 1]  & $12369$ &~~ \\
~~$J/\psi \Upsilon (12565)$   & 2  & [2; 1]  & $12565$ &$12369$~~  \\[2ex]
~~$\eta^8_c \eta^8_b$           & 3  & [1; 2]  & $12858$ &~~ \\
~~$J/\psi^8 \Upsilon^8$       & 4  & [2; 2]  & $12838$ &$12809$~~  \\[2ex]
~~$[cb]^{\bf 3_{\rm c}}[\bar{c}\bar{b}]^{\bf \bar{3}_{\rm c}}$   & 5  & [3; 3] & $12861$ &~~ \\
~~$[cb]^{\bf 6_{\rm c}}[\bar{c}\bar{b}]^{\bf \bar{6}_{\rm c}}$   & 6  & [3; 4] & $12868$ &~~ \\
~~$[cb]^{\bf 3_{\rm c}}[\bar{c}\bar{b}]^{\bf \bar{3}_{\rm c}}$   & 7  & [4; 3] & $12892$ &~~ \\
~~$[cb]^{\bf 6_{\rm c}}[\bar{c}\bar{b}]^{\bf \bar{6}_{\rm c}}$   & 8  & [4; 4]  & $12784$  &$12707$~~\\[2ex]
~~K$_1$   & 9 & [5; 5]   & $12854$ & \\
                 & 10  & [5; 6]  & $12771$ & \\
                 & 11  & [6; 5]  & $12872$ &  \\
                 & 12 & [6; 6]  & $12601$ & $12599$~~ \\[2ex]
~~K$_3$   & 13  & [9; 9]  & $12786$ & \\
                  & 14  & [9; 10]  & $12896$ & \\
                  & 15  & [10; 9]  & $12876$ & \\
                 & 16  & [10; 10]  & $12872$ &$12717$~~  \\[2ex]
\multicolumn{4}{l}{Lowest mass of the fully-coupled result:}  & $12369$~~ \\
\end{tabular}
\end{ruledtabular}
\end{table}

{\bf The $\bm{J^{P(C)}=1^{+(\pm)}}$ channel:} Table~\ref{GresultCC17} shows the 22 channels available in this case. The three meson-meson structures of $\eta_c \Upsilon$, $J/\psi \eta_b$ and $J/\psi \Upsilon$ in color-singlet channels are all unbound, and their calculated masses are 12431 MeV, 12503 MeV and 12565 MeV, respectively. Their hidden-color channels are around 12.86 GeV. As for the other excited channels in diquark-antidiquark and K-type structures, they lie in the energy range 12.75$-$12.90 GeV. The coupled-masses of each certain configurations are all above 12.6 GeV, and hence no bound state is found.

\begin{figure}[!t]
\includegraphics[clip, trim={3.0cm 1.0cm 3.0cm 1.0cm}, width=0.45\textwidth]{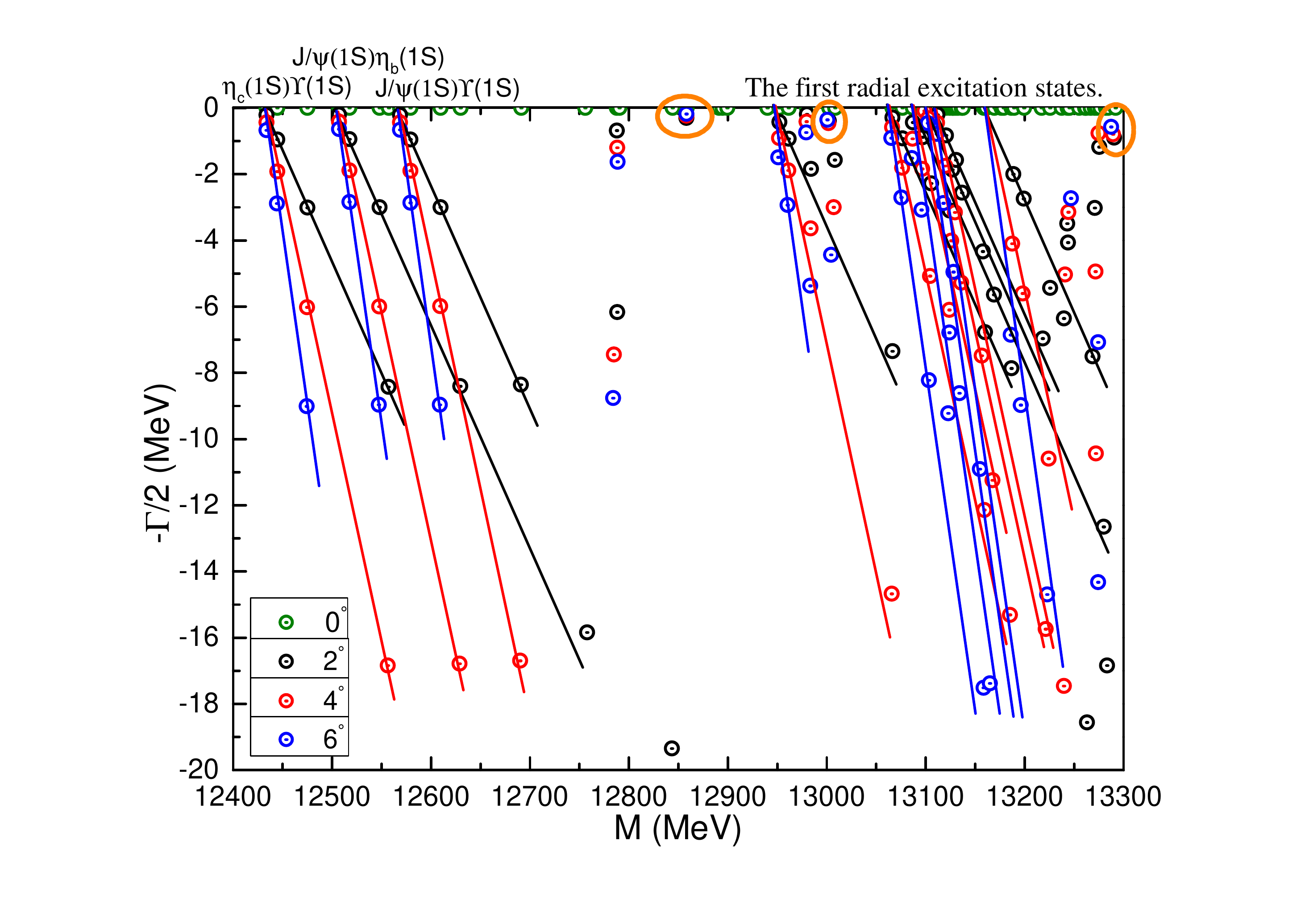} \\
\includegraphics[clip, trim={3.0cm 1.0cm 3.0cm 1.0cm}, width=0.45\textwidth]{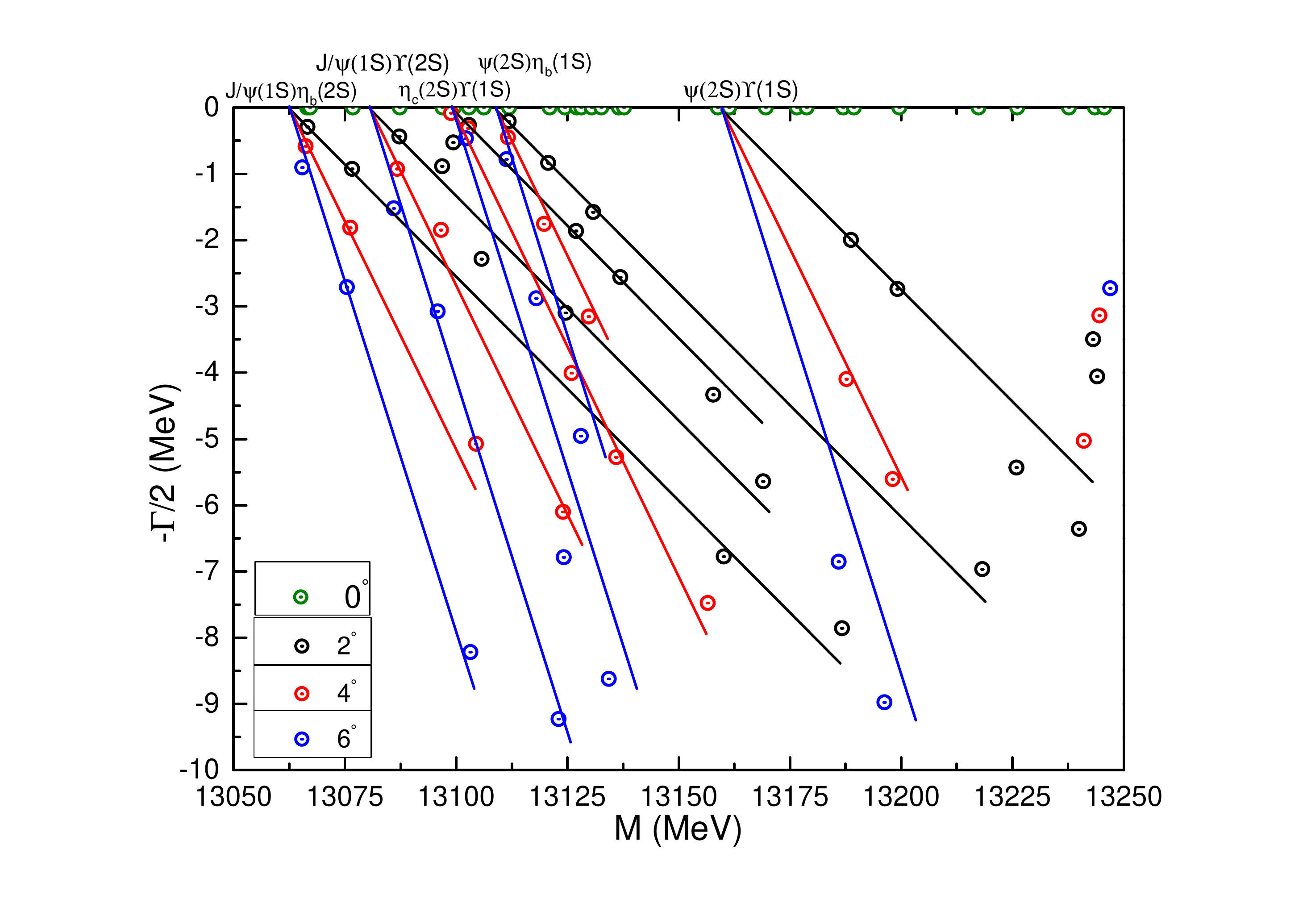}
\caption{\label{PP17} {\it Top panel:} Complex energy spectrum of the $\bar{c}c\bar{b}b$ system with $J^{P(C)}=1^{+(\pm)}$ from the complete coupled-channel calculation with CSM. The parameter $\theta$ varies from $0^\circ$ to $6^\circ$. {\it Bottom panel:} Enlarged top panel, with real values of energy ranging from $13.05\,\text{GeV}$ to $13.25\,\text{GeV}$.} 
\end{figure}

Nevertheless, three narrow resonances are obtained in a complete coupled-channels calculation by CSM, as shown in Fig.~\ref{PP17}. They are $J/\psi(1S)\Upsilon(1S)(12858)$, $\eta_c(1S)\Upsilon(2S)(13002)$ and $\psi(2S)\Upsilon(1S)(13290)$, with two-body strong decay widths 0.60 MeV, 0.94 MeV and 1.81 MeV, respectively. Additionally, an energy zoom shown in the bottom panel of Fig.~\ref{PP17} gives a more clear picture on the scattering nature for $J/\psi(1S)\eta_b(2S)$, $J/\psi(1S)\Upsilon(2S)$, $\eta_c(2S)\Upsilon(1S)$, $\psi(2S)\eta_b(1S)$ and $\psi(2S)\Upsilon(1S)$ and the fact that no resonance state is found.

\begin{table}[!t]
\caption{\label{GresultCC17} Calculation of the $J^{P(C)}=1^{+(\pm)}$ state of the $\bar{c}c\bar{b}b$ system in the real case ($\theta=0^\circ$). The first and second columns show all possible structures and channels. They are obtained by the necessary bases in spin and color degrees of freedom summarized in the third column. The values in the parentheses after the first three meson-meson structures give the theoretical results of the noninteracting meson-meson thresholds. The lowest energies of all channels without considering any channel-channel coupling are listed in the fourth column. The last column shows the lowest energies in a calculation taking into account the couplings among the channels belonging to each structure (configuration). The last row gives the lowest energy from the complete coupled-channel calculation. All results are in units of MeV.}
\begin{ruledtabular}
\begin{tabular}{lcccc}
~~Structure  & Channel & $\xi_1^{\alpha}$; $\varphi_\beta^{\phantom{\dag}}$ & $M$ & Mixed~~ \\
 & & $[\alpha;~\beta]$ &  &  \\[2ex]
~~$\eta_c \Upsilon (12431)$         & 1  & [1; 1] & $12431$ &~~ \\
~~$J/\psi \eta_b (12503)$         & 2  & [2; 1] & $12503$ &~~ \\
~~$J/\psi \Upsilon (12565)$        & 3 & [3; 1]   & $12565$ &$12431$~~ \\[2ex]
~~$\eta^8_c \Upsilon^8$         & 4  & [1; 2] & $12861$ &~~ \\
~~$J/\psi^8 \eta^8_b$         & 5  & [2; 2] & $12867$ &~~ \\
~~$J/\psi^8 \Upsilon^8$         & 6  & [3; 2]   & $12855$ &$12843$~~ \\[2ex]
~~$[cb]^{\bf 3_{\rm c}}[\bar{c}\bar{b}]^{\bf \bar{3}_{\rm c}}$   & 7  & [4; 3]  & $12888$ & ~~ \\
~~$[cb]^{\bf 3_{\rm c}}[\bar{c}\bar{b}]^{\bf \bar{3}_{\rm c}}$   & 8  & [6; 3]  & $12903$ & ~~ \\
~~$[cb]^{\bf 6_{\rm c}}[\bar{c}\bar{b}]^{\bf \bar{6}_{\rm c}}$   & 9  & [4; 4]  & $12858$ & ~~ \\
~~$[cb]^{\bf 6_{\rm c}}[\bar{c}\bar{b}]^{\bf \bar{6}_{\rm c}}$   & 10 & [6; 4]  & $12817$ & $12769$~~ \\[2ex]
~~K$_1$   & 11  & [7; 5]  & $12879$ & \\
                 & 12  & [8; 5]  & $12872$ & \\
                 & 13  & [9; 5]  & $12876$ & \\
                 & 14  & [7; 6]  & $12748$ & \\
                 & 15  & [8; 6]  & $12759$ & \\
                 & 16  & [9; 6]  & $12636$ & $12635$~~ \\[2ex]
~~K$_3$   & 17  & [13; 9]  & $12863$ & \\
                 & 18  & [14; 9]  & $12824$ & \\
                 & 19  & [15; 9]  & $12864$ & \\
                 & 20  & [13; 10]  & $12907$ & \\
                 & 21  & [14; 10]  & $12901$ & \\
                 & 22  & [15; 10]  & $12894$ & $12768$~~ \\[2ex]
\multicolumn{4}{l}{Lowest mass of the fully-coupled result:}    & $12431$~~  \\
\end{tabular}
\end{ruledtabular}
\end{table}

{\bf The $\bm{J^{P(C)}=2^{+(+)}}$ channel:} 8 channels characterize this case and they are listed in Table~\ref{GresultCC18}: one meson-meson structure, $J/\psi \Upsilon$, two diquark-antidiquark structures, $[cb]^{\bf 3_{\rm c}}[\bar{c}\bar{b}]^{\bf \bar{3}_{\rm c}}$ and $[cb]^{\bf 6_{\rm c}}[\bar{c}\bar{b}]^{\bf \bar{6}_{\rm c}}$, and four K-type configurations. The lowest computed mass, 12565 MeV, is at the theoretical threshold of $J/\psi \Upsilon$. The other excited channels locate within 12.77$-$12.93 GeV. Meanwhile, the coupled-channels effect is still too weak to form a bound state.

\begin{figure}[ht]
\includegraphics[clip, trim={3.0cm 1.0cm 2.0cm 1.0cm}, width=0.45\textwidth]{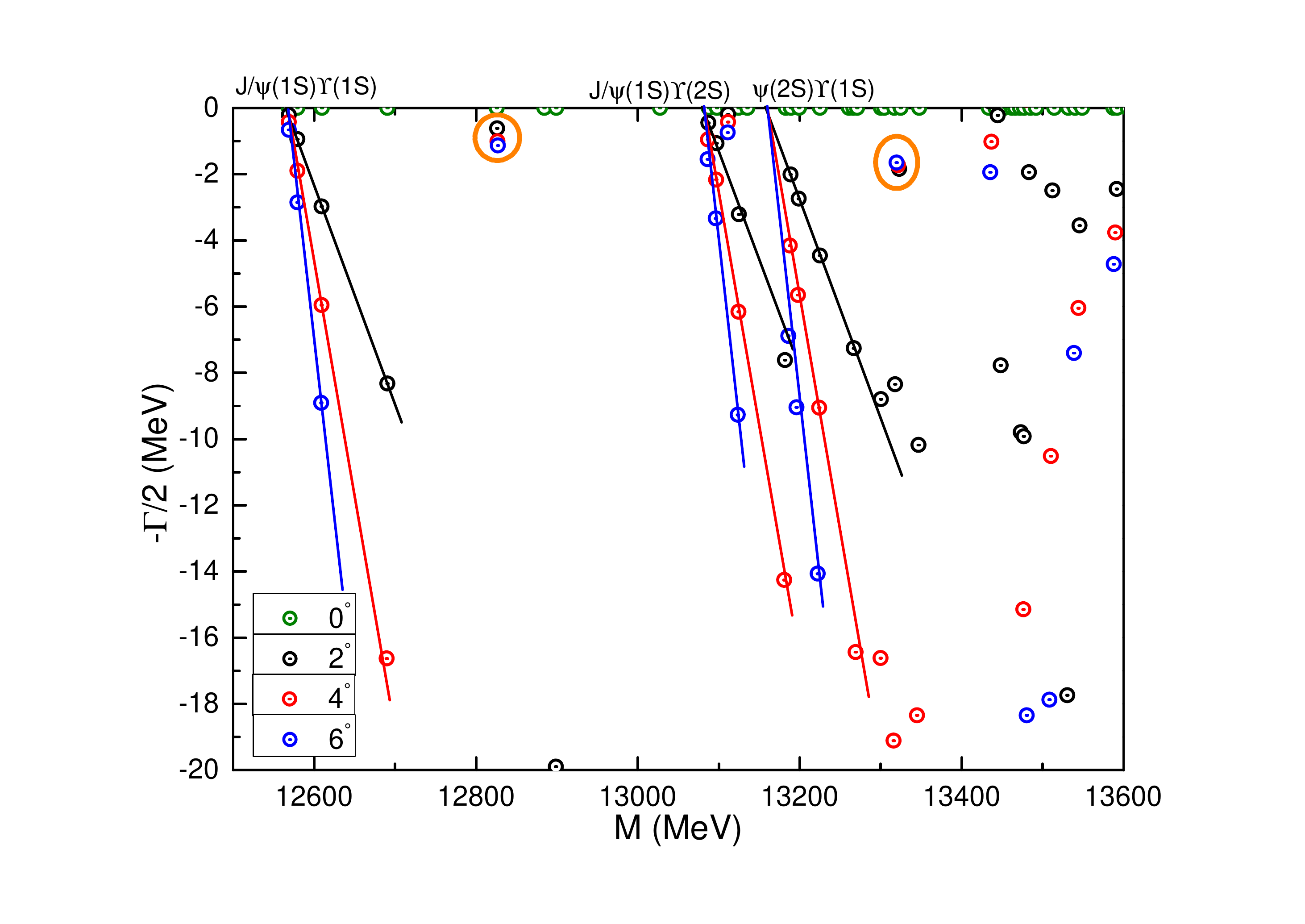} \\
\caption{Complex energy spectrum of the $\bar{c}c\bar{b}b$ system with $J^{P(C)}=2^{+(+)}$ from the complete coupled-channel calculation with CSM. The parameter $\theta$ varies from $0^\circ$ to $6^\circ$.} \label{PP18}
\end{figure}

However, two resonances are obtained in a fully coupled-channels investigation by CSM. In particular, Fig.~\ref{PP18} shows the distributions of scattering states of $J/\psi(1S)\Upsilon(1S)$, $J/\psi(1S)\Upsilon(2S)$ and $\psi(2S)\Upsilon(1S)$. Moreover, there are two stable poles marked with orange circles. One is identified as a $J/\psi(1S)\Upsilon(1S)(12826)$ resonance, and the other as a $\psi(2S)\Upsilon(1S)(13321)$ state, their widths are 1.25 MeV and 3.40 MeV, respectively.

\begin{table}[!t]
\caption{\label{GresultCC18} Calculation of the $J^{P(C)}=2^{+(+)}$ state of the $\bar{c}c\bar{b}b$ system in the real case ($\theta=0^\circ$). The first and second columns show all possible structures and channels. They are obtained by the necessary bases in spin and color degrees of freedom summarized in the third column. The value in the parenthesis after the first meson-meson structure gives the theoretical result of the noninteracting meson-meson threshold. The lowest energies of all channels without considering any channel-channel coupling are listed in the fourth column. The last column shows the lowest energies in a calculation taking into account the couplings among the channels belonging to each structure (configuration). The last row gives the lowest energy from the complete coupled-channel calculation. All results are in units of MeV.}
\begin{ruledtabular}
\begin{tabular}{lcccc}
~~Structure  & Channel & $\xi_2$; $\varphi_\beta^{\phantom{\dag}}$ & $M$ & Mixed~~ \\
 & & $[\beta]$ &  &  \\[2ex]
~~$J/\psi \Upsilon (12565)$       & 1  & 1  & $12565$ &$12565$~~  \\[2ex]
~~$J/\psi^8 \Upsilon^8$       & 2  & 2  & $12885$ &$12885$~~  \\[2ex]
~~$[cb]^{\bf 3_{\rm c}}[\bar{c}\bar{b}]^{\bf \bar{3}_{\rm c}}$   & 3  & 3   & $12926$  &~~\\
~~$[cb]^{\bf 6_{\rm c}}[\bar{c}\bar{b}]^{\bf \bar{6}_{\rm c}}$   & 4  & 4   & $12880$  &$12830$~~\\[2ex]
~~K$_1$   & 5  & 5  & $12900$ & \\
                 & 6  & 6  & $12771$ & $12771$~~ \\[2ex]
~~K$_3$   & 7 & 9  & $12887$ & \\
                 & 8 & 10  & $12933$ & $12844$~~ \\[2ex]
\multicolumn{4}{l}{Lowest mass of the fully-coupled result:}  & $12565$~~ \\
\end{tabular}
\end{ruledtabular}
\end{table}

\section{Summary}
\label{sec:summary}

The fully-heavy tetraquark system $QQ\bar{Q}\bar{Q}$ ($Q=c, b$) in its $S$-wave state, \emph{viz.} the $cc\bar{c}\bar{c}$, $bb\bar{b}\bar{b}$, $cb\bar{c}\bar{c}$, $cb\bar{b}\bar{b}$, $bb\bar{c}\bar{c}$ and $cb\bar{c}\bar{b}$ tetraquark lowest-lying states with quantum numbers $J^{P(C)}=0^{+(+)}$, $1^{+(\pm)}$, and $2^{+(+)}$, have been systemically investigated by using a potential model which mimics the lattice-QCD findings about the interaction between a heavy quark and antiquark pair. That is to say, the quark model involves the so-called Cornell potential supplemented by a spin-spin dependent interaction. Furthermore, the four-body problem is solved by a Gaussian expansion method where each relative motion is expanded by a Gaussian basis whose sizes are in geometric progression. The complex scaling method is used to identify bound, resonance and scattering states in the complex energy plane. And all four-body configurations, including meson-meson, diquark-antidiquark, and K-type structures, as well as the couplings among themselves, are comprehensively considered.
 
Several narrow resonances have been determined in any of the all-heavy tetraquark systems investigated. Tables~\ref{Rsum-cccc}, \ref{Rsum-bbbb}, \ref{Rsum-cccb}, \ref{Rsum-bbbc}, \ref{Rsum-cbcb} and \ref{Rsum-ccbb} summarize our findings. Amongst all the results presented, the following are of particular interest:
\begin{itemize}
\item For the narrow di-$J/\psi$ structure at 6.9 GeV recently seen by the LHCb collaboration, our investigation indicates that four resonances are consistent: $J/\psi(1S) \psi(2S) (6875)$ and $J/\psi(1S) \psi(2S) (6987)$ with $J^{P(C)}=0^{+(+)}$, $\eta_c(2S) J/\psi(1S) (6885)$ with $J^{P(C)}=1^{+(-)}$, and $J/\psi(1S) \psi(2S) (7007)$ with $J^{P(C)}=2^{+(+)}$.
\item Within the invariant-mass range 6.2$-$6.8 GeV where a broad structure was reported by the LHCb collaboration, our study provides four possible candidates. They are the resonances $\eta_c (1S)\eta_c (2S) (6640)$ and $\eta_c (1S)\eta_c (2S) (6762)$ with $J^{P(C)}=0^{+(+)}$ and the resonances $J/\psi (1S)J/\psi (1S)(6274)$ and $J/\psi (1S)J/\psi (1S) (6653)$ with $J^{P(C)}=1^{+(-)}$.
\item Related with the possible fully-heavy tetraquark structure around 7.2 GeV also reported by the LHCb collaboration, our investigation provides a possible candidate with mass $7195$ MeV and width $8$ MeV, in the $J/\psi(1S)\psi(2S)$ channel and with quantum numbers $J^{P(C)}=0^{+(+)}$.
%
%
%
\end{itemize}

The predicted resonances of the fully-heavy tetraquark systems can be investigated in future experiments such as the ATLAS, CMS, and LHCb at CERN. In particular, the CMS experiment has momentum and mass resolutions comparable in size with those of the LHCb; however, the CMS collaboration has already collected 20 times more data than the LHCb one. From a theoretical point of view, the fully-heavy tetraquark systems can be safely studied from a non-relativistic approximation and thus they constitute a nice platform to examine and improve the phenomenological potential model applied to multiquark systems.


\begin{acknowledgments}
G. Yang would like to thank J. Zhao, L. He and Q. Wang for constructive discussions. Work supported by: National Natural Science Foundation of China under grant nos. 11535005 and 11775118; Ministerio Espa\~nol de Ciencia e Innovaci\'on under grant no. PID2019-107844GB-C22; and Junta de Andaluc\'ia,  contract nos. P18-FRJ-1132 and Operativo FEDER Andaluc\'ia 2014-2020 UHU-1264517.
\end{acknowledgments}



\end{document}